\documentstyle[prd,aps]{revtex}

\begin{document} 
\draft

\title{Can the Universe Create Itself?}
\author{J. Richard Gott, III and Li-Xin Li}
\address{Department of
Astrophysical Sciences, Princeton University, Princeton, NJ 08544}
\date{December 29, 1997}
\maketitle

\begin{abstract}
The question of first-cause has troubled philosophers and cosmologists
alike. Now that it is apparent that our universe began in a Big Bang
explosion, the question of what happened before the Big Bang
arises. Inflation seems like a very promising answer, but as Borde and Vilenkin
have shown, the inflationary state preceding the Big Bang could not
have been infinite in duration --- it must have had a beginning
also. Where did it come from? Ultimately, the difficult question seems
to be how to make something out of nothing. This paper explores the
idea that this is the wrong question --- that {\em that} is not how the
Universe got here. Instead, we explore the idea of whether there is
anything in the laws of physics that would prevent the Universe from
creating itself. Because spacetimes can be curved and multiply
connected, general relativity allows for the possibility of closed
timelike curves (CTCs). Thus, tracing backwards in time through the original
inflationary state we may eventually encounter a region of CTCs 
--- giving {\em no} first-cause. This region of CTCs 
may well be over by now (being bounded toward the future
by a Cauchy horizon). We illustrate that such models --- with CTCs
--- are {\em not} necessarily inconsistent by demonstrating
self-consistent vacuums for Misner space and a multiply connected de
Sitter space in which the renormalized energy-momentum tensor 
does not diverge as one approaches the Cauchy horizon and solves
Einstein's equations. Some specific scenarios
(out of many possible ones) for this
type of model are described. For example: a metastable vacuum inflates
producing an infinite number of (Big-Bang-type) bubble universes. In
many of these, either by natural causes or by action of advanced
civilizations, a number of bubbles of metastable vacuum are created at
late times by high energy events. These bubbles will usually collapse
and form black holes, but occasionally one will tunnel to create an
expanding metastable vacuum (a baby universe) on the other side of the
black hole's Einstein-Rosen bridge as proposed by Farhi, Guth, and
Guven. One of the expanding metastable-vacuum baby universes produced
in this way simply turns out to be the original inflating metastable
vacuum we began with. We show that a Universe with CTCs can be stable
against vacuum polarization. And, it can be classically stable and
self-consistent if and only if the
potentials in this Universe are retarded --- which gives a natural
explanation of the arrow of time in our universe. Interestingly, the laws of physics
may allow the Universe to be its own mother.  
\end{abstract}

\pacs{PACS number(s): 98.80.-k}

\section{Introduction}
The question of first-cause has been troubling to philosophers and
scientists alike for over two thousand years. Aristotle found this
sufficiently troubling that he proposed avoiding it by having the
Universe exist eternally in both the past and future. That way, it was
always present and one would not have to ask what caused it to come
into being. This type of model has been attractive to modern scientists as
well. When Einstein developed general relativity and applied it to
cosmology, his first cosmological model was the Einstein static
universe, which had a static $S^3$ spatial geometry which lasted forever,
having no beginning and no end \cite{ein17}.

As we shall discuss, since the Big Bang model's success, models with a
finite beginning have taken precedence, even when inflation and quantum
tunneling are included. So the problem of first-cause reasserts
itself. The big question appears to be how to create the universe out
of nothing. In this paper we shall explore the idea that this is the
wrong question. A
remarkable property of general relativity is that it allows
solutions that have closed timelike curves (CTCs)
\cite{sto37,god49,tau51,new63,mis67,mor88,got91a}
(for review see \cite{tho93,vis95}). Often, the beginning of
the universe, as in Vilenkin's tunneling model \cite{vil82} and
Hartle and Hawking's no-boundary model \cite{har83},
is pictured as being like the
south pole of the earth and it is usually said that asking what
happened before that is like asking what is south of the south
pole \cite{haw88}. But, suppose the early universe contains a region of
CTCs. Then, asking what was the earliest point might be like
asking what is the easternmost point on the Earth. You can keep going
east around and around the Earth --- there is no eastern-most point. In
such a model every event in the early universe would have events that
preceded it. This period of CTCs could well have
ended by now, being bounded by a Cauchy horizon. Some initial calculations of vacuum
polarization in spacetimes with CTCs indicated that the renormalized
energy-momentum tensor diverged at the Cauchy
horizon separating the region with CTCs from the region without
closed causal curves, or at the polarized hypersurfaces
nested inside the Cauchy horizon
\cite{his82,fro91,kim91,kli92,gra93}. 
Some of these results motivated Hawking \cite{haw92a,haw92b}
to propose the chronology protection conjecture
which states that the laws of physics do not allow the appearance of
CTCs. But, a number of people have challenged the chronology protection
conjecture by giving counter-examples
\cite{kim91,tho93,bou92,li93,li94,low95,tan95,li96,kra96,sus97,vis97,li97}.
In particular, Li and Gott \cite{li97}
have recently found that there is a self-consistent vacuum in Misner
space for which the renormalized energy-momentum tensor of vacuum
polarization is zero everywhere. (Cassidy \cite{cas97a} has
independently given an existence proof that there should be a quantum state for a
conformally coupled scalar field in Misner space, for which the renormalized
energy-momentum tensor is zero everywhere, but he has not shown what
state it should be. Li and Gott \cite{li97} have found that it is the
``adapted'' Rindler vacuum.) 
In this paper we give some examples to
show how it is possible in principle to find self-consistent vacuum
states where the renormalized energy-momentum tensor does not blow up
as one approaches
the Cauchy horizon. To produce such a region of CTCs,
the universe must, at some later time, be able to reproduce conditions
as they were earlier, so that a multiply connected solution is
possible. Interestingly, inflation is well suited to this. A little
piece of inflationary state expands to produce a large volume of
inflationary state, little pieces of which resemble the starting
piece. Also there is the possibility of forming baby universes at late
times where new pieces of inflating states are formed. Farhi, Guth, and
Guven \cite{far90}, Harrison \cite{har95}, Smolin \cite{smo92a,smo97}, 
and Garriga and Vilenkin \cite{gar97} have
considered such models. If one of those later inflating pieces simply
turns out to be the inflating piece that one started out with, then
the Universe can be its own mother. Since an infinite number of
baby universes are created,  as
long as the probability of a particular multiple connection forming 
is not exactly zero, then such a connection
might be expected, eventually.
Then the Universe neither tunneled
from nothing, nor arose from a singularity; it created itself (Fig.~\ref{f1}).

Before discussing this approach to the first-cause problem, let us
review just how troublesome this problem has been. As we have noted,
Einstein \cite{ein17} initially tried to avoid it by siding with Aristotle in
proposing a model which had an infinite past and future.
The Einstein
static universe appears to be the geometry Einstein found {\em a priori} most
aesthetically appealing, thus presumably he started with this
preferred geometry and substituted it into the field equations to
determine the energy-momentum tensor required to produce it. He found a source
term that looks like dust (stars) plus a term that was proportional to
the metric which he called the cosmological constant. The cosmological constant,
because of its homogeneous large negative pressure, exerts a repulsive gravitational
effect offsetting the attraction of the stars for each other; allowing
a static model which could exist (ignoring instabilities, which he
failed to consider) to the infinite past and future. If
one did not  require a static model, there would be no need for the
cosmological constant. Friedmann \cite{fri22} calculated models without it, of
positive, negative or zero curvature, all of which were
dynamical.
When Hubble \cite{hub29} discovered the expansion of the universe,
Einstein pronounced the cosmological constant the biggest blunder of
his life. 

But now there was a problem: all three Friedmann models
($k=0$, $k=1$, and $k=-1$) that were expanding at the present epoch
had a beginning in the finite past (see e.g. \cite{wei72,haw73}). 
In the Friedmann models the
universe began in a singularly dense state at a finite time in the
past. The equations could not be pushed beyond that finite beginning
singularity. Furthermore, if today's Hubble constant is $H_0$, then all of
the Friedmann models had ages less than $t_H=H_0^{-1}$. The universe thus
began in a Big Bang explosion only a short time ago, a time which could
be measured in billions of years. The universe was not infinitely
old. Gamow \cite{gam48a,gam48b} and his colleagues Alpher and Herman
\cite{alp48} calculated the
evolution of such a Big Bang cosmology, concluding correctly that in
its early phases it should have been very dense and very hot, and that
the thermal radiation present in the early universe should still be
visible today as microwave radiation with a temperature of approximately
$5K$. Penzias and Wilson's discovery of the radiation with a
temperature of $2.7K$ \cite{pen65} cinched the case for the Big Bang model. The
COBE results which have shown a beautifully thermal spectrum
\cite{mat90,mat94} and small
fluctuations in the temperature $\delta T/T=10^{-5}$
\cite{smo92}, fluctuations that
are of approximately the right magnitude to grow into the galaxies and
clusters of galaxies that we see at the present epoch, have served to
make the Big Bang model even more certain. With the Big Bang model in
ascendancy, attention focused on the initial singularity. Hawking and
Penrose proved a number of singularity theorems \cite{haw67,haw70,haw73}
showing that, with
some reasonable constraints on the energy-momentum tensor, if
Einstein's equations are correct and the expansion of the universe is as
observed today, there is no way to
avoid an initial singularity in the model; that is, initial singularities
would form even in models that were not exactly uniform. So the
initial singularity was taken to be the first-cause of the
Universe. This of course prompted  questions of what caused the
singularity and what happened before the singularity. The standard answer
to what happened before the Big Bang singularity has been that time was
created at the singularity, along with space, and that there was
no time before the Big Bang. Asking what happened before the Big Bang
was considered to be like asking what is south of the south pole. But particularly
troublesome was the question of what caused the initial singularity to
have its almost perfect uniformity --- for otherwise the
microwave background radiation would be of vastly different
temperatures in different directions on the sky. Yet the initial
singularity could not be exactly uniform, for then we would have a perfect Friedmann
model with no fluctuations which would form no galaxies. It needed to
be almost, but not quite perfectly uniform --- a remarkable situation
--- how did it get that way? These seemed to be special initial
conditions with no explanation for how they got that way. 

Another problem was that singularities in physics are usually smeared by
quantum effects. 
As we extrapolated back toward the initial singularity
(of infinite density), we would first reach a surface where the
density was equal to the Planck density and at this epoch classical general
relativity would break down. We could not extrapolate confidently back
to infinite density, we could only say that we would eventually reach
a place where quantum effects should become important and where
classical general relativity no longer applied. Since we do not have a
theory of quantum gravity or a theory-of-everything we
could honestly say that the singularity theorems only told us that we
would find regions in the early universe where the density exceeded the
GUT or Planck densities beyond which we did not know what happened
--- rather much like the Terra Incognita of old maps. We could not
then say how our universe formed. 

So, questions about how the initial Big Bang singularity was formed and
what preceded it remained. The closed Friedmann model, popular because
it is compact and therefore needs no boundary conditions, re-collapses in
a finite time in the future to form a Big Crunch singularity at the
end. Singularity theorems tell us that in a collapsing universe the
final Big Crunch singularity cannot be avoided. Classical general
relativity tells us that a closed universe begins with a singularity
and ends with a singularity, with nothing before and nothing
after. Nevertheless, many people speculated that there could be more
than one connected cycle --- after all, the singularities only
indicated a breakdown of classical general relativity and the quantum
Terra Incognita at the Planck density might allow a cosmology
collapsing toward a Big Crunch to bounce and make another Big Bang
\cite{lem33,bek75,dur96}. In
support of this is the fact that de~Sitter space (representing the
geometry of a false vacuum --- an inflationary state as proposed by
Guth \cite{gut81} --- with a large
cosmological constant) looks like a spatially closed $S^3$ universe whose radius as a
function of proper time is
$a(t)=r_0\cosh({t/r_0})$, where $r_0=(3/\Lambda)^{1/2}$ is
the radius of the de~Sitter space and $\Lambda$ is the cosmological
constant (throughout the paper we use units $G=c=\hbar=k_{\rm B}=1$), 
which is a collapsing cosmology which bounces and turns into an
expanding one. Thus if quantum gravitational effects make the
geometry look like de~Sitter space once the density reaches the Planck
density as some have suggested \cite{fro90,bar96a,bar96b}, 
then a Big Crunch singularity might
be avoided as the closed universe bounced and began a Big Bang all
over again. This
bouncing model avoids the first-cause problem. The answer to what
caused our universe in this model is ``the collapse of the previous
universe'', and so on. An infinite number of expansion and contraction cycles
make up the Universe (note the capital U --- in this paper this
denotes the ensemble of causally connected universes) which consists
of an infinite number of closed Big Bang models laid out in time like
pearls on a string. The Universe (the
infinite string of pearls) has always been in existence and will
always be in existence, even though our cycle, our standard closed Big
Bang cosmology (our pearl) has a finite duration. So we are back to
Aristotle, with an eternal Universe, and close to Einstein with just
an oscillating (rather than static) closed Universe that has infinite
duration to the past and future. Thus in this picture there is no
first-cause because the Universe has existed infinitely far back in
the past.

The oscillating universe was thought to have some problems with
entropy \cite{tol34}. Entropy is steadily increasing with time, and so each cycle
would seem to be more disordered than the one that preceded
it. Since our universe has a finite entropy per baryon it was argued,
there could not be an infinite number of cycles preceding us. 
Likewise it was argued that each cycle of the universe
should be larger
than the preceding one, so if there were an infinite number preceding
us, our universe would have to look indistinguishable from flat (i.e., closed
but having an infinite radius of curvature). The real challenge in
this model is to produce initial conditions for our universe (our
pearl) that were as uniform and low entropy as observed. COBE tells us
that our universe  at early times was uniform to one part in a hundred
thousand \cite{smo92}. At late times we expect the universe at the Big Crunch to be
very non-uniform as black hole singularities combine to form the Big
Crunch. In the early universe the Weyl tensor is zero, whereas at the
Big Crunch it would be large \cite{pen79,pen89}. How
does the chaotic high-entropy state at the Big Crunch get recycled
into the low-entropy, nearly uniform, state of the next Big Bang? If it
does not, then after an infinite number of cycles, why are we not in a
universe with chaotic initial conditions?  

Entropy and the direction of time may be intimately tied up with this
difference between the Big Bang and the Big Crunch. Maxwell's equations
(and the field equations of general relativity) are time-symmetric, so
why do we see only retarded potentials? Wheeler and Feynman addressed
this with their absorber theory \cite{whe45}. They supposed that an electron shaken
today produces half-advanced-half-retarded fields. The half-advanced
fields propagate back in time toward the early universe where they are
absorbed (towards the past the universe is a perfect absorber) by
shaking charged particles in the early universe. These charged
particles in turn emit half-advanced-half-retarded fields; their
half-retarded fields propagate toward the future where they: (a)
perfectly cancel the half-advanced fields of the original electron,
(b) add to its retarded fields to produce the electron's full retarded
field, and (c) produce a force on the electron which is equal to the
classical radiative reaction force. Thus, the electron only experiences
forces due to fields from other charged particles. This is a
particularly ingenious solution. It requires only that the early
universe is opaque --- which it is --- and that the initial conditions are
low-entropy; that is, there is a cancelation of half-advanced fields from
the future by half-retarded fields from the past, leaving no
``signals'' in the early universe from later events --- a state of
low-entropy. (Note that this argument works equally well in an open
universe where the universe may not be optically thick toward the
future --- all that is required is that the universe be a perfect
absorber in the past, i.e., toward the state of low-entropy.) Wheeler
and Feynman noted that entropy is time-symmetric like Maxwell's
equations. If you find an ice cube on the stove, and then come back
and re-observe it a minute later, you will likely find it 
half-melted. Usually an ice cube gets on a stove by someone just putting it
there (initial conditions), but suppose we had a truly isolated system
so that the ice cube we found was just a statistical fluctuation.
Then if we asked what we would see if we had observed one minute {\em
before} our first observation, we will also be likely to see a
half-melted ice cube, for finding a still larger ice cube one minute
before would be unlikely because it would represent an even more
unlikely statistical fluctuation than the original ice cube. In an
{\em isolated} system, an (improbable) state of low-entropy is likely
to be both followed and preceded by states of higher-entropy in a
time-symmetric fashion. Given that the early universe represents a
state of high order, it is thus not surprising to find entropy
increasing after that. Thus, according to Wheeler and Feynman \cite{whe45}, the fact
that the retarded potentials arrow of time and the entropy
arrow of time point in the same direction is simply a reflection of
the low-entropy nature of the Big Bang. The Big Crunch is
high-entropy, so time follows from past to future between the Big Bang and
the Big Crunch.

Thus, in an oscillating universe scenario, we might expect entropy to
go in the opposite direction with respect to time, in the previous
cycle of oscillation. In that previous universe there would be 
only advanced potentials and observers there would sense a direction
of time opposite to ours (and would have a reversed definition of matter
and anti-matter because of CPT invariance). Thus the cycle previous to
us would, according to {\em our} definition of time, have advanced
potentials and would end with a uniform low-entropy Big Crunch and
begin with a chaotic high-entropy Big Bang (see Gott \cite{got74} for further
discussion). Thus, an infinite string of oscillating universes could
have alternating high and low-entropy singularities, with the
direction of the entropy (and causality --- via electromagnetic
potentials) time-reversing on each succeeding cycle. Every observer
using the entropy direction of time would see in his ``past'' a 
low-entropy singularity (which he would call a Big Bang) and in his
``future'' a high-entropy singularity (which he could call a Big
Crunch). Then the mystery is why the low-entropy Big Bangs exist ---
they now look improbable. An oscillating universe with chaotic bangs
and crunches and half-advanced-half-retarded potentials throughout
would seem more likely. At this point anthropic arguments \cite{car74} could be
brought in to say that only low-entropy Big Bangs might produce
intelligent observers and that, with an infinite number of universes in
the string, eventually there would be --- by chance --- a
sufficiently low-entropy Big Bang to produce intelligent
observers. Still, the uniformity of the early universe that we observe
seems to be more than that required to produce intelligent
observers, so we might wonder whether a random intelligent observer in
such a Universe would be expected to see initial conditions in his/her
Big Bang as uniform as ours. (Among intelligent observers, the
Copernican principle tells us that you should not expect to be
special. Out of all the places for intelligent observers to be there
are by definition only a few special places and many non-special
places, so you should expect to be in one of the many non-special
places \cite{got93}.)        

\section{Inflation as a Solution}
Guth's proposal of inflation  \cite{gut81} offered an 
explanation of why the initial
conditions in the Big Bang should be approximately, but not exactly
uniform. (For review of inflation see \cite{lin90,kol90}.) 
In the standard Big Bang cosmology this was always a puzzle
because antipodal points on the sky on the last scattering surface at
$1+z\simeq1000$ had not had time to be in communication with each
other. When we see two regions which are at the same temperature, the
usual explanation is that they have at some time in the past been in
causal communication and have reached thermal equilibrium with each
other. But there is not enough time to do this in the standard Big Bang model
where the expansion of the scale factor at early times is $a(t)\propto t^{1/2}$.
Grand unified theories (GUT) of particle physics suggest that at early times
there might have been a non-zero cosmological constant $\Lambda$,
which then decayed to the zero cosmological constant we see
today. This means that the early universe approximates de~Sitter space
with a radius $r_0=(3/\Lambda)^{1/2}$ whose expansion rate at late
times approaches $a(t)=r_0\exp({t/r_0})$.
Regions that start off very close together, and have time to thermally
equilibrate, end up very far apart. When they become separated by a
distance $r_0$, they effectively pass out of causal contact --- if
inflation were to continue forever, they would be beyond each other's
event horizons. But eventually the epoch of inflation ends, the energy
density of the cosmological constant is dumped into thermal radiation,
and the expansion then continues as $a(t)\propto t^{1/2}$ as in a
radiation-dominated Big Bang cosmology. As the regions slow their
expansion from each other, enough time elapses so that they are able
to interchange photons once again and they come back into effective
causal contact. As Bill Press once said, they say ``hello'',
``goodbye'', and ``hello again''. When they say ``hello again'' they
appear just like regions in a standard Big Bang cosmology that are
saying ``hello'' for the first time (i.e., are just coming within the
particle horizon) except that with inflation these regions are already
in thermal equilibrium with each other, because they have seen each
other in the past. Inflation also gives a natural explanation for why
the observed radius of curvature of the universe is so large
($a\geq cH_0^{-1}\simeq3000h^{-1}{\rm Mpc}$; here $H_0=100h$ km s$^{-1}$
Mpc$^{-1}$ is the Hubble constant). During the Big Bang phase,
as the universe expands,
the radius of the universe $a$ expands by the same factor as the
characteristic wavelength $\lambda$ of the microwave background
photons, so $a/\lambda= {\rm
costant}\geq e^{67}$. How should we explain this large observed
dimensionless number? Inflation makes this easy. The energy density
during the inflationary epoch is $\Lambda/8\pi$. Let $\lambda$ be the
characteristic wavelength of thermal radiation which would have that
density. Even if $a$ started out of the same order as $\lambda$, by
the end of the inflationary epoch $a\geq\lambda e^{67}$, providing that
the inflationary epoch lasts at least as long as $67r_0$, or $67$
$e$-folding times. At the end of the inflationary epoch when the
inflationary vacuum of density $\Lambda/8\pi$ decays and is converted
into an equivalent amount of thermal radiation, the wavelength of
that radiation will be $\lambda$ and the ratio of $a/\lambda$ is fixed
at a constant value which is a dimensionless constant $\geq e^{67}$,
retained as the universe continues to expand in the radiation
and matter-dominated epochs. Thus, even a short run of inflation, of
$67$ $e$-folding times or more, is sufficient to explain why the
universe is as large as it is observed to be.

Another success of inflation is that the
observed Zeldovich-Peebles-Yu-Harrison
fluctuation spectrum with index $n=1$ \cite{zel72,pee70,har70} has
been naturally predicted as the result of random quantum fluctuations 
\cite{bar83,gut82,haw82,sta82}. 
The inflationary power spectrum with CDM 
has been amazingly successful in explaining the qualitative
features of observed galaxy clustering
(cf.
\cite{bah83,lap86,got86a,got87,gel89,par90a,par90b,par91,got91,par92a}).
The amount of large scale power seen in the observations
suggests an inflationary CDM power spectrum with $0.2<\Omega h<0.3$
\cite{mad90,sau91,par92b,she95,vog94,got97}.

\section{Open Bubble Universes}
Gott \cite{got82} has shown how an open inflationary model might be
produced. The initial inflationary state approximates de~Sitter space,
which can be pictured by embedding it as the surface 
$W^2+X^2+Y^2+Z^2-V^2=r_0^2$
in a five-dimensional Minkowski space with metric
$ds^2=-dV^2+dW^2+dX^2+dY^2+dZ^2$ \cite{haw73,sch56}.
Slice de~Sitter space along surfaces of $V={\rm constant}$, then the
slices are three-spheres of positive curvature $W^2+X^2+Y^2+Z^2=a^2$
where $a^2=r_0^2+V^2$. If $t$ measures the proper time, 
then $V=r_0\sinh(t/r_0)$ and $a(t)=r_0\cosh(t/r_0)$. This is
a closed universe that contracts then re-expands --- at late times
expanding exponentially as a function of proper time. If slices of
$V+X={\rm constant}$ are chosen, the slices have a flat geometry and
the expansion is exponential with $a(t)=r_0\exp(t/r_0)$. If the slices are
vertical ($W={\rm constant}>r_0$), then the intersection with the
surface is $H^3$, a hyperboloid $X^2+Y^2+Z^2-V^2=-a^2$
living in a Minkowski space, where 
$a^2=W^2-r_0^2$. This is a negatively curved surface with
a radius of curvature $a$. Let $t$ be the proper time from the event E
($W=r_0,X=0,Y=0,Z=0,V=0$) in the de~Sitter space. Then the entire future
of E can be described as an open $k=-1$ cosmology where
$a(t)=r_0\sinh(t/r_0)$. At early times, $t\ll r_0$, near E,
$a(t)\propto t$, and the model resembles a Milne cosmology
\cite{mil32}, but at late times the model expands
exponentially with time as expected for inflation. This is a
negatively curved (open) Friedmann model with a cosmological constant and nothing
else. Note that the entire negatively curved hyperboloid ($H^3$), which
extends to infinity, is nevertheless causally connected because all
points on it have the event E in their past light cone. Thus, the
universe should have a microwave background that is isotropic, except
for small quantum fluctuations. At a proper time $\tau_1$ after the
event E, the cosmological constant would decay leaving us with a
hot Big Bang open ($k=-1$) cosmology with a
radius of curvature of $a=r_0\sinh(\tau_1/r_0)$ at the end of the
inflationary epoch. If $\tau_1=67r_0$, then $\Omega$ is a few tenths
today; if $\tau_1\gg67r_0$, then $\Omega\simeq 1$ today \cite{got82}. 

Gott \cite{got82} noted
that this solution looks just like the interior of a Coleman
bubble \cite{col77}. Coleman and de~Luccia \cite{col80} showed that if a metastable
symmetric vacuum (with the Higgs field $\phi=0$), with positive
cosmological constant $\Lambda$ were to decay by tunneling directly
through a barrier to reach the current vacuum with a zero cosmological
constant (where the Higgs field $\phi=\phi_0$), then it would do this
by forming a bubble of low-density vacuum of radius $\sigma$ around an
event E. The pressure inside the bubble is zero while the pressure
outside is negative (equal to $-\Lambda/8\pi$), so the bubble wall
accelerates outward, forming in spacetime a hyperboloid of one sheet
(a slice of de~Sitter space with $W={\rm constant}<r_0$). This bubble wall surrounds and is
asymptotic to the future light cone of E. If the tunneling is direct,
the space inside the bubble is Minkowski space (like a slice $W={\rm
constant}<r_0$ in the embedding space, which is flat). The inside of
the future light cone of E thus looks like a Milne cosmology with
$\Omega=0$ and $a(t)=t$. Gott \cite{got82} noted that what was needed to
produce a realistic open model with $\Omega$ of a few tenths today was to
have the inflation continue inside the bubble for about $67$
$e$-folding times. Thus, our universe was one of the bubbles and this
solved the problem of Guth's inflation that in general one expected the bubbles
not to percolate \cite{haw82a,gut83}. But, from inside one of the bubbles, our view could be
isotropic \cite{got82}. 

It was not long before a concrete mechanism to
produce such continued inflation inside the bubble was proposed. A
couple of weeks after Gott's paper appeared Linde's \cite{lin82} proposal of
new inflation appeared, followed shortly by Albrecht and
Steinhardt
\cite{alb82}. They proposed that the Higgs vacuum potential
$V(\phi)$ had a local
minimum at $\phi=0$ where $V(0)=\Lambda/8\pi$. Then there was a
barrier at $\phi=\phi_1$, followed by a long flat plateau
from $\phi_1$ to $\phi_0$ where it drops precipitately to zero at
$\phi_0$. The relation of this to the open bubble universe's geometry is outlined by
Gott \cite{got86} (see Fig.~1 and Fig.~2 in \cite{got86}). 
The de~Sitter space outside the bubble
wall has $\phi=0$. Between the bubble wall, at a spacelike separation
$\sigma$ from the event E, and the end of the inflation at the
hyperboloid $H^3$ which is the set of points at a future timelike separation
of $\tau_1$ from E, the Higgs field is between $\phi_1$ and $\phi_0$, and
$\tau_1$ is the time it takes the field (after tunneling) to roll
along the long plateau [where $V(\phi)$ is approximately equal
to $\Lambda/8\pi$ and the geometry is approximately de
Sitter]. After that epoch, $\phi=\phi_0$ where the energy density has
been dumped into thermal radiation and the vacuum density is zero
(i.e., a standard open Big Bang model). In order that inflation proceeds
and the bubbles do not percolate, it is required that the probability
of forming a bubble in de~Sitter space per four volume $r_0^4$ is
$\epsilon<\epsilon_{\rm cr}$ where $5.8\times10^{-9}<\epsilon_{\rm
cr}<0.24$ \cite{gut83}. In order that there be a greater
than $5\%$ chance that no bubble should have collided with our bubble
by now, so as to be visible in our past light cone, $\epsilon<0.01$ for
$\Omega=0.4$, $\Lambda=0$, $h=0.63$ today \cite{got97}, but this is no problem since we
expect tunneling probabilities through a barrier to be exponentially
small. This model has an event horizon, which is the
future light cone of an event E$^\prime$ ($W=-r_0, X=0,Y=0,Z=0,V=0$) which is
antipodal to E. Light from events within the future light cone of E$^\prime$
never reaches events inside the future light cone of E. So we are
surrounded by an event horizon. This produces Hawking radiation;
and, if $r_0$ is of order the Planck length, then the Gibbons-Hawking
thermal state \cite{gib77} (which looks like a cosmological constant
due to the trace anomaly \cite{pag82}) should be dynamically important
\cite{got82}.

If we observe $\Omega<1$ and $\Omega_\Lambda=0$, then $k=-1$ and we
need inflation more than ever --- we still need it to explain
the isotropy of the microwave background radiation and we would now
have a large but {\em finite} radius of curvature to explain, which $67$
$e$-folds of inflation could naturally produce. When Gott told this to
Linde in 1982, Linde said, yes, if we found that $\Omega<1$, he would
still have to believe in inflation but he would have a headache in the
morning! Why? Because one has to produce a particular amount of
inflation, approximately $67$ $e$-folds. If there were $670$ $e$-folds
or $670$ million $e$-folds, then $\Omega$ currently would be only
slightly less than 1. So there would be what is called a ``fine tuning
of parameters'' needed to produce the observed results.

The single-bubble open inflationary model \cite{got82} discussed above
has recently come back into fashion because of a number of
important developments. On the theoretical side, 
Ratra and Peebles \cite{rat94,rat95a}
have shown how to calculate quantum
fluctuations in the $H^3$ hyperbolic geometry with
$a(t)=r_0\sinh(t/r_0)$ during the inflationary epoch inside the 
bubble in the single bubble model. 
This allows predictions of fluctuations in the microwave
background. Bucher, Goldhaber, and Turok \cite{buc95a,buc95b} have
extended these
calculations, as well as Yamamoto, Sasaki and Tanaka
\cite{yam95}. Importantly, they have explained \cite{buc95a,buc95b} 
that the fine tuning in these
models is only ``logarithmic'' and, therefore, not so serious. Linde
and Mezhlmian \cite{lin95a,lin95b}
have shown how there are reasonable potentials which could produce such
bubble universes with different values of $\Omega$. In a standard
chaotic inflationary potential $V(\phi)$ \cite{lin83}, one could simply build in a
bump, so that one would randomly walk to the top of the curve via
quantum fluctuations and then roll down till one lodged behind the
bump in a metastable local minimum. One would then tunnel through the
bump, forming bubbles that would roll down to the bottom in a time
$\tau_1$. One could have a two-dimensional potential 
$V(\phi,\sigma)={1\over2}g^2\phi^2\sigma^2+V(\sigma)$,
where $g$ is a constant and there is a metastable trough at $\sigma=0$
with altitude $V(\phi,0)=\Lambda/8\pi$ with a barrier on both sides,
but one could tunnel through the barrier to reach $\sigma>0$ where
$V(\phi,\sigma)$ has a true minimum, and at fixed $\sigma$, is
proportional to $\phi^2$ \cite{lin95a,lin95b}. 
Then individual bubbles could tunnel across
the barrier at different values of $\phi$, and hence have different
roll-down times $\tau_1$ and thus different values of
$\Omega$. With a myriad of open universes being created, anthropic
arguments \cite{car74} come into play and if shorter roll-down
times were more probable than large ones, we might not be
surprised to find ourselves in a model which had $\Omega$ of a few
tenths, since if $\Omega$ is too small, no galaxies will form \cite{got75}.

A second reason for the renaissance of these open inflationary models is
the observational data. A number of recent estimates of $h$ (the
present Hubble constant in units of 100 km s$^{-1}$ Mpc$^{-1}$) have been
made (i.e., $h=0.65\pm0.06$ \cite{rie95}, $0.68
\leq h\leq 0.77$ \cite{mou96}, $0.55\leq h\leq0.61$
\cite{san96}, and $h=0.64\pm0.06$ \cite{kun97}). Ages of globular cluster
stars have a $2\sigma$ lower limit of about 11.6 billion years \cite{bol95}, we
require $h<0.56$ if $\Omega=1$, but a more acceptable $h<0.65$ if
$\Omega=0.4$, $\Omega_\Lambda=0$. Models with low $\Omega$ but
$\Omega+\Omega_\Lambda=1$ are also acceptable. Also, studies of large
scale structure have shown that with the inflationary CDM power
spectrum, the standard $\Omega=1$, $h=0.5$ model simply does not have
enough power at large scales. A variety of observational samples and
methods have suggested this: counts in cells, angular covariance
function on the sky, power spectrum analysis of 3D samples, and
finally topological analysis, all showing that $0.2<\Omega h<0.3$
\cite{par92a,mad90,sau91,par92b,she95,vog94,got97}. If $h>0.55$ this implies
$\Omega<0.55$, which also agrees with what one would deduce from the
age argument as well as the measured masses in groups and clusters of
galaxies \cite{got77}. With the COBE normalization there is also the problem that
with $\Omega=1$, $(\delta M/M)_{8h^{-1}{\rm Mpc}}=1.1-1.5$ and this
would require galaxies to be anti-biased [since for galaxies  $(\delta
M/M)_{8h^{-1}{\rm Mpc}}=1$] and would also lead to an excess of
large-separation gravitational lenses over those observed
\cite{cen94}. These things have forced even enthusiasts of $k=0$ models to
move to models with $\Omega<1$ and a cosmological constant so that
$\Omega+\Omega_{\Lambda}=1$ and $k=0$ \cite{ost95}. They then have to
explain the small ratio of the cosmological constant to the Planck
density ($10^{-120}$). Currently we do not have such a natural
explanation for a small yet finite $\Lambda$ as inflation naturally
provides for explaining why the radius of curvature should be a big
number in the $k=-1$ case.

Turner \cite{tur90} and Fukugita, Futamase, and Kasai \cite{fuk90} showed that a
flat $\Omega_{\Lambda}=1$ model produces about 10 times as many
gravitational lenses as a flat model with $\Omega=1$, and Kochanek
\cite{koc96} was able to set a $95\%$ confidence lower limit of
$0.34<\Omega$ in flat models where $\Omega+\Omega_{\Lambda}=1$, and a
$90\%$ confidence lower limit $0.15<\Omega$ in open models with
$\Omega_{\Lambda}=0$. Thus, extreme-$\Lambda$ dominated models are ruled
out by producing too many gravitational lenses.

Data on cosmic microwave background fluctuations for 
spherical harmonic modes from $l=2$
to $l=500$ will provide a strong test of these models. With
$\Omega_Bh^2=0.0125$, the $\Omega=1$, $\Omega_{\Lambda}=0$ model power
spectrum reaches its peak value at $l=200$; an $\Omega=0.3$,
$\Omega_{\Lambda}=0.7$  model reaches its peak value also at $l=200$
\cite{rat95c}; while an $\Omega=0.4$, $\Omega_{\Lambda}=0$ model reaches
its peak value at $l=350$ \cite{rat95b}. This should be decided
by the MAP and PLANCK
satellites which will measure this range with high accuracy \cite{par97}.

For the rest of this paper we shall usually assume single-bubble open inflationary
models for our Big Bang universe (while recognizing that chaotic
inflationary models and models with multiple epochs of inflation 
are also possible; it is interesting to note that Penrose also prefers
an open universe from the point of view of the complex-holomorphic
ideology of his twister theory \cite{haw96}). If the inflation within the
bubble is of order 67 $e$-folds, then we can have $\Omega$ of a few
tenths; but if it is longer than that, we will usually see $\Omega$ near 1
today. In any case, we will be assuming an initial metastable vacuum
which decays by forming bubbles through barrier penetration. The
bubble formation rate per unit four volume $r_0^4$ is thus expected to
be exponentially small so the bubbles do not percolate. Inflation is
thus eternal to the future \cite{vil83,sta86,lin86,gon87}. 
Borde and Vilenkin have proved that if the Universe were
infinitely old (i.e., if
the de~Sitter space were complete) then the bubbles would percolate
immediately and inflation would never get started (see \cite{bor94,bor96}
and references cited therein). Recall that a
complete de~Sitter space may be covered with an $S^3$ coordinate
system (a $k=1$ cosmology) whose radius varies as
$a(t)=r_0\cosh(t/r_0)$ so that for early times ($t<0$) the universe
would be contracting and bubbles would quickly collide preventing the
inflation from ever reaching $t=0$. Thus Borde and Vilenkin have proved that in the
inflationary scenario the universe must have a beginning. If it starts
with a three-sphere of radius $r_0$ at time $t=0$, and after that
expands like $a(t)=r_0\cosh(t/r_0)$, the bubbles do not percolate
(given that the bubble formation rate per four volume $r_0^4$ is $\epsilon\ll
1$) and the inflation continues eternally to $t=\infty$ producing an
infinite number of open bubble universes.
Since the number of bubbles forming
increases exponentially with time without limit, our universe is expected to form
at a finite but arbitrarily large time after the beginning of the
inflationary state. In this picture our universe (our bubble) is
only 12 billion years old, but the Universe as a whole (the entire
bubble forming inflationary state) is of a finite but arbitrarily old
age. 

\section{Vilenkin's Tunneling Universe and Hartle-Hawking's
No-Boundary Proposal}
But how to produce that initial spherical $S^3$ universe? Vilenkin \cite{vil82}
suggested that it could be formed from quantum tunneling. 
Consider the embedding diagram for de~Sitter space. De~Sitter
space can be embedded as the surface
$W^2+X^2+Y^2+Z^2-V^2=r_0^2$ in a five-dimensional Minkowski space with
metric $ds^2=-dV^2+dW^2+dX^2+dY^2+dZ^2$. This can be seen as an $S^3$
cosmology with radius $a(t)=r_0\cosh(t/r_0)$ where $V=r_0\sinh(t/r_0)$
and $a^2=W^2+X^2+Y^2+Z^2$ gives the geometry of $S^3$. This
solution represents a classical trajectory with a turning point at
$a=r_0$. But just as it reaches this turning point it could tunnel to
$a=0$ where the trajectory may be shown as a hemisphere of the Euclidean four-sphere
$W^2+X^2+Y^2+Z^2+V^2=r_0^2$ embedded in a flat Euclidean space
with the metric $ds^2=dV^2+dW^2+dX^2+dY^2+dZ^2$ and $a(t_E)=r_0\cos(t_E/r_0)$
where $a^2=W^2+X^2+Y^2+Z^2$ and $V=r_0\sin(t_E/r_0)$. The
time-reversed 
version of this process would show tunneling from a point at
$(V=-r_0, W=0, X=0, Y=0, Z=0)$ to a three sphere at $V=0$ of radius
$r_0$ which then
expands with proper time like $a(t)=r_0\cosh(t/r_0)$ giving a normal
de~Sitter space --- thus Vilenkin's universe created from nothing
is obtained \cite{vil82}.

Hawking has noted that in this
case, in Hartle and Hawking's formulation,
the point $(V=-r_0, W=0, X=0, Y=0, Z=0)$ is not special, the
curvature does not blow up there: it is like other points in the
Euclidean hemispherical section \cite{haw88}. 
However, this point is still the
earliest point in Euclidean time since it is at the center of the
hemisphere specified by the Euclidean boundary at $V=0$. 
So the beginning point in the Vilenkin model is indeed like
the south pole of the Earth \cite{haw88}.

Vilenkin's tunneling universe was based on an
analogy between quantum creation of universes and tunneling in
ordinary quantum mechanics \cite{vil82}. In ordinary quantum mechanics, a particle
bounded in a well surrounded by a barrier has a finite probability to
tunnel through the barrier to the outside if the height of the barrier
is finite (as in the $\alpha$-decay of radioactive nuclei 
\cite{gam28a,gam28b,gur29}). The wave
function outside the barrier is an outgoing wave, the wave function in
the well is the superposition of an outgoing wave and an ingoing wave
which is the reflection of the outgoing wave by the barrier. Due to
the conservation of current, there is a net outgoing current in the
well. The probability for
the particle staying in the well is much greater than the probability for
the particle running out of the barrier. The energy of the particle in
the well {\em cannot} be zero, otherwise the uncertainty principle is
violated. Thus there is always a finite zero-point-energy. 
The Vilenkin universe was supposed to be created from
``nothing'', where according to Vilenkin 
``nothing'' means ``a state with no classical
spacetime'' \cite{vil84}. Thus this is essentially different from
tunneling in ordinary quantum mechanics since in ordinary quantum
mechanics tunneling always takes place from one classically allowed
region to another classically allowed region where the current and the
probability are conserved. But creation from ``nothing'' is supposed
to take place
from a classically forbidden (Euclidean) region to a classically
allowed (Lorentzian) region, so the conservation of current is obviously
violated. Vilenkin obtained his tunneling universe by choosing a so-called
``tunneling boundary condition'' for the Wheeler-DeWitt equation 
\cite{vil84,vil88}. His
``tunneling from nothing'' boundary condition demands that when the
universe is big ($a^2\Lambda/3>1$ where $\Lambda$ is the cosmological
constant and $a$ is the scale factor of the universe) there is only
an outgoing wave in the superspace \cite{vil84,vil88}. If the probability
and current are conserved (in fact there does exist a
conserved current for the Wheeler-DeWitt equation \cite{hal91}, and a
classically allowed solution with $a=0$ and zero ``energy''),
there must be a finite probability for the universe being in the state
before tunneling (i.e., $a=0$) 
and this probability is much bigger than the probability for
tunneling. This implies that there must be ``something'' instead of
``nothing'' before tunneling. This becomes more clear if matter fields are
included in considering the creation of universes. In the case of a
cosmological constant $\Lambda$ 
and a conformally coupled scalar field $\phi$ (conformal fields are
interesting not only for their simplicity but also because
electromagnetic fields are conformally invariant) as the source terms in
Einstein's equations,
in the mini-superspace model (where the configurations are the
scale factor $a$ of the $S^3$ Robertson-Walker metric and a homogeneous
conformally coupled scalar field $\phi$) the
Wheeler-DeWitt equation separates \cite{har83,haw84}
\begin{eqnarray}
   {1\over2}\left(-{d^2\over d\chi^2}+\chi^2\right)\Phi(\chi)=E\Phi(\chi),
   \label{EA1a}
\end{eqnarray}
\begin{eqnarray}
   {1\over2}\left[-{1\over a^p}{d\over da}\left(a^p{d\over da}\right)
   +\left(a^2-
   {\Lambda\over3}a^4\right)\right]\Psi(a)=E\Psi(a),
   \label{EA1}
\end{eqnarray}
where $\Psi(a)\Phi(\chi)$ is the wave function of the universe
[$\chi\equiv(4\pi/3)^{1/2}\phi a$], $E$ is the ``energy level'' of the
conformally coupled scalar field , (we use quotes
because for radiation the conserved quantity is $E=4\pi\rho a^4/3$ instead of the
energy $4\pi\rho a^3/3$ where $\rho$ is the energy density), 
and $p$ is a constant determining the
operator ordering. Eq.~(\ref{EA1a}) is just the Schr\"{o}dinger
equation of a harmonic oscillator with unit mass and unit frequency
and energy $E$, the eigenvalues of $E$ are $n+{1\over2}$ where $n=0,1,2,...$
Eq.~(\ref{EA1}) is equivalent to the Schr\"{o}dinger equation for a unit
mass particle with total energy $E=n+{1\over2}$ in the one-dimensional
potential
\begin{eqnarray}
   U(a)={1\over2}\left(a^2-{\Lambda\over3}a^4\right).
   \label{EA2}
\end{eqnarray}
It is clear that in the case of $n<{1\over2}({3\over4\Lambda}-1)$,
there exist one classically forbidden region $a_1<a<a_2$ and two classically
allowed regions $0\leq a<a_1$ and $a>a_2$ where
$a^2_{1,2}\equiv{3\over2\Lambda}\left[1\mp\sqrt{1-{4\over3}(2n+1)\Lambda}
\right]$ (Fig.~\ref{f2}).
Because $U(a)$ is regular at $a=0$, we expect that the wave function
$\Psi(a)$ is also regular at $a=0$. If $\Lambda\ll 1$ and the
conformally coupled scalar field is in the ground state with $n=0$, we
have $a_1\simeq 1$, $a_2\simeq(3/\Lambda)^{1/2}$ and the potential in
region $0\leq a<a_1$ is $U(a)\simeq{1\over2}a^2$ like a
harmonic oscillator. The quantum behavior of the universe in region
$0\leq a<a_1$ is like a quantum harmonic oscillator. This may describe
a quantum oscillating (Lorentzian) universe without Big Bang or Big
Crunch singularities, which has a finite (but small) probability
[$\simeq\exp(-1/\Lambda)$] to tunnel through the barrier to form a de
Sitter-type inflating universe. The existence of this tiny oscillating
universe is due to the existence of a finite ``zero-point-energy'' ($1/2$) of a
conformally coupled scalar field and this ``zero-point-energy'' is 
required by the uncertainty principle. Since a conformally
coupled scalar field has an equation of state like that of radiation, the
Friedmann equation for $k=+1$ is
\begin{eqnarray}
   \left({da\over dt}\right)^2={C\over a^2}+{\Lambda\over3}a^2-1,
   \label{EA3}
\end{eqnarray}
where $C=8\pi\rho a^4/3={\rm constant}$ and $\rho$ is the energy
density of the conformally coupled scalar field. Eq.~(\ref{EA3}) is
equivalent to the energy-conservation equation for a classical unit
mass particle with zero total energy moving in the potential
\begin{eqnarray}
    V(a)={1\over2}\left(1-{\Lambda\over3}a^2-{C\over a^2}\right).
    \label{EA4}
\end{eqnarray}
The difference between $U(a)$ and $V(a)$ is caused by the fact that in
the integral of action the volume element contains a factor $a^3$ which is
also varied when one makes the variation to obtain the dynamical
equations. The potential $V(a)$ is singular at $a=0$ and near $a=0$ we
have $V(a)\simeq-{C\over2a^2}$. For $\Lambda\ll 1$ and $n=0$ (we take
$C=2E=2n+1$), the classical universe in region $0\leq a< a_1$ is
radiation dominated. This universe expands from a Big Bang
singularity, reaches a maximum radius, then re-collapses to a Big Crunch
singularity: $a=0$ is a singularity in the classical picture. But from the
above discussion, the Wheeler-DeWitt equation gives a regular wave
function at $a=0$. In such a case near $a=0$ the quantum behavior of the
universe is different from classical behavior.
This implies that, near $a=0$, classical general
relativity breaks down and quantum gravity may remove
singularities. This case is like that of a hydrogen atom where the
classical instability (according to classical electrodynamics, an
electron around a hydrogen nucleus will fall into the nucleus due to
electromagnetic radiation) is cured by quantum mechanics. Anyway,
it is {\em not} nothing at $a=0$. There is a small classically
allowed, oscillating,
radiation dominated, closed, quantum (by ``quantum'' we mean that its quantum
behavior deviates significantly from its classical behavior)
Friedmann universe near $a=0$, which
has a small probability to tunnel through the barrier to form an
inflationary universe. (If $\Lambda>0.75$ there is no classically
forbidden region and thus no tunneling.) 

So in this model the universe did not come from a point (nothing) but
from a tiny classically allowed,
oscillating, quantum Friedmann universe whose radius is of order the
Planck 
magnitude. But where did this oscillating universe come from? Because
it has a finite probability to tunnel (each time it reaches maximum
radius) to a de~Sitter space, it has a finite ``half-life'' for
decay into the de~Sitter phase and cannot last forever. It could, of
course, originate by tunneling from a collapsing de~Sitter phase (the
time-reversed version of the creation of a de~Sitter state from the
oscillating state), but then we are back where we started. In fact,
starting with a collapsing de~Sitter phase one is more likely to
obtain an expanding de~Sitter phase by simply re-expanding at the
classical turning point rather than tunneling into and then out of
the tiny oscillating universe state. An
alternative might be to have the original tiny oscillating universe
created via a quantum fluctuation (since it has just the
``zero-point-energy'') but here we are basically returning to the idea of
Tryon \cite{try73} that
you could get an entire 
Friedmann universe of any size directly via quantum fluctuation.
But quantum fluctuation of what? You have to have laws
of physics and a potential etc. 

Hartle and Hawking \cite{har83} made their no-boundary proposal and obtained a
model of the universe similar to Vilenkin's tunneling universe.
The no-boundary proposal is expressed in terms of a
Euclidean path integral of the wave function of the universe
\begin{eqnarray}
   \Psi(h_{ab},\phi_1,\partial M)=\sum_M\int{\cal D}g_{ab}{\cal D}\phi
   \exp[-I(g_{ab}, \phi, M)],
   \label{EA5}
\end{eqnarray}
where the summation is over compact manifolds $M$ with the prescribed
boundary $\partial M$ (being a compact three-manifold representing the
shape of the universe at a given epoch) as the {\em only} boundary;
$g_{ab}$ is the Euclidean metric on the
manifold $M$ with induced three-metric $h_{ab}$ on $\partial M$,
$\phi$ is the matter field with induced value $\phi_1$ on $\partial
M$; $I$ is the Euclidean action obtained from the Lorentzian action
$S$ via Wick rotation: $I=-iS(t\rightarrow-i\tau)$. In the
mini-superspace model the configuration space is taken to include the
$k=+1$ Robertson-Walker metric and a homogeneous matter field. In the
WKB approximation the wave function is (up to a normalization factor)
\begin{eqnarray}
   \Psi\simeq\sum_MB_M\exp[-I_{\rm cl}(g_{ab}, \phi,M)],
   \label{EA5a}
\end{eqnarray}
where $I_{\rm cl}$ is the Euclidean action for the solutions of the
Euclidean field equations (Einstein's equations and matter field equations).
The factor $B_M$ is the determinant of
small fluctuations around solutions of the field equations
\cite{har83}. If the matter field is a conformally coupled
scalar field $\phi\equiv(3/4\pi)^{1/2}\chi/a$ (which is the case that
Hartle and Hawking \cite{har83} discussed), $\rho a^4$ is conserved
where $\rho$ is the energy density of $\phi$ satisfying the
field equations. Then the Friedmann equation is given by
Eq.~(\ref{EA3}). The corresponding
Euclidean equation is obtained from Eq.~(\ref{EA3}) via
$t\rightarrow-i\tau$
\begin{eqnarray}
   \left({da\over d\tau}\right)^2=1-{\Lambda\over3}a^2-{C\over a^2}.
   \label{EA6}
\end{eqnarray}
The solution to Eq.~(\ref{EA6}) is (for the case ${4\over3}\Lambda C<1$)
\begin{eqnarray}
   a(\tau)=H^{-1}\left[{1\over2}+{1\over2}\left(1-4H^2C\right)^{1/2}
   \cos(2H\tau)\right]^{1/2},
   \label{EA7}
\end{eqnarray}
where $H=({\Lambda\over3})^{1/2}$. This is a Euclidean bouncing space
with a maximum radius
$a_{\max}=H^{-1}\left[{1\over2}+{1\over2}(1-4H^2C)^{1/2}\right]^{1/2}$ and a
minimum radius
$a_{\min}=H^{-1}\left[
{1\over2}-{1\over2}(1-4H^2C)^{1/2}\right]^{1/2}$ (Fig.~\ref{f3}).
If $C=0$,
we have $a_{\max}=H^{-1}$, $a_{\min}=0$, and
$a(\tau)=H^{-1}\cos(H\tau)$, one copy of this bouncing space is a
four-sphere with the Euclidean de~Sitter metric
$ds^2=d\tau^2+H^{-2}\cos^2(H\tau)[d\chi^2+\sin^2\chi(d\theta^2+\sin^2\theta
d\phi^2)]$ --- which is just a four-sphere embedded in a
five-dimensional Euclidean space $(V,W,X,Y,Z)$ with metric $ds^2=dV^2+
dW^2+dX^2 +dY^2+dZ^2$ --- this is the solution that Hartle and Hawking
used \cite{har83}. But, as we have argued above, according to Hartle and
Hawking \cite{har83} and Hawking \cite{haw84}, the Wheeler-DeWitt
equation for $\Phi(\chi)$ [Eq.~(\ref{EA1a})] gives rise to a
``zero-point-energy'' for the conformally coupled scalar field:
$C_0=2E(n=0)=1$ (the state with $C=0$ violates the uncertainty
principle). One copy
of this bouncing Euclidean space is {\em not} a compact
four-dimensional manifold with no boundaries, but has two boundaries
with $a=a_{\min}$ (see Fig.~\ref{f3}). 
If $H\ll 1$ (i.e. $\Lambda\ll 1$), we have
$a_{\max}\simeq H^{-1}$, $a_{\min}\simeq 1$. 

Penrose \cite{haw96} has criticized Hawking's no-boundary
proposal and the model obtained by gluing a de~Sitter space onto a four-sphere
hemisphere by pointing out that there are
only very few spaces for which one can
glue a Euclidean and a Lorentzian solution together since it is
required that they have both a Euclidean and a Lorentzian solution, but
the generic case is certainly very far from that. Here ``with a
zero-point-energy'' we have
have both a Euclidean solution and a Lorentzian
solution, and they can be glued together. But the Euclidean solution is
not closed in any way; that is, it does not enforce the no-boundary
proposal. Hartle and Hawking argued that there should be a constant
$\epsilon_0$ in $E$ which arises from the renormalization of the matter
field, i.e., $E$ should be $n+{1\over2}+\epsilon_0$ \cite{har83}. But
there is {\em no} reason that $\epsilon_0$ should be $-{1\over2}$ to exactly
cancel the ``zero-point-energy'' ${1\over2}$. (As in the case of a quantum
harmonic oscillator, we have no reason to neglect the zero-point-energy.)
In fact, since $\epsilon_0$ comes from the renormalization of
the matter field (without quantization of gravity), it should be much less
than the Planck magnitude, i.e., $\epsilon_0\ll 1$,  and thus 
$\epsilon_0$ is negligible compared with ${1\over2}$. In fact in
\cite{haw84} Hawking has dropped $\epsilon_0$.

In \cite{har83}
Hartle and Hawking have realized that for excited states ($n>0$),
there are two kinds of classical solutions: one represents universes
which expand from zero volume, to reach a maximum radius, and then
re-collapse (like our tiny oscillating universe); 
the other represents the de~Sitter-type state of continual
expansion. There are probabilities for a universe to tunnel from one
state to the other. Here we argue that for the ground
state ($n=0$), there are also two such kinds of Lorentzian
universes. One is a tiny quantum oscillating universe (having
a maximum radius with Planck magnitude). Here ``quantum''
just means that the classical description fails (so singularities
might be removed). The other is a big
de~Sitter-type universe. These two universes can be joined to
one another through a Euclidean section, which describes quantum
tunneling from a tiny oscillating universe to an inflating universe
(or from a contracting de~Sitter-type universe to a tiny
oscillating universe). During the tunneling, the radius of
the universe makes a jump (from the Planck length to $H^{-1}$ or
{\sl vice versa}).

As Hartle and Hawking \cite{har83} calculated the wave
function of the universe for the ground state, they argued that,
for the conformally coupled scalar field case, the path integral over $a$
and $\chi=(4\pi/3)^{1/2}\phi a$ separates since ``not only the
action separates into a sum of a gravitational part and a matter
part, but the boundary condition on the $a(\eta)$ and $\chi(\eta)$
summed over do not depend on one another'' where $\eta$ is the
conformal time. The critical point for the variable's
separation in the path integral is that ``the ground state boundary
conditions imply that geometries in the sum are conformal to
half of a Euclidean-Einstein static universe; i.e., the range
of $\eta$ is $(-\infty,0)$. The boundary conditions at infinite
$\eta$ are that $\chi(\eta)$ and $a(\eta)$ vanish. The boundary
conditions at $\eta=0$ are that $a(0)$ and $\chi(0)$ match the
arguments of the wave function $a_0$ and $\chi_0$'' \cite{har83}.
But this holds only for some specific cases, such as de~Sitter space.
Our solution (\ref{EA7}) does not obey Hartle and Hawking's assumption
that $\eta$ ranges from $-\infty$ to $0$. 
For a general $k=+1$ (Euclidean) Robertson-Walker metric, $\eta=
\int{d\tau\over a}$ is a functional of $a$, and the action of
matter (an integral over $\eta$) is a functional of $a$. Therefore, the
action {\em cannot} be separated into a sum of a gravitational
part and a matter part as Hartle and Hawking did. The failure of
Hartle and Hawking's path integral calculation is also
manifested in the fact that
de~Sitter space is {\em not} a solution of the Friedmann equation if
the ``zero-point-energy'' of the conformally coupled scalar field is
considered, whereas the semiclassical approximation implies that the
principal contribution to the path integral of the
wave function comes from the
configurations which solve Einstein's equations. One may hope to overcome
this difficulty by introducing a scalar field with a flat potential $V(\phi)$
(as in the inflation case). But this does not apply to the quantum cosmology
case since as $a\rightarrow0$ the universe always becomes radiation-dominated
unless the energy density of radiation is exactly zero (but the uncertainty
principle does not allow this case to occur).

\section{CTCs and the Chronology Protection Conjecture}
From the arguments in the last section, we find that the Universe does
{\em not} seem to be created from nothing. On the other hand, if the Universe
is created from {\em something}, that something could have been {\em itself}.
Thus it is possible
that the Universe is its own mother. In such a case, if we trace
the history of the Universe backward, inevitably we will enter a region of CTCs.
Therefore CTCs may play an important role in the creation of the Universe.
It is interesting to note that Hawking and Penrose's singularity
theorems do not apply if the Universe has had CTCs. And, it has been
shown that, if a compact Lorentzian spacetime undergoes topology
changes, there must be CTCs in this spacetime \cite{ger67,haw92a,haw92b}.
[Basically there are two type of spacetimes with
CTCs: for the first type, there 
are CTCs everywhere (G\"{o}del
space belongs to this type); for the second type, 
the CTCs are confined within some regions and there exists at
least one region where there are no closed causal (timelike or null)
curves, and the regions with CTCs are separated from the regions
without closed causal curves by Cauchy horizons
(Misner space belongs to this type). 
In this paper, with the word ``spacetimes with CTCs'' we always refer 
to the second type unless otherwise specified.]

While in classical general relativity there exist many solutions with
CTCs, some calculations of vacuum polarization of quantum fields in
spacetimes with CTCs indicated that the
energy-momentum tensor (in this paper when we deal with quantum fields,
with the word ``the energy-momentum tensor'' we always refer to ``the
renormalized energy-momentum tensor'' because ``the unrenormalized
energy-momentum tensor'' has no physical meaning) diverges as one
approaches the Cauchy horizon separating the region with CTCs from
the region without closed causal curves. This means that spacetimes with CTCs may be
unstable against vacuum polarization since when the energy-momentum
tensor is fed back to the semiclassical Einstein's equations (i.e. 
Einstein's equations with quantum corrections to the energy-momentum
tensor of matter fields) 
the back-reaction may distort the spacetime geometry so strongly that
a singularity may form and CTCs may be destroyed. Based on some of these
calculations, Hawking \cite{haw92a,haw92b} has proposed the
chronology protection
conjecture which states that the laws of physics do not allow the
appearance of CTCs. (It should be mentioned that the chronology
protection conjecture does {\em not} provide any restriction on
spacetimes with CTCs but no Cauchy horizons since there is 
{\em no} any indication that
this type of spacetime is unstable against vacuum
polarization. In the next section we will show a simple example of
a spacetime with CTCs but no Cauchy horizons, where the
energy-momentum tensor is finite
everywhere.)

But, on the other hand, Li, Xu, and Liu \cite{li93} have pointed out
that even if the energy-momentum tensor of vacuum polarization diverges
at the Cauchy horizon, it does {\em not} mean that CTCs must be
prevented by physical laws because: (1) Einstein's equations are local
equations and the energy-momentum tensor may diverge only at the
Cauchy horizon (or at the polarized hypersurfaces) and be well-behaved
elsewhere within the region with CTCs; 
(2) the divergence of the energy-momentum tensor at the
Cauchy horizon does {\em not} mean that the Cauchy horizon must be
destroyed by the back-reaction of vacuum polarization, {\em but}
instead means
that near the Cauchy horizon the usual quantum field theory on a
prescribed classical spacetime background cannot be used and the
quantum effect of gravity must be considered. (This is like the case
that Hawking and Penrose's singularity theorems do {\em not} mean that
the Big Bang cosmology is wrong but mean that near the Big Bang
singularity quantum gravity effects become important \cite{haw88}.) When
Hawking proposed his chronology protection conjecture,
Hawking \cite{haw92b} and Kim
and Thorne \cite{kim91} had a controversy over whether quantum gravity can save
CTCs. Kim and Thorne claimed that quantum gravitational effects would
cut the divergence off when an observer's proper time from crossing
the Cauchy horizon was the Planck time, and this would only give such a
small perturbation on the metric that the Cauchy horizon could not be
destroyed. But, Hawking \cite{haw92b} noted that one would expect the quantum
gravitational cut-off to occur when the invariant distance from the
Cauchy horizon was of order the Planck length, and this would give a
very strong perturbation on the metric so that the Cauchy horizon
would be destroyed. Since there does not exist a self-consistent
quantum theory of gravity at present, we cannot judge who (Hawking or
Kim and Thorne) is right. But in any case, these arguments imply that in
the case of a spacetime with CTCs where the energy-momentum tensor of
vacuum polarization diverges at the Cauchy horizon, quantum gravity
effects should become important near the Cauchy horizon. Li, Xu, and
Liu \cite{li93} have argued that if the effects of quantum gravity are
considered, in a spacetime with CTCs the region with CTCs and the
region without closed causal curves may be separated by a {\em quantum
barrier} (e.g. a region where components of the metric have 
complex values) instead of a Cauchy horizon
generated by closed null geodesics. By quantum processes, a time
traveler may tunnel from the region without closed causal curves to
the region with CTCs (or {\sl vice versa}), and the spacetime itself
can also tunnel from one side to the other side of the quantum barrier
\cite{li93}. In classical general relativity, a region with CTCs and a
region without closed causal curves must be separated by a Cauchy
horizon (compactly generated or non-compactly generated) which usually
contains closed null geodesics if it is compactly generated
\cite{haw92b}. But if quantum gravity effects are considered (e.g. in
quantum cosmology), they can be separated by a complex geometric
region (as a quantum barrier) instead of a Cauchy horizon
\cite{li93}. (In the path integral approach to quantum cosmology,
complex geometries are {\em required} in order to make the path
integral convergent and to overcome the difficulty that in general
situations a Euclidean space {\em cannot} be directly joined to a
Lorentzian space \cite{hal90}). And, using a simple example of a space
with a region with CTCs separated from a region without closed causal
curves by a complex geometric region, Li, Xu, and Liu \cite{li93} have
shown that in such a space the energy-momentum tensor of vacuum
polarization is finite everywhere and the chronology protection
conjecture has been challenged.

Without appeal to quantum gravity, counter-examples to the chronology
protection conjecture also exist. By introducing a spherical reflecting
boundary between two mouths of a wormhole,
Li \cite{li94} has shown that with some boundary conditions
for geodesics (e.g. the reflection boundary condition) closed null
{\em geodesics} [usually the ``archcriminal'' for the divergence of
the energy-momentum tensor as the Cauchy horizon is approached
(see e.g. \cite{kim91})] may be removed from
the Cauchy horizon separating the region with CTCs and the region
without closed causal curves. In such a case the spacetime contains
neither closed null {\em geodesics} nor closed timelike {\em
geodesics}, though it contains both closed timelike {\em non-geodesic}
curves and closed null {\em non-geodesic} curves.
Li \cite{li94} has shown that in this spacetime the
energy-momentum tensor is finite everywhere. Following Li \cite{li94},
Low \cite{low95} has given another example of spacetime with CTCs but
without closed causal {\em geodesics}. 

Recently, with a very general argument, Li \cite{li96} has shown that
the appearance of an absorber in a spacetime with CTCs may make the
spacetime stable against vacuum polarization. Li
\cite{li96} has given some examples to show that there exist many
collision processes in high energy physics for which the total
cross-sections increase (or tend to a constant) as the frequency of
the incident waves increases. Based on these examples, Li \cite{li96} has argued
that material will become opaque for waves (particles) with 
extremely high frequency or energy, since in such cases the absorption
caused by various types of scattering processes becomes very
important. Based on calculation of the renormalized
energy-momentum tensor and the fluctuation in the metric, Li \cite{li96} has
argued that if an absorbing material with appropriate density is
introduced, vacuum polarization may be smoothed out near the Cauchy
horizon so that the metric perturbation caused by vacuum
fluctuations will be very small and a spacetime with CTCs can be
stable against vacuum polarization.

Boulware \cite{bou92} and Tanaka and Hiscock \cite{tan95} have found
that for sufficiently massive fields in Gott space \cite{got91a,hea94} and
Grant space \cite{gra93} respectively, the energy-momentum tensor
remains regular on the Cauchy horizon. Krasnikov \cite{kra96} has
found some two-dimensional spacetimes with CTCs for which the
energy-momentum tensor of vacuum polarization is bounded on the Cauchy
horizon. Sushkov \cite{sus97} has found that for an automorphic
complex scalar field in Misner space there is a vacuum state for which
the energy-momentum tensor is zero everywhere. More recently, Cassidy
\cite{cas97a} and Li and Gott \cite{li97} have independently found that
for the real conformally coupled scalar field in Misner space there
exists a quantum state for which the energy-momentum tensor is zero
everywhere. Li and Gott \cite{li97} have found that this quantum
state is the ``adapted'' Rindler vacuum (i.e. the usual Rindler vacuum
with multiple images) and it is a self-consistent vacuum state because
it solves the semiclassical Einstein's equations exactly. Li and Gott
\cite{li97} have also found that for this ``adapted'' Rindler vacuum
in Misner space, an inertial particle detector perceives nothing.
In this paper, we find that for a
multiply connected de~Sitter space there also exists a self-consistent
vacuum state for a conformally coupled scalar field (see section \ref{IX}).

Thorne \cite{tho93} has noted that, even if Hawking's argument that a quantum
gravitational cut-off would occur when the geometric invariant
distance from the Cauchy horizon is of order the Planck length is correct, by
using two wormholes the metric fluctuations near the Cauchy horizon can
be made arbitrarily small so a spacetime with CTCs created from two
wormholes can be stable against vacuum polarization. Recently Visser
\cite{vis97} has generalized this result to the Roman-ring case.

The above arguments indicate that the back-reaction of
vacuum polarization may {\em not} destroy the Cauchy horizon in
spacetimes with CTCs, and thus such spacetimes can be stable against
vacuum polarization.

In a recent paper, Cassidy and Hawking \cite{cas97b} have admitted
that ``back-reaction does not enforce chronology protection''. On the
other hand, Cassidy and Hawking \cite{cas97b} have argued that
the ``number of states'' may enforce the 
chronology protection conjecture since ``this quantity will always
tend to zero as one tries to introduce CTCs''. Their arguments are
based on the fact that for the particular spacetime with CTCs they constructed
[which is the product of a multiply connected (via a boost) 
three-dimensional de~Sitter space and $S^1$]
the entropy of a massless scalar field diverges to minus infinity when
the spacetime develops CTCs \cite{cas97b}. However, whether this
conclusion holds for general spacetimes with CTCs remains an open
question and further research is required. And, from ordinary
statistical thermodynamics we know that entropy is always positive,
so the physical meaning of a {\em negative} entropy is unclear. The
number of states in phase space is given by $N=\Delta p\Delta q/(2\pi\hbar)^s$
where $\Delta q=\Delta q_1\Delta q_2...\Delta q_s$, 
$\Delta p=\Delta p_1\Delta p_2...\Delta p_s$, $q_i$ ($i=1,2,...s$) 
is a canonical coordinate, $p_i$ is a canonical momentum,
and $s$ is the number of degrees of freedom.
The uncertainty principle demands that $\Delta p_i\Delta q_i\geq2\pi\hbar$ and thus
we should always have $N\geq 1$. Thus the ``fact'' that the number of
states tends to zero as one tries to develop CTCs (i.e. as one approaches
the Cauchy horizon) may simply imply that near the Cauchy horizon quantum
effects of gravity cannot be neglected, which is consistent with Li,
Xu, and Liu's argument \cite{li93}. The entropy is defined by $k_{\rm B}\ln N$
where $N$ is the number of states and $k_{\rm B}$ is the Boltzmann constant.
When $N$ is small, quantization of the entropy becomes important
(remember that the number of states $N$ is always an integer). The
entropy cannot {\em continuously} tend to negative infinity; it
should {\em jump} from $k_{\rm B}\ln3$ to $k_{\rm B}\ln2$, {\em jump}
from $k_{\rm B}\ln2$
to zero (but in Cassidy and Hawking's arguments \cite{cas97b} we
have not seen such a jump),
then  the uncertainty principle demands that the entropy
should stand on the zero value as one approaches the Cauchy horizon.
On the other hand, ordinary continuous thermodynamics holds only for the
case with $N\gg1$. Thus, as
one approaches the Cauchy horizon the thermodynamic limit has
already been violated and ordinary thermodynamics should be revised
near the Cauchy horizon. In other words, Cassidy and 
Hawking's results \cite{cas97b} cannot be extended to the Cauchy horizon.
Based on the fact that the effective action density diverges at the
polarized hypersurfaces of spacetimes with CTCs \cite{cas97a},
Cassidy and Hawking \cite{cas97b} have argued that the effective
action ``would provide new insight into issues of chronology
protection''. But we should note that the effective action is
only a {\em tool} for computing some physical quantities (such
as the energy-momentum tensor) and the effective action itself has
not much physical meaning. The divergence of the effective action
may imply that the effective action is not a good {\em tool} as
the polarized hypersurfaces are approached. Our argument is supported
by the fact that there exist many examples for which the
energy-momentum tensor is finite everywhere, as mentioned above.

Recently, Kay, Radzikowski, and Wald \cite{kay97} have proved
two theorems which demonstrate that some fundamental quantities
such as Hadamard functions and energy-momentum tensors must be
ill-defined on a compactly generated Cauchy horizon
in a spacetime with CTCs, as one extends the {\em usual} quantum
field theory in a global hyperbolic spacetime to an acausal spacetime
with a compactly generated Cauchy horizon. Basically speaking,
their theorems imply that the {\em usual} quantum field theory
cannot be {\em directly} extended to a spacetime with CTCs
\cite{kay97}. Their theorems tell us that serious difficulties
arise when attempting to {\em define} quantum field theory on a
spacetime with a compactly generated Cauchy horizon \cite{kay97}.
The ordinary quantum field theory must be significantly changed
or some new approach must be introduced when one tries to do
quantum field theory on a spacetime with CTCs. A candidate procedure
for overcoming this difficulty is the Euclidean quantization
proposed by Hawking \cite{haw78,haw79}. Quantum field theory is well-defined
in a Euclidean space because there are no CTCs in a Euclidean
space \cite{haw95}. In fact, even in simply connected Minkowski
spacetime, quantum field theory is {\em not} well-defined
since the path integral does not converge. To overcome this
difficulty, the technique of Wick-rotation (which is essentially
equivalent to Euclidean quantization) is used. Kay, Radzikowski,
and Wald \cite{kay97} have also argued that their results may be
interpreted as indicating that in order to create CTCs it would
be necessary to enter a regime where quantum effects of gravity
will be dominant (see also the discussions of Visser
\cite{vis97a,vis97b});
this is also consistent with Li, Xu, and Liu's
arguments \cite{li93}. Cramer and Kay \cite{cra96,cra97} have shown that Kay,
Radzikowski, and Wald's theorems \cite{kay97} also apply to Misner space
(for Sushkov's automorphic field case \cite{sus97} and
Krasnikov's two-dimensional case \cite{kra96}, respectively) where the
Cauchy horizon is not compactly generated, in the sense
that the energy-momentum tensor must be ill-defined on the
Cauchy horizon itself. But we note that this only happens in a set
of measure zero which does not make much sense in physics for if the
renormalized energy-momentum tensor is zero everywhere except on a set
of measure zero where it is formally ill-defined, then continuity
would seem to require setting it to zero there also \cite{li97}.

Perhaps a conclusion on the chronology protection conjecture can only be reached
after we have a quantum theory of gravity. However, we can
conclude that the back-reaction of vacuum polarization does {\em not}
enforce the chronology protection conjecture, a point Hawking himself also
admits \cite{cas97b}. (Originally the back-reaction of
vacuum polarization was supposed to be the strongest candidate
for chronology protection \cite{haw92a,haw92b}.)

\section{Multiply Connected Minkowski Spacetimes with CTCs}
\label{VI}
A simple spacetime with CTCs is obtained from Minkowski spacetime by
identifying points that are related by time translation. Minkowski
spacetime is $(R^4, \eta_{ab})$. In Cartesian coordinates
$(t, x, y, z)$ the Lorentzian metric $\eta_{ab}$ is given by
\begin{eqnarray}
   ds^2=-dt^2+dx^2+dy^2+dz^2.
\label{E1}
\end{eqnarray}
Now we identify points $(t, x, y, z)$ with points $(t+nt_0, x, y, z)$
where $t_0$ is a positive constant and $n$ is any integer. Then we obtain
a spacetime with topology $S^1\times R^3$ and the Lorentzian metric. Such a
spacetime is closed in the time direction and has no Cauchy horizon.
All events in this spacetime are threaded by CTCs.
(This is the only acausal spacetime without a Cauchy horizon considered in this paper.)
Minkowski spacetime $(R^4,\eta_{ab})$
is the covering space of this spacetime.

Usually there is no well-defined quantum field theory in a spacetime
with CTCs. (Kay-Radzikowski-Wald's theorems \cite{kay97} enforce this
claim, though they do not apply directly to an acausal spacetime
without a Cauchy horizon.)
However, in the case where a
covering space exists, we can
do it in the covering space with the method of images. In fact in
most cases where the energy-momentum tensor
in spacetimes with CTCs has been calculated,
this method has been
used (for the
theoretical basis for the method of images see Ref. \cite{fro91} and
references cited therein). The method of images is sufficient for our
purposes in this paper (computing the energy-momentum tensor and the
response function of particle detectors). Thus in this
paper we use this method to deal with quantum field theory
in spacetimes with CTCs.

For any point $(t, x, y, z)$ in $(S^1\times R^3, \eta_{ab})$, there
are an
infinite number of images of points $(t+nt_0, x, y, z)$ in the covering
space $(R^4, \eta_{ab})$. For the Minkowski vacuum $\vert 0_{\rm M}\rangle$ of
a conformally coupled scalar field (by ``conformally coupled'' we mean
that the mass of the scalar field is zero and the coupling between the
scalar field $\phi$ and the gravitational field is given by
${1\over6}R\phi^2$ where $R$ is the Ricci scalar curvature) 
in the Minkowski spacetime, the Hadamard function is
\begin{eqnarray}
   G_{\rm M}^{(1)}(X,X^\prime)={1\over2\pi^2}~
   {1\over
   -(t-t^\prime)^2+(x-x^\prime)^2+(y-y^\prime)^2+(z-z^\prime)^2},
   \label{E2}
\end{eqnarray}
here $X=(t,x,y,z)$ and $X^\prime=(t^\prime,x^\prime,y^\prime,
z^\prime)$.
With the method of images, the Hadamard function of the ``adapted''
Minkowski vacuum (which is the Minkowski vacuum with multiple images)
in the spacetime $(S^1\times R^3, \eta_{ab})$ is given by 
the summation of the Hadamard function in (\ref{E2}) for all images
\begin{eqnarray}
   G^{(1)}(X,X^\prime)={1\over2\pi^2}\sum_{n=-\infty}^{\infty}
   {1\over
   -(t-t^\prime+nt_0)^2+(x-x^\prime)^2+(y-y^\prime)^2+(z-z^\prime)^2}.
   \label{E3}
\end{eqnarray}
The regularized Hadamard function is usually taken to be
\begin{eqnarray}
   G_{\rm
   reg}^{(1)}(X,X^\prime)&=&G^{(1)}(X,X^\prime)-G_{\rm M}^{(1)}(X,X^\prime)
   \nonumber\\
   &=&{1\over2\pi^2}\sum_{n\not=0}
   {1\over
   -(t-t^\prime+nt_0)^2+(x-x^\prime)^2+(y-y^\prime)^2+(z-z^\prime)^2}.
   \label{E4}
\end{eqnarray}
The renormalized energy-momentum tensor is given by \cite{wal78,bir82}
\begin{eqnarray}
   \langle T_{ab}\rangle_{\rm ren}={1\over2}\lim_{X^\prime\rightarrow
   X}\left({2\over3}\nabla_a\nabla_{b^\prime}
   -{1\over3}\nabla_a\nabla_b-{1\over6}\eta_{ab}
   \nabla_c\nabla^{c^\prime}\right)G^{(1)}_{\rm reg}.
   \label{E5}
\end{eqnarray}
Inserting Eq.~(\ref{E4}) into Eq.~(\ref{E5}) we get
\begin{eqnarray}
   \langle T_{\mu}^{~\nu}\rangle_{\rm
   ren}={\pi^2\over90t_0^4}
   \left(\begin{array}{cccc}
   -3 & 0 & 0 & 0 \\
   0  & 1 & 0 & 0 \\
   0  & 0 & 1 & 0 \\
   0  & 0 & 0 & 1
   \end{array}\right).
   \label{E6}
\end{eqnarray}
We find that this energy-momentum tensor is constant and finite
everywhere and has the form of radiation. Thus CTCs
do not mean that the energy-momentum tensor must
diverge.  

Now let us consider a particle detector \cite{bir82,unr76} 
moving in this spacetime. The particle
detector is coupled to the field $\phi$ by the
interaction Lagrangian $cm(\tau)\phi[X(\tau)]$, where $c$ is a small
coupling constant, $m$ is the detector's monopole moment, $\tau$ is the proper
time of the detector's worldline, and $X(\tau)$ is the trajectory of the
particle detector \cite{bir82}. Suppose initially the detector is in its
ground state with energy $E_0$ and the field $\phi$ is in some quantum
state $\vert\rangle$. Then the transition probability for the detector to
all possible excited states with energy
$E>E_0$ and the field $\phi$ to all possible
quantum states is given by \cite{bir82}
\begin{eqnarray}
   P=c^2\sum_{E>E_0}\vert\langle E\vert m(0)\vert E_0\rangle\vert^2
   {\cal F}(\Delta E),
   \label{E14a}
\end{eqnarray}
where $\Delta E=E-E_0>0$ and ${\cal F}(\Delta E)$ is the
response function 
\begin{eqnarray}
   {\cal F}(\Delta E)=\int_{-\infty}^{\infty}d\tau\int_{-\infty}^\infty
   d\tau^\prime
   e^{-i\Delta E(\tau-\tau^\prime)}G^+(X(\tau),X(\tau^\prime)),
   \label{E7}
\end{eqnarray}
which is independent of the details of the particle detector and
is determined by the positive frequency Wightman function $G^+(X,X^\prime)
\equiv\langle\vert\phi(X)\phi(X^\prime)\vert\rangle$ (while the
Hadamard function is defined by
$G^{1}(X,X^\prime)\equiv\langle\vert\phi(X)\phi(X^\prime)+
\phi(X^\prime)\phi(X)\vert\rangle$). The response
function represents the bath of particles that the detector effectively
experiences \cite{bir82}. The remaining factor in Eq.~(\ref{E14a}) represents the
selectivity of the detector to the field and depends on the internal structure
of the detector \cite{bir82}.
The Wightman
function for the Minkowski vacuum is
\begin{eqnarray}
   G_{\rm M}^+(X,X^\prime)=
   {1\over4\pi^2}~
   {1\over
   -(t-t^\prime-i\epsilon)^2+(x-x^\prime)^2+(y-y^\prime)^2+(z-z^\prime)^2},
   \label{E8}
\end{eqnarray}
where $\epsilon$ is an infinitesimal positive real number which is introduced
to indicate that $G^+$ is the boundary value of a function which is
analytic in the lower-half of the complex $\Delta t\equiv t-t^\prime$
plane. For the
adapted Minkowski vacuum in our spacetime $(S^1\times R^3, \eta_{ab})$,
the Wightman function is
\begin{eqnarray}
   G^+(X,X^\prime)={1\over4\pi^2}\sum_{n=-\infty}^{\infty}
   {1\over
   -(t-t^\prime+nt_0-i\epsilon)^2+(x-x^\prime)^2+(y-y^
   \prime)^2+(z-z^\prime)^2}.
   \label{E9}
\end{eqnarray}
Assume that the detector moves along the geodesic $x=\beta t$ $(\beta<1)$,
$y=z=0$, then the proper time is $\tau=t/\zeta$ with 
$\zeta=1/\sqrt{1-\beta^2}$. On the geodesic, the Wightman
function is reduced to
\begin{eqnarray}
   G^+(\tau,\tau^\prime)&=&{1\over4\pi^2}\sum_{n=-\infty}^
   {\infty}{1\over-(t-t^\prime+nt_0-i\epsilon)^2+\beta^2(t-t^\prime)^2}
   \nonumber\\
   &=&-{1\over4\pi^2\zeta^2}\sum_{n=-\infty}^
   {\infty}{1\over(\tau-\tau^\prime+nt_0/\zeta-i\epsilon/\zeta)^2
   -\beta^2(\tau-\tau^\prime)^2}.
   \label{E10}
\end{eqnarray}
Inserting Eq.~(\ref{E10}) into Eq.~(\ref{E7}), we obtain
\begin{eqnarray}
   {\cal F}(\Delta E)=-{1\over4\pi^2\zeta^2}\sum_{n=-\infty}^
   {\infty}\int_{-\infty}^{\infty}dT\int_{-\infty}^{\infty}d\Delta\tau
   e^{-i\Delta E\Delta\tau}{1\over(\Delta\tau+nt_0/\zeta-i\epsilon/\zeta)^2
   -\beta^2(\Delta\tau)^2},
   \label{E11}
\end{eqnarray}
where $\Delta\tau=\tau-\tau^\prime$ and $T=(\tau+\tau^\prime)/2$.
The integration over $\Delta\tau$ is taken along a contour closed in
the lower-half plane of complex $\Delta\tau$. Inspecting the poles of
the integrand, we find that all poles are in the upper-half plane of
complex $\Delta\tau$ (remember that $\beta<1$). Therefore according to
the residue theorem we have
\begin{eqnarray}
   {\cal F}(\Delta E)=0.
   \label{E12}
\end{eqnarray}
Such a particle detector perceives no particles, though the renormalized
energy-momentum tensor of the field has the form of radiation. 

Another simple space with CTCs constructed from
Minkowski space is Misner space \cite{mis67}. In Cartesian coordinates
$(t,x,y,z)$ in Minkowski spacetime, a boost transformation in the
$(t,x)$ plane (we can always adjust the coordinates so that the boost
is in this plane) takes point $(t,x,y,z)$ to point $(t\cosh b+x\sinh
b, x\cosh b+t\sinh b, y, z)$ where $b$ is the boost parameter. In
Rindler coordinates $(\eta, \xi, y, z)$, defined by
\begin{eqnarray}
   \left\{\begin{array}{l}
   t=\xi\sinh\eta,\\
   x=\xi\cosh\eta,\\
   y=y,\\
   z=z, 
   \end{array}  
   \right.
   \label{E55}
\end{eqnarray}
the Minkowski metric can then be written in the Rindler form
\begin{eqnarray}
   ds^2=-\xi^2d\eta^2+d\xi^2+dy^2+dz^2.
   \label{E56}
\end{eqnarray}
The Rindler coordinates $(\eta,\xi,y,z)$ only cover the right quadrant
of Minkowski space (i.e. the region R defined by $x>|t|$). By
a reflection $(t,x,y,z)\rightarrow(-t,-x,y,z)$ [or
$(\eta,\xi,y,z)\rightarrow$ $(\eta,-\xi,y,z)$], the Rindler
coordinates
and the Rindler metric
can be extended to the left quadrant (L, defined by $x<-|t|$).
By the transformation
\begin{eqnarray}
   \eta\rightarrow\tilde{\xi}-i{\pi\over2},~~~
   \xi\rightarrow\pm i\tilde{\eta},~~~
   y\rightarrow y,~~~
   z\rightarrow z,
   \label{E56a}
\end{eqnarray}
the Rindler coordinates can be extended to the future quadrant 
(F, defined by $t>|x|$) and the past quadrant (P, defined by
$t<-|x|$). In region L the Rindler metric has the same form as the
metric in region R, which is given by
Eq.~(\ref{E56}). But in F and P the
Rindler metric is extended to be
\begin{eqnarray}
   ds^2=-d\tilde{\eta}^2+\tilde{\eta}^2d\tilde{\xi}^2+dy^2+dz^2.
   \label{E56b}
\end{eqnarray}
Misner space is obtained by identifying $(t,x,y,z)$ with $(t\cosh
nb+x\sinh nb,x\cosh nb+t\sinh nb,y,z)$.
Under such an identification, point
$(\eta,\xi,y,z)$ in R (or L) is identified with points
$(\eta+nb,\xi,y,z)$ in R (or L), point $(\tilde{\eta},
\tilde{\xi},y,z)$ in F (or P) is identified with points  $(\tilde{\eta},
\tilde{\xi}+nb,y,z)$ in F (or P). Clearly there are CTCs
in R and L but there are no closed causal curves in F and P, and these regions
are separated by
the Cauchy horizons $x=\pm t$, generated by closed null geodesics.

Misner space is not a manifold at the intersection of $x=t$ and
$x=-t$. However, as Hawking and Ellis \cite{haw73} have pointed out,
if we consider the bundle of linear frames over
Minkowski space, the corresponding induced bundle of linear frames over Misner 
space is a Hausdorff manifold and therefore well-behaved everywhere. 

The energy-momentum tensor of a conformally coupled scalar
field in Misner space has been studied in \cite{his82,li97}.
Hiscock and Konkowski \cite{his82} have calculated the energy-momentum tensor
of the adapted Minkowski vacuum. In Rindler coordinates their
results can be written as
\begin{eqnarray}
   \langle T_\mu^{~\nu}\rangle_{\rm M,ren}={A\over12\pi^2\xi^4}
   \left(\begin{array}{cccc}
   -3&0&0&0\\
   0&1&0&0\\
   0&0&1&0\\
   0&0&0&1
   \end{array}
   \right),
   \label{E67}
\end{eqnarray}
where the constant $A$ is
\begin{eqnarray}
   A=\sum_{n=1}^{\infty}{2+\cosh nb\over(\cosh nb-1)^2}.
   \label{E68}
\end{eqnarray}
Eq.~(\ref{E67}) holds only in region R [because Rindler coordinates
defined by Eq.~(\ref{E55}) only cover R], but it can be
analytically extended to other regions by writing $\langle
T_\mu^{~\nu}\rangle_{\rm M,ren}$ in Cartesian
coordinates or by the transformations mentioned above. Obviously for
any finite $b$, $\langle T_\mu^{~\nu}\rangle_{\rm
M,ren}$ diverges as one approaches the Cauchy horizon ($\xi\rightarrow0$). 
This divergence is coordinate independent since
$\langle T^{\mu\nu}\rangle_{\rm M,ren}\langle T_{\mu\nu}\rangle_{\rm
M,ren}$ also diverges as $\xi\rightarrow0$. This indicates that 
though the Minkowski vacuum is a good and 
self-consistent vacuum for simply connected Minkowski space, the
adapted Minkowski vacuum is {\em not} self-consistent for Misner space
(i.e. it does not solve Einstein's equations given the Misner space geometry). This
result has led Hawking \cite{haw92a,haw92b} to conjecture that the
laws of physics do not allow the
appearance of CTCs (i.e., his chronology protection
conjecture). 

Li and Gott \cite{li97} have studied 
the adapted Rindler vacuum in Misner space.
The Hadamard function for the Rindler vacuum is
\cite{dow78}
\begin{eqnarray}
   G_{\rm R}^{(1)}(X,X^\prime)={1\over2\pi^2}{\gamma\over\xi\xi^\prime
   \sinh\gamma~[-(\eta-\eta^\prime)^2+\gamma^2]},
   \label{E76}
\end{eqnarray}
where $X=(\eta,\xi,y,z)$, $X^\prime=(\eta^\prime,\xi^\prime,y^\prime,
z^\prime)$, and $\gamma$ is defined by
\begin{eqnarray}
   \cosh\gamma={\xi^2+{\xi^\prime}^2+(y-y^\prime)^2+(z-z^\prime)^2\over
   2\xi\xi^\prime}.
   \label{E77}
\end{eqnarray}
The Hadamard function for the adapted Rindler vacuum in Misner space is
\begin{eqnarray}
   G^{(1)}(X,X^\prime)={1\over2\pi^2}\sum_{n=-\infty}^{\infty}
   {\gamma\over\xi\xi^\prime
   \sinh\gamma[-(\eta-\eta^\prime+nb)^2+\gamma^2]}.
   \label{E78}
\end{eqnarray}
Though $G_{\rm R}^{(1)}$ and $G^{(1)}$ given by Eq.~(\ref{E76})
and Eq.~(\ref{E78}) are defined only in region R, 
they can be analytically
extended to regions L, F, and P in Minkowski and Misner space. The
regularized Hadamard function for the adapted Rindler vacuum is
$G_{\rm reg}^{(1)}(X,X^\prime)=G^{(1)}(X,X^\prime)-G_{\rm
M}^{(1)}(X,X^\prime)$, where $G_{\rm M}^{(1)}$ is the Hadamard
function for the Minkowski vacuum given by Eq.~(\ref{E2}). Inserting
this together with 
Eq.~(\ref{E78}) and Eq.~(\ref{E2}) into Eq.~(\ref{E5}), we obtain the
energy-momentum tensor for a conformally coupled scalar field in the
adapted Rindler vacuum \cite{li97}
\begin{eqnarray}
   \langle T_\mu^{~\nu}\rangle_{\rm R,ren}={1\over1440\pi^2\xi^4}
   \left[\left({2\pi\over b}\right)^4-1\right]
   \left(\begin{array}{cccc}
   -3&0&0&0\\
   0&1&0&0\\
   0&0&1&0\\
   0&0&0&1
   \end{array}
   \right),
   \label{E69}
\end{eqnarray}
which is expressed in Rindler coordinates and thus holds only in
region R but can be analytically extended to other regions with the
method mentioned above for the case of the adapted Minkowski
vacuum. We \cite{li97} have found that
unless $b=2\pi$, $\langle
T_\mu^{~\nu}\rangle_{\rm R,ren}$ blows up as one approaches the
Cauchy horizon ($\xi\rightarrow0$) (as also does$\langle T^{\mu\nu}
\rangle_{\rm R,ren}\langle T_{\mu\nu}\rangle_{\rm
R,ren}$). But, if $b=2\pi$, we have
\begin{eqnarray}
   \langle T_\mu^{~\nu}\rangle_{\rm R,ren}=0,
   \label{E70}
\end{eqnarray}
which is regular as one approaches the Cauchy horizon and can be regularly extended
to the whole Misner space, where it is also zero. In such a case, the
vacuum Einstein's equations without cosmological constant are
automatically satisfied. Thus this is an example of a spacetime with
CTCs at the semiclassical quantum gravity level. We \cite{li97} have
called this vacuum the
{\em self-consistent vacuum} for Misner space, and $b=2\pi$ is the
{\em self-consistent condition}. (Cassidy \cite{cas97a} has also
independently proven that for a conformally coupled scalar field in
Misner space there should exist a quantum state for which the
energy-momentum tensor is zero everywhere. But he has not shown what
quantum state it should be. We \cite{li97} have shown that it is the
adapted Rindler vacuum.)

Another way to deal with quantum fields in spacetimes with CTCs is to
do the quantum field theory in the Euclidean section and then
analytically extend the results to the Lorentzian section \cite{haw95}. For Misner
space the Euclidean section is obtained by taking $\eta$ and $b$ to be
$-i\bar{\eta}$ and $-i\bar{b}$. The resultant space is the Euclidean
space with metric $ds^2=\xi^2d\bar{\eta}^2+d\xi^2+dy^2+dz^2$ 
and $(\bar{\eta},\xi,y,z)$ and $(\bar{\eta}+n\bar{b},\xi,y,z)$ are
identified where $(\bar{\eta},\xi,y,z)$ are cylindrical polar
coordinates with $\bar{\eta}$ the angular polar coordinate and $\xi$ the
radial polar coordinate. The geometry at the 
hypersurface $\xi=0$ is conical singular
unless $\bar{b}=2\pi$. When extending that case to the Lorentzian section, we
get $b=2\pi$ which is just the self-consistent condition. This may be
the geometrical explanation of the self-consistent condition. By doing
quantum field theory in the Euclidean space, then analytically
extending the results to the Lorentzian section, we obtain the
renormalized energy-momentum tensor in R (or L) region of the Misner
space. Then we can extend the renormalized energy-momentum tensor in R (or L) to
regions F (or P). The results are the same as that obtained with
the method of images.  

Let us consider a particle detector moving in Misner space with the
adapted Rindler vacuum.
Suppose the detector moves along a geodesic with $x=a$, $y=\beta t$, 
and $z=0$ ($a$ and $\beta$ are constants and $a$ is positive), which 
goes through the P, R, and F regions. The proper time of the detector is
$\tau=t/\zeta$ with $\zeta=1/\sqrt{1-\beta^2}$. On
this geodesic, the Hadamard function in (\ref{E78}) is reduced to
\begin{eqnarray}
   G^{(1)}(t,t^\prime)={1\over2\pi^2}{\gamma\over
   \sinh\gamma\sqrt{(a^2-t^2)(a^2-{t^\prime}^2)}}
   \sum
   _{n=-\infty}^{\infty}{1\over-(\eta-\eta^\prime+nb)^2+\gamma^2},
   \label{I1}
\end{eqnarray}
where $\gamma$ is given by
\begin{eqnarray}
   \cosh\gamma={2a^2-t^2-{t^\prime}^2+\beta^2(t-t^\prime)^2\over
   2\sqrt{(a^2-t^2)(a^2-{t^\prime}^2)}},
   \label{I2}
\end{eqnarray}
and $\eta-\eta^\prime$ is given by
\begin{eqnarray}
   \sinh(\eta-\eta^\prime)={a(t-t^\prime)\over
   \sqrt{(a^2-t^2)(a^2-{t^\prime}^2)}}.
   \label{I3}
\end{eqnarray}
Though this Hadamard function is originally defined only in R, it can be
analytically extended to F, P, and L. The Wightman function is equal to
$1/2$ of the Hadamard function with $t$ replaced by
$t-i\epsilon/2$ and $t^\prime$ replaced by
$t^\prime+i\epsilon/2$,  
where $\epsilon$ is an infinitesimal positive
real number. Then the response function is \cite{li97}
\begin{eqnarray}
   &&{\cal F}(E)={1\over4\pi^2}\sum_{n=-\infty}^{\infty}
   \int_{-\infty}^{\infty}dT\int_{-\infty}^\infty d\Delta\tau \nonumber\\
   &&{\gamma^+e^{-iE\Delta\tau}\over\sinh\gamma^+\sqrt{[a^2-\zeta^2
   (T+{\Delta\tau\over2}-{i\epsilon\over2\zeta})^2]
   [a^2-\zeta^2(T-{\Delta\tau\over2}+{i\epsilon\over2\zeta})^2]}~\left\{-
   [(\eta-\eta^\prime)^++nb]^2+{\gamma^+}^2\right\}},
   \label{I4}
\end{eqnarray}
where $T\equiv(\tau+\tau^\prime)/2$, $\Delta\tau\equiv\tau-\tau^\prime$; 
$\gamma^+$ and $(\eta-\eta^\prime)^+$ are given by (\ref{I2}) and (\ref{I3})
with $t$ replaced by $t-i\epsilon/2$ and $t^\prime$ replaced by
$t^\prime+i\epsilon/2$. The
integral over $\Delta\tau$ can be worked out by the residue theorem where
we choose the integration contour to close in the lower-half
complex-$\Delta\tau$ plane. The result is zero since there are no poles
in the lower-half plane. Therefore such a detector
cannot be excited and so it detects nothing \cite{li97}. 
We \cite{li97} have also calculated the response
functions for detectors on worldlines with constant $\xi$, $y$,
and $z$ and worldlines with constant $\tilde{\xi}$, $y$,
and $z$ --- both are zero.

\section{Vacuum Polarization in Vilenkin's Tunneling  Universe}
In order to compare our model for the creation of the universe with
Vilenkin's tunneling universe, in this section we calculate the vacuum
fluctuation of a conformally coupled scalar field in Vilenkin's
tunneling universe.
The geometry of Vilenkin's tunneling universe has been described in
section IV. 
Such a universe is described by a Lorentzian-de~Sitter space
joined to a Euclidean de~Sitter space \cite{vil82}. The Lorentzian section
has the topology $R^1\times S^3$ and the metric
\begin{eqnarray}
   ds^2=-d\tau^2+ r_0^2\cosh^2{\tau\over r_0}[d\chi^2+\sin^2\chi(
   d\theta^2+ \sin^2\theta d\phi^2)].
   \label{E13}
\end{eqnarray}
The
Euclidean section has the topology $S^4$ and the metric
\begin{eqnarray}
   ds^2=d\tau^2+ r_0^2\cos^2{\tau\over r_0}[d\chi^2+\sin^2\chi(
   d\theta^2+ \sin^2\theta d\phi^2)].
   \label{E14}
\end{eqnarray}
The Lorentzian section and the Euclidean section 
are joined at the boundary $\Sigma$
defined by $\tau=0$. $\Sigma$
is a three-sphere with the minimum radius in de~Sitter space
and the maximum radius in the Euclidean four-sphere. The boundary
condition for a conformally coupled scalar field $\phi$ is \cite{hay92,hay93}
\begin{eqnarray}
   \left.{\partial\phi\over\partial\tau}\right|_\Sigma=0,
   \label{E15}
\end{eqnarray}
which is a kind of Neumann boundary condition and indicates that the
boundary $\Sigma$ is like a kind of reflecting boundary. 
The Green functions (including both the Hadamard function and the
Wightman function) should also satisfy this boundary condition
\begin{eqnarray}
   \left.{\partial G(\tau,\chi,\theta,\phi;\tau^\prime,\chi^\prime,
   \theta^\prime, \phi^\prime)\over\partial\tau}\right\vert_\Sigma=0.
   \label{E16}
\end{eqnarray}

The vacuum state of a conformally coupled scalar field in de
Sitter space is usually taken to be that obtained from the Minkowski
vacuum by the conformal transformation according to which de~Sitter
space is conformally flat. (The quantum state so obtained is usually
called the conformal vacuum \cite{bir82}.) Such a vacuum is de~Sitter invariant and we
call it the conformal Minkowski vacuum. 
The Hadamard function for this de~Sitter
vacuum (i.e. the conformal Minkowski vacuum) is \cite{bun78}
\begin{eqnarray}
   G_{\rm CM}^{(1)}(X,X^\prime)={1\over4\pi^2 r_0^2}~{1\over1-Z(X,X^\prime)},
   \label{E17}
\end{eqnarray}
where $X=(\tau,\chi,\theta,\phi)$, $X^\prime=(\tau^\prime,\chi^\prime,
\theta^\prime,\phi^\prime)$, and $Z(X,X^\prime)$ is defined by
\begin{eqnarray}
   Z(X,X^\prime)&=&-\sinh{\tau\over r_0}\sinh{\tau^\prime\over r_0}+
   \cosh{\tau\over r_0}\cosh{\tau^\prime\over r_0}\{\cos\chi\cos\chi^\prime
   \nonumber\\
   &&+\sin\chi\sin\chi^\prime[\cos\theta\cos\theta^\prime+
   \sin\theta\sin\theta^\prime\cos(\phi-\phi^\prime)]\}.
   \label{E18}
\end{eqnarray}
In Vilenkin's tunneling universe, the Hadamard function satisfying 
the boundary condition (\ref{E16}) is given by
\begin{eqnarray}
   G^{(1)}(X,X^\prime)&=&G_{\rm CM}^{(1)}(X,X^\prime) +
   G_{\rm CM}^{(1)}(X^-,X^\prime)
   \nonumber\\
   &=&{1\over4\pi^2 r_0^2}\left[{1\over1-Z(X,X^\prime)}+{1\over1-
   Z(X^-,X^\prime)}\right],
   \label{E19}
\end{eqnarray}
where $X^-=(-\tau,\chi,\theta,\phi)$ is the image of 
$X=(\tau,\chi,\theta,\phi)$ with respect to the reflecting boundary $\Sigma$.

There are various schemes for obtaining the renormalized
energy-momentum tensor for de~Sitter space (e.g. \cite{bun78,ber86}).
They all are equivalent to subtracting from the Hadamard function
a reference term $G^{(1)}_{\rm ref}$ to obtain a regularized Hadamard
function and then calculating the renormalized energy-momentum tensor
by \cite{wal78,bir82}
\begin{eqnarray}
   \langle T_{ab}\rangle_{\rm ren}={1\over2}\lim_{X^\prime\rightarrow X}
   {\cal D}_{ab^\prime}(X,X^\prime)G_{\rm reg}^{(1)}(X,X^\prime).
   \label{E21}
\end{eqnarray}
For the conformally coupled scalar field, 
the differential operator ${\cal D}_{ab^\prime}$ is
\begin{eqnarray}
   {\cal D}_{ab^\prime}={2\over3}\nabla_a\nabla_{b^\prime}-{1\over6}
   g_{ab^\prime}g_{dd^\prime}\nabla^d\nabla^{d^\prime}-{1\over3}
   \nabla_{a^\prime}\nabla_{b^\prime}+{1\over3}g_{ab^\prime}\nabla_
   {d^\prime}\nabla^{d^\prime}+{1\over6}\left(R_{ab}-{1\over2}Rg_{ab}\right),
   \label{E22}
\end{eqnarray}
where $g_{ab^\prime}$ is the geodesic parallel displacement bivector
\cite{dew65}. [It is easy to show that if $R_{ab}=0$ Eq.~(\ref{E21})
and Eq.~(\ref{E22}) are reduced to Eq.~(\ref{E5}).]
The regularized Hadamard function for the adapted conformal Minkowski
vacuum in Vilenkin's tunneling universe is 
\begin{eqnarray}
   G_{\rm
   reg}^{(1)}(X,X^\prime)=G^{(1)}(X,X^\prime)-G_{\rm ref}^{(1)}(X,X^\prime)
   =\left[G_{\rm CM}^{(1)}(X,X^\prime)-G_{\rm ref}^{(1)}\right] +
   G_{\rm CM}^{(1)}(X^-,X^\prime).
   \label{E20}
\end{eqnarray}
(In this paper the exact form of $G^{(1)}_{\rm ref}$ is not important for us.)
Substituting Eqs.~(\ref{E17}-\ref{E19}) and Eq.~(\ref{E20})
into Eq.~(\ref{E21}), we find that
$\lim_{X^\prime\rightarrow X}{\cal D}_{ab^\prime}G_{\rm CM}^{(1)}(X^-,X^\prime)=0,
$ which shows that the boundary condition (\ref{E15}) does not 
produce any renormalized energy-momentum tensor;
but the action of ${\cal D}_{ab^\prime}$ on $G_{\rm CM}^{(1)}
(X,X^\prime)-G_{\rm ref}^{(1)}$ should give the
energy-momentum tensor for the 
conformal Minkowski vacuum in an eternal de~Sitter space \cite{bun78,ber86}
\begin{eqnarray}
   {1\over2}\lim_{X^\prime\rightarrow X}
   {\cal D}_{ab^\prime}\left[G_{\rm CM}^{(1)}(X,X^\prime)-
   G_{\rm ref}^{(1)}\right]=-
   {1\over960\pi^2 r_0^4} g_{ab}.
   \label{E24}
\end{eqnarray}
Therefore, the energy-momentum tensor of a conformally
coupled scalar field in the adapted Minkowski vacuum in
Vilenkin's tunneling universe is 
\begin{eqnarray}
   \langle T_{ab}\rangle_{\rm ren}=-
   {1\over960\pi^2 r_0^4} g_{ab},
   \label{E25}
\end{eqnarray}
which is the same as that for an eternal de~Sitter space.

Now consider a particle detector moving along a geodesic with
$\chi,\theta,\phi={\rm constants}$. The response function is given
by Eq.~(\ref{E7}) but with the integration over $\tau$ and $\tau^\prime$
ranging from $0$ to $\infty$. The Wightman function is obtained from the
corresponding Hadamard function by the relation
\begin{eqnarray}
   G^+(\tau,\chi,\theta,\phi;\tau^\prime,\chi^\prime,\theta^\prime,
   \phi^\prime)={1\over2}G^{(1)}\left(\tau-i\epsilon/2,\chi,\theta,
   \phi;\tau^\prime+i\epsilon/2,\chi^\prime,\theta^\prime,
   \phi^\prime\right),
   \label{E26}
\end{eqnarray}
where $\epsilon$ is an infinitesimal positive real number. Along the
worldline of the detector, we have
\begin{eqnarray}
   &&Z(\tau,\tau^\prime)=-\sinh{\tau\over r_0}
   \sinh{\tau^\prime\over r_0^\prime}+\cosh{\tau\over r_0}
   \cosh{\tau^\prime\over r_0^\prime}=\cosh{\tau-\tau^\prime\over r_0},
   \label{E27}\\
   &&Z(-\tau,\tau^\prime)=+\sinh{\tau\over r_0}
   \sinh{\tau^\prime\over r_0^\prime}+\cosh{\tau\over r_0}
   \cosh{\tau^\prime\over r_0^\prime}=\cosh{\tau+\tau^\prime\over r_0},
   \label{E28}
\end{eqnarray}
and
\begin{eqnarray}
   G^+(X,X^\prime)={1\over8\pi^2 r_0^2}\left({1\over1-\cosh
   {\tau-\tau^\prime-i\epsilon\over r_0}}+{1\over1-\cosh{\tau+
   \tau^\prime\over r_0}}\right).
   \label{E29}
\end{eqnarray}
Then the response function is
\begin{eqnarray}
   {\cal F}(\Delta E)={1\over8\pi^2}\int_0^{\infty} dT\int_{-\infty}
   ^{\infty}d\Delta\tau e^{-i\Delta Er_0\Delta\tau}\left[{1\over1-\cosh
   (\Delta\tau-i\epsilon)}+{1\over1-\cosh 2T}\right],
   \label{E30}
\end{eqnarray}
where $\Delta\tau=(\tau-\tau^\prime)/ r_0$ and $T=
(\tau+\tau^\prime)/2 r_0$. It is easy to calculate the contour
integral over $\Delta\tau$. We find that the integration of the second
term is zero and therefore, the result is the same as that for an inertial particle
detector in an eternal de~Sitter space \cite{gib77,bir82}. Thus we have
\begin{eqnarray}
   {d{\cal F}\over dT}={r_0\over2\pi}{\Delta E\over e^{2\pi r_0 \Delta E}-1},
   \label{E31}
\end{eqnarray}
which is just the response function for a detector in 
a thermal radiation with the Gibbons-Hawking temperature \cite{gib77}
\begin{eqnarray}
   T_{\rm G-H}={1\over2\pi r_0}.
   \label{E32}
\end{eqnarray}
[The factor $r_0$ over $2\pi$ in Eq.~(\ref{E31}) is due to the fact
that by definition $T=(\tau+\tau^\prime)/2r_0$ is dimensionless.] 
Therefore such a detector perceives a thermal bath of radiation with
the temperature $T_{\rm G-H}$.

Though the boundary between the Lorentzian section and the 
Euclidean section behaves as a reflecting boundary, a
particle detector cannot distinguish Vilenkin's tunneling
universe from an eternal
de~Sitter space, and they have the same energy-momentum
tensor for the conformally coupled scalar field.

\section{A Time-Nonorientable de~Sitter Space}
\label{VIII}
A time-nonorientable de~Sitter space can be constructed from de
Sitter space by identifying antipodal points \cite{cal62,haw73}. Under such an
identification, point $X=(\tau,\chi,\theta,\phi)$ is identified
with $-X=(-\tau,\pi-\chi,\pi-\theta,\pi+\phi)$. Friedman and Higuchi
\cite{fri95,fri97} have 
described this space as a ``Lorentzian universe from nothing''
(without any Euclidean section),
although one could also describe it as always existing. Friedman and
Higuchi have studied quantum field theory in this space but have not
calculated the renormalized energy-momentum tensor \cite{fri95}.

De~Sitter space is the covering space of this
time-nonorientable model.
Using the method of images, the Hadamard function of a conformally
coupled scalar field in the time-nonorientable de~Sitter space with
the ``adapted'' conformal Minkowski vacuum can be constructed as
\begin{eqnarray}
   G^{(1)}(X,X^\prime)&=&G_{\rm CM}^{(1)}(X,X^\prime)+G_{\rm CM}^{(1)}(-X,X^\prime)
   ={1\over4\pi^2 r_0^2}\left[{1\over1-Z(X,X^\prime)}+{1\over1-Z(-X,X^\prime)}\right]
   \nonumber\\
   &=&{1\over4\pi^2 r_0^2}\left[{1\over1-Z(X,X^\prime)}+
   {1\over1+Z(X,X^\prime)}\right].
   \label{E33}
\end{eqnarray}
The regularized Hadamard function is
\begin{eqnarray}
   G_{\rm reg}^{(1)}(X,X^\prime)&=&G^{(1)}(X,X^\prime)-G_
   {\rm ref}^{(1)}(X,X^\prime)\nonumber\\
   &=&\left[G_{\rm CM}^{(1)}(X,X^\prime)-G_
   {\rm ref}^{(1)}(X,X^\prime)\right]+G_{\rm CM}^{(1)}(-X,X^\prime).
   \label{E34}
\end{eqnarray}
Inserting Eq.~(\ref{E33}) and Eq.~(\ref{E34}) into Eq.~(\ref{E21}), we find
that the contribution of $G_{\rm CM}^{(1)}(-X,X^\prime)$ to the
energy-momentum tensor is zero. Therefore
the renormalized energy-momentum tensor is the same as that in an eternal
de~Sitter space, which is given by Eq.~(\ref{E25}).

Suppose a particle detector moves along a worldline with $\chi,\theta,
\phi={\rm constants}$. The response function is given by
Eq.~(\ref{E7}). The Wightman function is obtained from the Hadamard
function through Eq.~(\ref {E26}). On the worldline of the
particle detector, we have
\begin{eqnarray}
   G^+(\tau,\tau^\prime)={1\over8\pi^2 r_0^2}\left({1\over1-\cosh{\tau-\tau^\prime-i
   \epsilon\over r_0}}+
   {1\over1+\cosh{\tau-\tau^\prime-i\epsilon\over r_0}}\right).
   \label{E35}
\end{eqnarray}
Inserting this into Eq.~(\ref{E7}) we get
\begin{eqnarray}
   {d{\cal F}\over dT}={r_0\over2\pi}{\Delta E\over e^{\pi r_0 \Delta E}-1},
   \label{E36}
\end{eqnarray}
which represents a thermal spectrum with a temperature equal to 
twice the Gibbons-Hawking temperature.
Therefore a particle detector moving along such a geodesic in this
time-nonorientable spacetime
perceives thermal radiation with temperature $T=2T_{\rm G-H}$.

For this time-nonorientable de~Sitter space, the area of the event
horizon is one half that of an eternal de~Sitter space. This together
with $T=2T_{\rm G-H}$ tells us that the first thermodynamic law of event
horizons $\delta M_c=T\delta A$ is preserved, where $M_c$ is the mass
within the horizon, and $A$ is the area of the horizon \cite{gib77}.

\section{A Multiply Connected de~Sitter Space with CTCs}
\label{IX}
\subsection{Construction of a Multiply Connected de~Sitter Space}
\label{IX.A}
De~Sitter space is a solution of the vacuum Einstein's equations with a
positive cosmological constant $\Lambda$, which is one of the maximally
symmetric spacetimes (the others being Minkowski space and anti-de~Sitter
space) \cite{wei72,haw73}. De~Sitter space can be represented
by a timelike hyperbolic hypersurface 
\begin{eqnarray}
   W^2+X^2+Y^2+Z^2-V^2= r_0^2,
   \label{E39}
\end{eqnarray}
embedded in a five-dimensional Minkowski
space $(V, W, X, Y, Z)$ with the metric
\begin{eqnarray}
   ds^2=-dV^2+dW^2+dX^2+dY^2+dZ^2,
   \label{E40}
\end{eqnarray}
where $r_0=(3/\Lambda)^{1/2}$ \cite{haw73,sch56}.
De~Sitter space has ten killing vectors ---
four of them are boosts, and the other six are
rotations. The global coordinates $(\tau,\chi,\theta,\phi)$ have been
described in previous sections. 
Static coordinates $(t,r,\theta,\phi)$ on de~Sitter
space are defined by
\begin{eqnarray}
   \left\{\begin{array}{l}
   V=( r_0^2-r^2)^{1/2}\sinh{t\over r_0},\\
   W=( r_0^2-r^2)^{1/2}\cosh{t\over r_0},\\
   X=r\sin\theta\cos\phi,\\
   Y=r\sin\theta\sin\phi,\\
   Z=r\cos\theta,
   \end{array}\right.
   \label{E41}
\end{eqnarray}
where $-\infty<t<\infty$, $0\leq r<r_0$, $0<\theta<\pi$, and $0\leq\phi<2\pi$.  
In these coordinates the de~Sitter metric is written as
\begin{eqnarray}
   ds^2=-\left(1-{r^2\over r_0^2}\right)dt^2+\left(1-{r^2\over r_0^2}
   \right)^{-1}dr^2+r^2
   (d\theta^2+\sin\theta^2 d\phi^2).
   \label{E42}
\end{eqnarray}
We divide de~Sitter space $dS$ into four regions
\begin{eqnarray}
   &&{\cal R}\equiv\{p\in dS\vert W>\vert V\vert\},\\
   \label{E43}
   &&{\cal L}\equiv\{p\in dS\vert W<-\vert V\vert\},\\
   \label{E44}
   &&{\cal F}\equiv\{p\in dS\vert V>\vert W\vert\},\\
   \label{E45}
   &&{\cal P}\equiv\{p\in dS\vert V<-\vert W\vert\},
   \label{E46}
\end{eqnarray}
which are separated by horizons where $W=\pm V$ and
$X^2+Y^2+Z^2=r_0^2$. (See Fig.~\ref{f4}). It is
obvious that the static
coordinates defined by Eq.~(\ref{E41}) only cover region
${\cal R}$. However, similar to the Rindler coordinates, these static 
coordinates 
can be extended to region ${\cal F}$ by the
complex transformation
\begin{eqnarray}
   t\rightarrow l-i{\pi\over2} r_0,~~~
   r\rightarrow\tilde{t},~~~
   \theta\rightarrow\theta,~~~
   \phi\rightarrow\phi,
   \label{E47}
\end{eqnarray}
where $-\infty<l<\infty$ and $\tilde{t}>2r_0$. 
In region ${\cal F}$, with the coordinates $(\tilde{t},l,\theta,\phi)$,
the de~Sitter metric can be written as
\begin{eqnarray}
   ds^2=-\left({\tilde{t}^2\over r_0^2}-1\right)^{-1}d\tilde{t}^2+\left(
   {\tilde{t}^2\over r_0^2}-1\right)dl^2+
   \tilde{t}^2(d\theta^2+\sin\theta^2 d\phi^2).
   \label{E48}   
\end{eqnarray}
Transforming the coordinate $\tilde{t}$ to the proper time $\tau$ by
\begin{eqnarray}
   \tilde{t}= r_0\cosh{\tau\over r_0},
   \label{E49}
\end{eqnarray}
the de~Sitter metric in ${\cal F}$ is written as
\begin{eqnarray}
   ds^2=-d\tau^2+ \sinh^2{\tau\over r_0}d l^2+
    r_0^2\cosh^2{\tau\over r_0}(d\theta^2+\sin^2\theta d\phi^2).
   \label{E50}
\end{eqnarray}
(See Fig.~\ref{f4}.) The coordinates 
$(\tau,l,\theta,\phi)$ are related to $(V,W,X,Y,Z)$ by
\begin{eqnarray}
   \left\{\begin{array}{l}
   V=r_0\sinh{\tau\over r_0}\cosh{l\over r_0},\\
   W=r_0\sinh{\tau\over r_0}\sinh{l\over r_0},\\
   X=r_0\cosh{\tau\over r_0}\sin\theta\cos\phi,\\
   Y=r_0\cosh{\tau\over r_0}\sin\theta\sin\phi,\\  
   Z=r_0\cosh{\tau\over r_0}\cos\theta.
   \end{array}\right.
   \label{E50a}
\end{eqnarray} 
The universe with metric (\ref{E50}) 
is a type of Kantowski-Sachs universe \cite{kan66}. Any hypersurface
of $\tau={\rm constant}$ has topology $R^1\times S^2$ and has four
killing vectors. Similarly, the static coordinates can also be extended to
${\cal P}$ and ${\cal L}$. 

Another coordinate system which will be used in this paper is the steady-state
coordinate system $(\tau,x,y,z)$, defined by
\begin{eqnarray}
   \left\{\begin{array}{l}
   \tau= r_0\ln{W+V\over r_0},\\
   x={ r_0 X\over W+V},\\
   y={ r_0 Y\over W+V},\\
   z={ r_0 Z\over W+V}.
   \end{array}\right.
   \label{E51}
\end{eqnarray} 
These coordinates cover regions ${\cal R}+{\cal F}$ 
and the horizon at $W=V>0$. With
these steady-state coordinates, the de~Sitter metric can be written in the
steady-state form
\begin{eqnarray}
   ds^2=-d\tau^2+e^{2\tau/ r_0}(dx^2+dy^2+dz^2).
   \label{E52}
\end{eqnarray}
Introducing the conformal time
\begin{eqnarray}
   \overline{\eta}=- r_0 e^{-\tau/ r_0}=-{r_0^2\over W+V},
   \label{E53}
\end{eqnarray}
and spherical coordinates $(\rho,\theta,\phi)$ defined by
$x=\rho\sin\theta\cos\phi$, $y=\rho\sin\theta\sin\phi$, and
$z=\rho\cos\theta$, 
the de~Sitter metric can be written as
\begin{eqnarray}
   ds^2={ r_0^2\over{\overline{\eta}}^2}\left[-d\overline{\eta}^2+d\rho^2+
   \rho^2(d\theta^2 +\sin^2\theta d\phi^2)\right].
   \label{E54}
\end{eqnarray}

The de~Sitter metric is invariant under the action of the de~Sitter
group. Because the boost group in de~Sitter space is a sub-group of
the de
Sitter group, the de~Sitter metric is also invariant under the action
of the boost group. A boost transformation
in the $(V,W)$ plane in the embedding five-dimensional Minkowski space
induces a boost transformation in the de~Sitter space. Under such a
transformation, point $(V,W,X,Y,Z)$ is taken to $(V\cosh b+W\sinh 
b, W\cosh b+V\sinh b, X, Y, Z)$. In static coordinates
in ${\cal R}$, point $(t,r,\theta,\phi)$ is taken to 
$(t+\beta,r,\theta,\phi)$ where $\beta=br_0$. In coordinates
$(\tilde{t}, l, \theta, \phi)$ in ${\cal F}$, point $(\tilde{t},
l, \theta, \phi)$ is taken to $(\tilde{t}, l+\beta,
\theta, \phi)$. Similar to Misner space, our 
multiply connected de~Sitter space is constructed  
by identifying points $(V,W,X,Y,Z)$ with 
$(V\cosh nb+W\sinh nb,
W\cosh nb+V\sinh nb, X, Y, Z)$ on de~Sitter space $dS$.
In regions ${\cal R}$, points
$(t,r,\theta,\phi)$ are identified 
with $(t+n\beta,r,\theta,\phi)$; in region ${\cal F}$, points $(\tilde{t},
l, \theta, \phi)$ are identified with $(\tilde{t},
l+n\beta, \theta, \phi)$. We denote the multiply connected de
Sitter space so obtained by $dS/B$, where $B$ denotes the 
boost group. Under the identification generated by the boost transformation,
clearly $dS/B$ has CTCs in regions 
${\cal R}$ and ${\cal L}$, but has no closed
causal curves in regions ${\cal F}$ and ${\cal P}$. 
The boundaries at $W=\pm V$ and $X^2+Y^2+Z^2=r_0^2$ are
the Cauchy horizons which separate the causal regions ${\cal F}$ and
${\cal P}$ from the acausal regions ${\cal R}$ and ${\cal L}$ and are generated
by closed null geodesics (Fig.~\ref{f4}).

Similar to
the case of Misner space, $dS/B$ is not a manifold at the two-sphere
defined by $W=V=0$ and $X^2+Y^2+Z^2= r_0^2$.
However, as in Hawking and Ellis's arguments for Misner space \cite{haw73}, the
quotient of the bundle of linear frames over de~Sitter space by the boost group is
a Hausdorff manifold and thus is well-behaved everywhere. It may not
be a serious problem in physics that $dS/B$ is not a
manifold at the two-sphere mentioned above since this is a set of
measure zero.

\subsection{Conformal Relation between Our Multiply Connected de
Sitter Space and Misner Space}
\label{IX.B}
It is well known that de~Sitter space is conformally flat. The de
Sitter metric is
related to the Minkowski metric by the conformal transformation
\begin{eqnarray}
   g_{ab}=\Omega^2\eta_{ab}.
   \label{E58}
\end{eqnarray}
It is easy to show this relation by writing the steady-state de~Sitter
metric using conformal time [see Eq.~(\ref{E54})]. However, in this
paper it is more convenient to show this conformal relation by writing the
de~Sitter metric in the static form and the Minkowski metric in the
Rindler form, and using the transformation \cite{can79}
\begin{eqnarray}
   \left\{\begin{array}{l}
   \eta={t\over r_0},\\
   \xi={\sqrt{1-r^2/ r_0^2}\over1-r\cos\theta/ r_0},\\
   y={r\sin\theta\cos\phi/ r_0\over 1-r\cos\theta/ r_0},\\
   z={r\sin\theta\sin\phi/ r_0\over 1-r\cos\theta/ r_0},
   \end{array}
   \right.
   \label{E59}
\end{eqnarray}
then the conformal factor $\Omega^2$ is
\begin{eqnarray}
   \Omega^2= r_0^2(1-r\cos\theta/ r_0)^2.
   \label{E60}
\end{eqnarray}
The conformal relations given by Eq.~(\ref{E59}) and Eq.~(\ref{E60})
define a {\em conformal map} between the static de~Sitter space and
the Rindler space. The horizon at $r=r_0$ in
the static de~Sitter space coordinates
corresponds to the horizon $\xi=0$ in Rindler space, and the
worldline $r=0$ in de~Sitter space corresponds to the worldline with
$\xi=1$ and $y=z=0$ in Rindler space. This conformal relation can also be extended
to region ${\cal F}$ in de~Sitter space and region F in Minkowski
space, where we have
\begin{eqnarray}
   \left\{\begin{array}{l}
   \tilde{\eta}=\pm{\sqrt{\tilde{t}^2/ r_0^2-1}\over1-\tilde{t}
   \cos\theta/ r_0},\\
   \tilde{\xi}={l\over r_0},\\
   y={\tilde{t}\sin\theta\cos\phi/ r_0\over 1-\tilde{t}\cos\theta/ r_0},\\
   z={\tilde{t}\sin\theta\sin\phi/ r_0\over 1-\tilde{t}\cos\theta/ r_0},
   \end{array}
   \right.
   \label{E59a}
\end{eqnarray}
and
\begin{eqnarray}
   \Omega^2= r_0^2(1-\tilde{t}\cos\theta/ r_0)^2.
   \label{E60a}
\end{eqnarray}
Eq.~(\ref{E59a}) and Eq.~(\ref{E60a}) give a {\em locally} conformal
map in the sense that in ${\cal F}$ in de Sitter space, the map given
by Eq.~(\ref{E59a}) and Eq.~(\ref{E60a}) with a ``$+$'' sign only covers
$\theta_0<\theta<\pi$, where $\theta_0={\rm Arccos}(r_0/\tilde{t})$;
the map given
by Eq.~(\ref{E59a}) and Eq.~(\ref{E60a}) with a ``$-$'' sign only covers
$0<\theta<\theta_0$. (Remember that in F in Rindler space we 
have $\tilde{\eta}>0$.) This conformal map is singular at
$\theta=\theta_0$. However, since
the hypersurfaces $\tilde{t}={\rm constant}$ and $\tilde{\eta}={\rm
constant}$ are homogeneous, in a
neighborhood of any point in region F, we can always adjust coordinates
$(\theta,\phi)$ so that Eq.~(\ref{E59a}) and Eq.~(\ref{E60a}) hold,
except for the points lying in region O defined by
$\tilde{\eta}^2\geq 1+y^2+z^2$ (i.e. $t^2-x^2-y^2-z^2\geq1$) in F; 
because as $\tilde{t}\rightarrow\infty$
we have $\tilde{\eta}^2/(1+y^2+z^2)\rightarrow 1$. This means that
there always exists a {\em locally} conformal map between ${\cal
F}$ and F-O (defined by $t^2-x^2-y^2-z^2<1$ in F), 
and future infinity ($\tilde{t}\rightarrow\infty$) in
${\cal F}$ corresponds to the hyperbola 
$\tilde{\eta}^2= 1+y^2+z^2$ (i.e. $t^2-x^2-y^2-z^2=1$) in F. 

With the above conformal transformation, Misner space is 
naturally transformed to the
multiply connected de~Sitter space $dS/B$ with
\begin{eqnarray}
   \beta= br_0.
   \label{E61}
\end{eqnarray}

For a conformally coupled scalar field in a conformally flat
spacetime, the Green function $G(X,X^\prime)$ of the conformal vacuum
is related to the corresponding Green function $\overline{G}(X,X^\prime)$ in
the flat spacetime by \cite{bir82}
\begin{eqnarray}
   G(X,X^\prime)=\Omega^{-1}(X)\overline{G}(X,X^\prime)\Omega^{-1}(X^\prime),
   \label{E62}
\end{eqnarray}
the renormalized energy-momentum tensors are related by \cite{bir82}
\begin{eqnarray}
   \langle T_a^{~b}\rangle_{\rm ren}=\Omega^{-4}\langle\overline
   {T}_a^{~b}\rangle_{\rm ren}+{1\over16\pi^2}\left[{1\over9}
   a_1 ~^{(1)}H_a^{~b}+2a_3 ~^{(3)}H_a^{~b}\right],
   \label{E63}
\end{eqnarray}
where
\begin{eqnarray}
   &&^{(1)}H_{ab}=2\nabla_a\nabla_bR-2g_{ab}\nabla^c\nabla_cR
   -{1\over2}R^2g_{ab}+2RR_{ab}, \\
   &&^{(3)}H_{ab}=R_a^{~c}R_{cb}-{2\over3}RR_{ab}-{1\over2}R_{cd}R^{cd}
   g_{ab}+{1\over4}R^2g_{ab},
   \label{E63a}
\end{eqnarray}
and for scalar field we have $a_1={1\over120}$ and $a_3=-{1\over360}$
\cite{bir82}. [The sign before $1/16\pi^2$ is positive here because we
are using signature $(-,+,+,+)$]. For
de~Sitter space we have $R_{ab}=\Lambda g_{ab}$, $R=4\Lambda$,
and thus $^{(1)}H_{ab}=0$, $^{(3)}H_{ab}={1\over3}\Lambda^2g_{ab}
={3\over r_0^4}g_{ab}$.
Inserting them into Eq.~(\ref{E63}), we have
\begin{eqnarray}
   \langle T_a^{~b}\rangle_{\rm ren}=\Omega^{-4}\langle\overline
   {T}_a^{~b}\rangle_{\rm ren}-{1\over960\pi^2 r_0^4}\delta_a^{~b}.
   \label{E65}
\end{eqnarray}
Since the renormalized energy-momentum tensor for Minkowski space in
the Minkowski vacuum is zero, we have $\langle\overline
{T}_a^{~b}\rangle_{\rm ren}=0$, and thus for a conformally
coupled scalar field in the conformal Minkowski vacuum in a simply
connected de~Sitter space $dS$
\begin{eqnarray}
   \langle T_{ab}\rangle_{\rm ren}=-{1\over960\pi^2 r_0^4}g_{ab},
   \label{E66}
\end{eqnarray}
which is just the expected result
[see Eq.~(\ref{E25})].

If we insert the energy-momentum tensor in Eq.~(\ref{E66}) into the
semiclassical Einstein's equations 
\begin{eqnarray}
   G_{ab}+\Lambda g_{ab}=8\pi\langle T_{ab}\rangle_{\rm ren},
   \label{E75a}
\end{eqnarray}
and recall that for de~Sitter space we have $G_{ab}=R_{ab}-{1\over2}R
g_{ab}=-{3\over r_0^2}g_{ab}$, we find that the semiclassical 
Einstein's equations are satisfied if and only if
\begin{eqnarray}
   \Lambda-{3\over r_0^2}+{1\over120\pi r_0^4}=0.
   \label{E75b}
\end{eqnarray}
If $\Lambda=0$, the solutions to Eq.~(\ref{E75b}) are
$r_0=(360\pi)^{-1/2}$ and $r_0=\infty$ \cite{got82}. Gott \cite{got82} has 
called the vacuum state in de~Sitter space with $r_0=(360\pi)^{-1/2}$
the self-consistent vacuum state (it has a Gibbons-Hawking
thermal temperature $T_{\rm G-H}=1/2\pi r_0$) \cite{gib77}. In this self-consistent
case, $\langle T_{ab}\rangle_{\rm ren}=-g_{ab}/960\pi^2r_0^4$ itself
is the source term producing the de~Sitter geometry
\cite{got82}. This may give rise to inflation at the Planck scale
\cite{got82}. (In a recent paper of Panagiotakopoulos and Tetradis
\cite{pan97}, inflation at the Planck scale has been suggested to lead
to homogeneous initial conditions for a second stage inflation at the GUT
scale.) The second solution
$r_0=\infty$ corresponds to Minkowski
space. These perhaps supply a possible reason that the effective cosmological
constant is either of order unity in Planck units or exactly zero.
That is interesting because we observe $\Lambda_{\rm eff}=0$ today and
a high $\Lambda_{\rm eff}$ is needed for inflation.
If $\Lambda\not=0$, we find that the solutions to
Eq.~(\ref{E75b}) are
\begin{eqnarray}
   r_0^2={3\over2\Lambda}\left(1\pm\sqrt{1-{\Lambda\over270\pi}}\right).
   \label{E75c}
\end{eqnarray}
A de~Sitter space with $r_0$ given by Eq.~(\ref{E75c}) automatically
satisfies the semiclassical 
Einstein's equations (\ref{E75a}). Such a de~Sitter space and its
corresponding vacuum are thus self-consistent.

\subsection{Renormalized Energy-Momentum Tensor in Multiply Connected de
Sitter Space}
From Eq.~(\ref{E65}) we find that if we know the
energy-momentum tensor of a conformally coupled scalar field in some
vacuum state in Misner space, we can get the 
energy-momentum tensor in the corresponding conformal vacuum in the
multiply connected de~Sitter space. 

Two fundamental
vacuums in Minkowski space are the Minkowski vacuum and
the Rindler vacuum \cite{bir82,wal94}. The energy-momentum tensor of the
conformally coupled scalar field in the adapted Minkowski vacuum in
Misner space has been worked out by Hiscock and Konkowski
\cite{his82}; their results are given by Eq.~(\ref{E67}).
Inserting Eq.~(\ref{E67}) into Eq.~(\ref{E65}), and using
Eqs.~(\ref{E59}-\ref{E61}), we obtain the
energy-momentum tensor of a conformally coupled scalar field in the
adapted conformal Minkowski vacuum in our multiply connected de~Sitter
space $dS/B$. In static coordinates $(t,r,\theta,\phi)$, it is
written as
\begin{eqnarray}  
   \langle T_\mu^{~\nu}\rangle_{\rm CM,ren}={\tilde{A}\over12\pi^2
    r_0^4\left(1-r^2/ r_0^2\right)^2}
   \left(\begin{array}{cccc}
   -3&0&0&0\\
   0&1&0&0\\
   0&0&1&0\\
   0&0&0&1
   \end{array}
   \right)-{1\over960\pi^2 r_0^4}\delta_{\mu}^{~\nu},
   \label{E71}
\end{eqnarray}
where
\begin{eqnarray}
   \tilde{A}=\sum_{n=1}^{\infty}{2+\cosh {n\beta\over r_0}
   \over\left(\cosh {n\beta\over r_0}-1\right)^2}.
   \label{E72}
\end{eqnarray}
This result is defined in region ${\cal R}$, but it can be extended to region
${\cal F}$ through 
the transformation in Eq.~(\ref{E47}), and can also be extended to
region ${\cal L}$ and ${\cal P}$ through similar transformations. Similar
to Misner space, this
energy-momentum tensor diverges at the Cauchy horizon as $r\rightarrow r_0$ for
any finite $\beta$; and the divergence is coordinate independent
since $\langle T^{\mu\nu}\rangle_{\rm CM,ren}\langle T_{\mu\nu}\rangle_{\rm
CM,ren}$ also diverges there. Though the conformal Minkowski vacuum is
a good vacuum for simply connected de~Sitter space \cite{bun78,ber86}, it (in the
adapted version) is not self-consistent 
for the multiply connected de~Sitter space
$dS/B$. (That is, it does not solve the semiclassical Einstein's equations.)

In the case of an eternal Schwarzschild black
hole, there are the Boulware vacuum \cite{bou75} and the Hartle-Hawking vacuum
\cite{har76}. The globally defined Hartle-Hawking vacuum bears essentially the
same relationship to the Boulware vacuum as the Minkowski vacuum does
to the Rindler vacuum \cite{sci81}. For the Boulware vacuum, the
energy-momentum tensor diverges at the event horizon of the
Schwarzschild black hole, which means that this
state is {\em not} a good vacuum for the Schwarzschild black hole
because, when one inserts this energy-momentum tensor back into
Einstein's equations, the back-reaction will seriously alter the
Schwarzschild geometry near the event horizon. For the Hartle-Hawking vacuum,
however, the energy-momentum tensor is finite everywhere and a
static observer outside the horizon sees Hawking radiation \cite{can80}. People
usually regard the Hartle-Hawking vacuum as the reasonable vacuum
state for an eternal Schwarzschild black hole because, when its
energy-momentum tensor is fed back into Einstein's equations, the
Schwarzschild geometry is only altered slightly \cite{yor85}. Therefore, in the
case of Misner space, Li and Gott \cite{li97} have tried to find a vacuum which is
also self-consistent and found that the adapted Rindler vacuum is
such a vacuum if $b=2\pi$. 

Here we also try to find a
self-consistent vacuum for our multiply connected de~Sitter space.
Let us consider the adapted conformal Rindler vacuum in $dS/B$. 
The energy-momentum tensor of a conformally coupled scalar field in
the adapted Rindler vacuum in Misner space is given by Eq.~(\ref{E69}).
Inserting Eq.~(\ref{E69}) into Eq.~(\ref{E65}) and using
Eqs.~(\ref{E59}-\ref{E61}), we obtain the
energy-momentum tensor for the adapted conformal Rindler vacuum 
of a conformally coupled scalar field in our multiply connected de
Sitter space
\begin{eqnarray}  
   \langle T_\mu^{~\nu}\rangle_{\rm CR,ren}={1\over1440\pi^2
   r_0^4\left(1-r^2/ r_0^2\right)^2}\left[\left({2\pi r_0
   \over\beta}\right)^4-1\right]
   \left(\begin{array}{cccc}
   -3&0&0&0\\
   0&1&0&0\\
   0&0&1&0\\
   0&0&0&1
   \end{array}
   \right)-{1\over960\pi^2 r_0^4}\delta_{\mu}^{~\nu},
   \label{E73}
\end{eqnarray} 
where the coordinate system is the static coordinate system
$(t,r,\theta,\phi)$. Similarly, this result
can also be analytically extended to the whole $dS/B$, though the
static coordinates only cover region ${\cal R}$. We find that, if
\begin{eqnarray}
   \beta=2\pi r_0,
   \label{E74}
\end{eqnarray}
this energy-momentum tensor is regular on the whole space.
[Eq.~(\ref{E74}) corresponds to $b=2\pi$ via Eq.~(\ref{E61})]. 
Otherwise both $\langle T_\mu^{~\nu}\rangle_{\rm CR,ren}$ and
$\langle T^{\mu\nu}\rangle_{\rm CR,ren}\langle T_{\mu\nu}\rangle_{\rm
CR,ren}$ diverge as the Cauchy horizon is approached. For the case
$\beta=2\pi r_0$, the energy-momentum tensor is
\begin{eqnarray}
   \langle T_{ab}\rangle_{\rm CR,ren}=-{1\over960\pi^2 r_0^4}g_{ab},
   \label{E75}
\end{eqnarray}
which is the same as the energy-momentum tensor for the 
conformal Minkowski vacuum in the simply connected de~Sitter space. 

The Euclidean section of our multiply connected de~Sitter space is a
four-sphere $S^4$ embedded in a five dimensional flat Euclidean space
with those points related by an azimuthal rotation with angle $\beta/r_0$ being
identified. There are conical
singularities unless $\beta/r_0=2\pi$. This may be regarded as a
geometrical explanation of the self-consistent condition in (\ref{E74}). 

Similarly, our multiply connected de~Sitter space solves the
semiclassical Einstein's
equations with a cosmological constant $\Lambda$ and the energy-momentum
tensor in Eq.~(\ref{E75}) (and thus it is self-consistent) if
$r_0^2={3\over2\Lambda}\left(1\pm\sqrt{1-{\Lambda\over 270\pi}}\right)$ (if
$\Lambda=0$, we have the two solutions
$r_0^2=1/360\pi$ and $r_0=\infty$ \cite{got82}).

\subsection{Particle Detectors in the Multiply Connected de~Sitter Space}
It is well known that in the simply connected de~Sitter space, an
inertial particle detector perceives thermal radiation with the
Gibbons-Hawking temperature [Eq.~(\ref{E32})] if the conformally
coupled scalar field is in the conformal Minkowski vacuum \cite{gib77,bir82}. Now we
want to find what a
particle detector perceives in the adapted conformal Rindler vacuum
in our multiply connected de~Sitter space.

The response function of the particle detector is still given 
by Eq.~(\ref{E7}). The Wightman function is obtained from the
corresponding Hadamard function by Eq.~(\ref{E26}). The Hadamard
function for the conformally coupled scalar field in multiply
connected de~Sitter space is related to that in Misner space via 
Eq.~(\ref{E62}) [with $G(X,X^\prime)$ replaced by $G^{(1)}(X,X^\prime)$].
The Hadamard function for the adapted Rindler vacuum in Misner space
is given by Eq.~(\ref{E78}).
Inserting Eq.~(\ref{E78}) [as $\overline{G}_R^{(1)}$] 
into Eq.~(\ref{E62}) and using Eqs.~(\ref{E59}-\ref{E61}), 
we obtain the Hadamard function for the adapted conformal
Rindler vacuum of the conformally coupled scalar field in our multiply
connected de~Sitter space
\begin{eqnarray}
   G_{\rm CR}^{(1)}(X,X^\prime)={1\over2\pi^2}\sum_{n=-\infty}^{\infty}
   {\gamma\over\sinh\gamma\sqrt{(1-r^2/ r_0^2)(1-{r^\prime}^2/ r_0^2)}
   ~\left[-(t-t^\prime+n\beta)^2+ r_0^2\gamma^2\right]},
   \label{E79}
\end{eqnarray}
where $X=(t,r,\theta,\phi)$, $X^\prime=(t^\prime,r^\prime,
\theta^\prime,\phi^\prime)$, and $\gamma$ is written in $(t,r,
\theta,\phi)$ as
\begin{eqnarray}
   \cosh\gamma={1\over\sqrt{(1-r^2/ r_0^2)(1-{r^\prime}^2/ r_0^2)}}\left\{
   1-{rr^\prime\over r_0^2}[\cos\theta\cos\theta^\prime+\sin\theta
   \sin\theta^\prime\cos(\phi-\phi^\prime)]\right\}.
   \label{E80}
\end{eqnarray}
The Wightman function is obtained from Eq.~(\ref{E79}) via Eq.~(\ref{E26}). The
Hadamard function given by Eq.~(\ref{E79}) and the Wightman function
obtained from that are defined in region ${\cal R}$ 
in the multiply connected de
Sitter space, but they can be analytically extended to region ${\cal F}$ via
the transformation in Eq.~(\ref{E47}). However, it should be noted that
as we make the continuation from ${\cal R}$ to ${\cal F}$,
$\sqrt{(1-r^2/r_0^2)(1-{r^\prime}^2/r_0^2)}$ should be continued to be
$-\sqrt{(\tilde{t}^2/r_0^2-1)(\tilde{t}^{\prime2}/r_0^2-1)}$ instead of
$+\sqrt{(\tilde{t}^2/r_0^2-1)(\tilde{t}^{\prime2}/r_0^2-1)}$. This is
because if we take $\sqrt{1-z^2}=i\sqrt{z^2-1}$, we should also take 
$\sqrt{1-{z^\prime}^2}=i\sqrt{{z^\prime}^2-1}$ (instead of 
$-i\sqrt{{z^\prime}^2-1}$) ($z$ and $z^\prime$ should be continued
along the same path), thus $\sqrt{(1-z^2)({z^\prime}^2-1)}=
(i\sqrt{z^2-1}) (i\sqrt{{z^\prime}^2-1})=-\sqrt{(z^2-1)({z^\prime}^2-1)}$.
Using similar transformations, the results can also be continued to
regions ${\cal P}$ and ${\cal L}$ (we do not write them out 
because we do not use them here). 

We consider particle detectors moving along three kinds of worldlines
in our multiply connected de~Sitter space:

{\sl 1. A particle detector moving along a worldline with
$r,\theta,\phi={\rm constants}$ in ${\cal R}$.} In such a case, on the
worldline of the particle detector, $\gamma$ is zero and the Hadamard
function is reduced to
\begin{eqnarray}
   G_{\rm CR}^{(1)}(\tau,\tau^\prime)&=&-{1\over2\pi^2(1-r^2/r_0^2)}
   \sum_{n=-\infty}^{\infty}{1\over(t-t^\prime+n\beta)^2}\nonumber\\
   &=&-{1\over2\pi^2}\sum_{n=-\infty}^{\infty}{1\over\left(\tau-\tau^\prime+n
   \beta\sqrt{1-r^2/r_0^2}\right)^2},
   \label{E81}
\end{eqnarray}
where $\tau=t\sqrt{1-r^2/r_0^2}$ is the proper time of the particle
detector. The corresponding Wightman function obtained from
Eq.~(\ref{E26}) is
\begin{eqnarray}
   G_{\rm CR}^+(\tau,\tau^\prime)
   =-{1\over4\pi^2}\sum_{n=-\infty}^{\infty}{1\over\left(\tau-\tau^\prime+n
   \beta\sqrt{1-r^2/r_0^2}-i\epsilon\right)^2},
   \label{E82}
\end{eqnarray}
where $\epsilon$ is an infinitesimal positive real number. Inserting
it into Eq.~(\ref{E7}), obviously the integration over
$\Delta\tau=\tau-\tau^\prime$ is zero since all poles of the integrand
are in the upper-half plane of complex $\Delta\tau$ while the
integration contour is closed in the lower-half plane. Therefore the
response function ${\cal F}(\Delta E)$ is zero and no particles are detected.
All of these worldlines are accelerated, except for the one at $r=0$.

{\sl 2. A particle detector moving along a geodesic with
$l,\theta,\phi={\rm constant}$ in region ${\cal F}$.} In this region
the Hadamard function is
\begin{eqnarray}
   G_{\rm CR}^{(1)}(X,X^\prime)=-{1\over2\pi^2}\sum_{n=-\infty}^{\infty}
   {\tilde{\gamma}\over\sinh\tilde{\gamma}\sqrt{(\tilde{t}^2/r_0^2-1)
   (\tilde{t}^{\prime2}/r_0^2-1)}~
   \left[-(l-l^\prime+n\beta)^2+ r_0^2\tilde{\gamma}^2\right]},
   \label{E83}
\end{eqnarray}
where $\tilde{\gamma}$ is given by
\begin{eqnarray}
   \cosh\tilde{\gamma}={1\over\sqrt{(\tilde{t}^2/ r_0^2-1)({\tilde{t^
   \prime}}^2/ r_0^2-1)}}\left\{
   -1+{\tilde{t}\tilde{t}^\prime\over r_0^2}
   \left[\cos\theta\cos\theta^\prime+\sin\theta
   \sin\theta^\prime\cos(\phi-\phi^\prime)\right]\right\}.
   \label{E84}
\end{eqnarray}
[Eq.~(\ref{E83}) and Eq.~(\ref{E84}) are obtained from Eq.~(\ref{E79})
and Eq.~(\ref{E80}) via the transformation in Eq.~(\ref{E47}) respectively.]
On the worldline of the particle detector, the Hadamard function is
reduced to
\begin{eqnarray}
   G_{\rm CR}^{(1)}(\tilde{t},\tilde{t}^\prime)=-
   {1\over2\pi^2}\sum_{n=-\infty}^{\infty}
   {\tilde{\gamma}\over\sinh\tilde{\gamma}\sqrt{(\tilde{t}^2/r_0^2-1)
   (\tilde{t}^{\prime2}/r_0^2-1)}~
   \left(-n^2\beta^2+ r_0^2\tilde{\gamma}^2\right)},
   \label{E85}
\end{eqnarray}
and $\cosh\tilde{\gamma}$ is reduced to
\begin{eqnarray}
   \cosh\tilde{\gamma}={\tilde{t}\tilde{t}^\prime/r_0^2-
   1\over\sqrt{(\tilde{t}^2/ r_0^2-1)({\tilde{t^
   \prime}}^2/ r_0^2-1)}}.
   \label{E86}
\end{eqnarray}
Using the proper time $\tau$ defined by Eq.~(\ref{E49}), on the 
worldline of the particle detector $\cosh\tilde{\gamma}$ and $G_{\rm CR}^{(1)}$ can be
written as
\begin{eqnarray}
   \cosh\tilde{\gamma}={\cosh2T+\cosh\Delta\tau-2\over\cosh2T-\cosh\Delta\tau},
   \label{E88} 
\end{eqnarray}
and 
\begin{eqnarray}
   G_{\rm CR}^{(1)}(T,\Delta\tau)=
   {1\over\pi^2r_0^2}\sum_{n=-\infty}^{\infty}
   {\tilde{\gamma}\over\sinh\tilde{\gamma}(\cosh2T-\cosh\Delta\tau)
   (n^2b^2- \tilde{\gamma}^2)},
   \label{E89}
\end{eqnarray}
where $\tau>0$, $\tau^\prime>0$, $\Delta\tau=(\tau-\tau^\prime)/r_0$,
$T=(\tau+\tau^\prime)/2r_0$, and $b=\beta/r_0$. The Wightman
function is equal to one half of the Hadamard function with
$\Delta\tau$ replaced by $\Delta\tau-i\epsilon$ [Eq.~(\ref{E26})]. 
Thus the response function is
\begin{eqnarray}
   {\cal F}(\Delta E)=\sum_{n=-\infty}^\infty{\cal F}_n(\Delta E),
   \label{E90}
\end{eqnarray}
where
\begin{eqnarray}
  {\cal F}_n(\Delta E)&=&{1\over2\pi^2 r_0^2}\int_{0}^{\infty}
  dT\int_{-\infty}^\infty d\Delta\tau
  e^{-i\Delta Er_0\Delta\tau}\times\nonumber\\
  &&\left[{\tilde{\gamma}\over
  \sinh\tilde{\gamma}~(\cosh2T-\cosh\Delta\tau)(n^2b^2-
  \gamma^2)}\right]_{\Delta\tau\rightarrow\Delta\tau-i\epsilon}.
  \label{E91}
\end{eqnarray}
Now we consider the poles in the complex $\Delta\tau$ plane of the
integrand in the integral of ${\cal F}_n(\Delta E)$. The poles are given by
the equation
\begin{eqnarray}
   \tilde\gamma=\pm nb.
   \label{E92}
\end{eqnarray}
(It is easy to check that $\cosh2T=\cosh\Delta\tau$ does not give any
poles.) From Eq.~(\ref{E92}) and Eq.~(\ref{E88}), we have (we neglect
the term
$i\epsilon$, and at the end of the calculation we return it back to
the expressions)
\begin{eqnarray}
   \cosh2T+\cosh\Delta\tau-2=\cosh nb ~(\cosh2T-\cosh\Delta\tau).
   \label{E93}
\end{eqnarray}
Solutions to Eq.~(\ref{E93}) are
\begin{eqnarray}
   \Delta\tau=\Delta\tau_n+i2m\pi\equiv\Delta\tau_{nm},
   \label{E94}
\end{eqnarray}
where
\begin{eqnarray}
   \Delta\tau_n=\pm{\rm Arccosh}{(\cosh nb-1)\cosh2T+2\over
   \cosh nb+1}=\pm2{\rm Arcsinh}\left(\sinh T\tanh{nb\over2}\right),
   \label{E95}
\end{eqnarray}
where ${\rm Arccosh}z$ is the principal value of ${\rm arccosh}z$,
and here it is real (similarly for ${\rm Arcsinh}z$). We need to check
if all $\Delta\tau_{nm}$ are roots of
Eq.~(\ref{E92}), because the number of
roots might increase as we go from Eq.~(\ref{E92}) to
Eq.~(\ref{E93}). [E.g., for any integer $m$, $x_m=\pm2+im\pi$ solves
the equation $\cosh(2x)=\cosh4$; but, only $x_0=+2$ solves the equation
$2x=4$.] $\Delta\tau_n$ is obviously a root of
Eq.~(\ref{E92}). The question is: as $\Delta\tau$ goes from
$\Delta\tau_n$ to $\Delta\tau_n+i2m\pi$, does Eq.~(\ref{E88})
give the same $\tilde{\gamma}$ which is a real value [$\pm nb$; see
Eq.~(\ref{E92})]?
(Remember that ${\rm arccosh}z$ is a multi-valued complex function.) To
answer this question, let $\Delta\tau=\Delta\tau_n+i\theta$ (where $\theta$ is
real). Then from Eq.~(\ref{E88}) we have
\begin{eqnarray}
   \tilde{\gamma}&=&{\rm arccosh}{\cosh2T+\cosh\Delta\tau-
   2\over\cosh2T-\cosh\Delta\tau}=\ln{\sinh T+
   \sinh{\Delta\tau\over2}\over\sinh T-
   \sinh{\Delta\tau\over2}}\nonumber\\
   &=&\ln{\sinh T+
   \sinh{\Delta\tau_n\over2}\cos{\theta\over2}+i\cosh{\Delta\tau_n\over2}
   \sin{\theta\over2}\over\sinh T-
   \sinh{\Delta\tau_n\over2}\cos{\theta\over2}-i\cosh{\Delta\tau_n\over2}
   \sin{\theta\over2}}\equiv\ln{z_1\over z_2},
   \label{E96}
\end{eqnarray}
where we have used ${\rm arccosh}z=\ln(z+\sqrt{z^2-1})$.
The real components of $z_1$ and $z_2$ are respectively
\begin{eqnarray}
   \Re (z_1)=\sinh T+\sinh{\Delta\tau_n\over2}\cos{\theta\over2},
   \label{E97}
\end{eqnarray}
\begin{eqnarray}
   \Re (z_2)=\sinh T-\sinh{\Delta\tau_n\over2}\cos{\theta\over2}.
   \label{E98}
\end{eqnarray}
By Eq.~(\ref{E95}), we find that $\Re(z_1)$ and $\Re(z_2)$ are always
positive for any real $\theta$. This means that as $\Delta\tau$
goes from $\Delta\tau_n$ to $\Delta\tau_n+i2m\pi$,
the arguments (the argument of a complex number $z=\vert z\vert
e^{i\alpha}$ is $\alpha$)
of $z_1$ and $z_2$ do not change, neither does the argument of
$z_1/z_2$. The value of $\tilde{\gamma}$ remains in the same
branch of $\ln z$ as $\theta$ varies. Thus, for all
$\Delta\tau_{nm}=\Delta\tau_n+i2m\pi$, we have $\tilde{\gamma}=\pm nb$
and Eq.~(\ref{E92}) is satisfied. Therefore all $\Delta\tau_{nm}$ in
Eq.~(\ref{E94}) are poles.

The residues of the integrand in (\ref{E91}) at poles $\Delta\tau_{nm}$ are
(here $i\epsilon$ is returned to the expressions)
\begin{eqnarray}
   &&{\rm Res}(\Delta\tau=i2m\pi+i\epsilon,n=0)={iEr_0\over4\pi^2}
   e^{2m\pi \Delta Er_0},
   \label{E99}\\
   &&{\rm Res}(\Delta\tau=\Delta\tau_n+i2m\pi+i\epsilon,n\not=0)=
   -{1\over4\pi^2}
   {e^{2m\pi \Delta Er_0-i\Delta Er_0\Delta\tau_n}\over(\cosh nb+1)
   \sinh\Delta\tau_n}.
   \label{E100}
\end{eqnarray}
Then by the residue theorem (the contour for the integral is 
closed in the lower-half plane of complex $\Delta\tau$) we have
\begin{eqnarray}
   {d{\cal F}_0\over dT}={r_0\over2\pi}{\Delta E\over
   e^{2\pi \Delta Er_0}-1},
   \label{E101}
\end{eqnarray}
and
\begin{eqnarray}
   {d{\cal F}_{n\not=0}\over
   dT}={\sin(\Delta Er_0\vert\Delta\tau_n\vert)
   \over\pi(\cosh nb+1)\sinh\vert\Delta\tau_n\vert}
   ~{1\over
   e^{2\pi \Delta Er_0}-1}.
   \label{E102}
\end{eqnarray}
The $\sin(\Delta Er_0\vert\Delta\tau_n\vert)$ 
factor in Eq.~(\ref{E102}) indicates that the $n\not=0$ terms'
contribution can be both positive (absorption by the detector) and
negative (emission from the detector).
We see that the contribution of the $n=0$ term is just the Hawking
radiation with the Gibbons-Hawking
temperature $T_{\rm G-H}=1/2\pi r_0$ in the simply connected de~Sitter
space. The contribution of the $n\not=0$ terms is a kind
of ``grey-body'' Hawking radiation: the temperature is $T_{\rm G-H}$, but its
density or flux decreases as the universe expands
($\vert\Delta\tau_n\vert$ increases as the universe expands). The sum of
all $n\not=0$ contributions is
\begin{eqnarray}
   \sum_{n\not=0}{d{\cal F}_n\over dT}={1\over\pi^2}{1\over
   e^{2\pi r_0\Delta E}-1}\sum_{n\not=0}{\sin(\Delta Er_0\vert\Delta\tau_n\vert)
   \over(\cosh nb+1)\sinh\vert\Delta\tau_n\vert}.
   \label{E103}
\end{eqnarray}
In the case of $b=2\pi$ (the self-consistent case), we have
$\cosh nb\simeq \exp(\vert n\vert b)/2\gg1$ ($
n\not=0$) and thus $\Delta\tau_n\simeq\pm2T$. Then
\begin{eqnarray}
   \sum_{n\not=0}{d{\cal F}_n\over dT}\simeq{1\over2\pi}
   {A\over
   e^{2\pi r_0\Delta E}-1}{\sin(2\Delta Er_0T)
   \over\sinh2T},
   \label{E104}
\end{eqnarray}
where $A=4\sum_{n=1}^\infty (\cosh2n\pi+1)^{-1}\simeq0.015$. As
$T\rightarrow\infty$, the contribution of all $n\not=0$ terms decreases
exponentially to zero. Thus, at events far from the Cauchy horizon in
${\cal F}$, the
particle detector perceives pure Hawking radiation given by the
$n=0$ term. As 
$T\rightarrow0$ (near the Cauchy horizon), we have
\begin{eqnarray}
   \sum_{n\not=0}{d{\cal F}_n\over dT}\simeq{Ar_0\over2\pi}{\Delta E\over
   e^{2\pi r_0 \Delta E}-1}.
   \label{E105}
\end{eqnarray}
This is a ``grey-body'' Hawking radiation with $A\simeq1.5\%$. Near the
Cauchy horizon the total radiation is the sum of a pure Hawking
radiation (given by the $n=0$ term) and a ``grey-body'' Hawking
radiation (given by all $n\not=0$ terms). The total
intensity of the radiation near the Cauchy horizon is a factor of
$\simeq101.5\%$ that of regular Hawking radiation, but
its spectrum is the same as the usual Hawking radiation.

{\sl 3. A particle detector moving along a co-moving worldline in
the steady-state coordinate system}. 
Suppose the detector moves along the geodesic
$\rho,\theta,\phi={\rm constants}$ (such a worldline is a timelike
geodesic passing through ${\cal R}$ and into ${\cal F}$) where
$\rho\equiv(x^2+y^2+z^2)^{1/2}$ and the proper time
$\tau$ are related to the static radius $r$ by
\begin{eqnarray}
   r=-r_0\rho/\overline{\eta}=\rho e^{\tau/r_0}.
   \label{E106}
\end{eqnarray}
The Cauchy horizon is at $r=r_0$, or $\rho=-\overline{\eta}=r_0
e^{-\tau/r_0}$. On the worldline of the detector the Hadamard function is
\begin{eqnarray}
   G_{\rm CR}^{(1)}(T,\Delta\tau)={1\over2\pi^2r_0^2}~
   {\gamma\over2L\sinh{\Delta\tau\over2}}\sum_{n=-\infty}^\infty
   {1\over\gamma^2-\left({t-t^\prime\over r_0}+n b\right)^2},
   \label{E107}
\end{eqnarray}
where $\Delta\tau=(\tau-\tau^\prime)/r_0$, $T=(\tau+\tau^\prime)/2r_0$,
$L=\rho e^T/r_0\equiv r(T)/r_0$, $\gamma$ is given by
\begin{eqnarray}
   \cosh\gamma={1-L^2\over\sqrt{1+L^4-2L^2\cosh\Delta\tau}},
   \label{E108}
\end{eqnarray}
and $t-t^\prime$ is related to $T$ and $\Delta\tau$ by
\begin{eqnarray}
   \cosh{t-t^\prime\over r_0}={\cosh\Delta\tau
   -L^2\over\sqrt{1+L^4-2L^2\cosh\Delta\tau}}.
   \label{E109}
\end{eqnarray}
By analytical continuation, Eqs.~(\ref{E107}-\ref{E109}) hold in the whole region
covered by the steady-state coordinates in de~Sitter space.
The Wightman function $G^+$ is equal to one half of $G^{(1)}$ with
$\Delta\tau$ replaced by $\Delta\tau-i\epsilon$ [Eq.~(\ref{E26})]. 
The response function is
${\cal F}(\Delta E)=\sum_{n=-\infty}^\infty{\cal F}_n(\Delta E)$ where
\begin{eqnarray}
   {\cal F}_n(\Delta E)&=&{1\over4\pi^2}\int_{-\infty}^\infty dT
   \int_{-\infty}^\infty d\Delta\tau e^{-i\Delta Er_0\Delta\tau}\times
   \nonumber\\
   &&\left\{{\gamma\over2L\sinh{\Delta\tau\over2}\left[\gamma^2-
   \left({t-t^\prime\over r_0}+n b\right)^2\right]}
   \right\}_{\Delta\tau\rightarrow\Delta
   \tau-i\epsilon}.
   \label{E110}
\end{eqnarray}
The poles of the integrand in the complex-$\Delta\tau$ plane are given by
\begin{eqnarray}
   {t-t^\prime\over r_0}+n b=\pm\gamma.
   \label{E111}
\end{eqnarray}
This together with Eq.~(\ref{E108}) and Eq.~(\ref{E109}) leads to
\begin{eqnarray}
   (\cosh\Delta\tau-L^2)\cosh n b+\sinh\Delta\tau\sinh n b
   =1-L^2.
   \label{E112}
\end{eqnarray}
The roots of Eq.~(\ref{E112}) are
\begin{eqnarray}
   \Delta\tau=\Delta\tau_n^\pm+i2m\pi\equiv\Delta\tau_{nm}^\pm,
   \label{E113}
\end{eqnarray}
where
\begin{eqnarray}
   \Delta\tau_n^\pm=\ln\left(1+2\mu^2\pm2\mu\sqrt{1+\mu^2}\right)-n b,
   \label{E114}
\end{eqnarray}
where $\mu\equiv\sinh(n b/2)$. By carefully checking
$\Delta\tau_{nm}^\pm$ in 
Eq.~(\ref{E113}), as we did in case 2, we find that: (1) For $L<1$ (or
$\rho e^T<r_0$, i.e., in region ${\cal R}$), only $\Delta
\tau_{n0}^\pm=\Delta\tau_n^\pm$ solve
Eq.~(\ref{E111}); (2) for $L>1$ (or $\rho e^T>r_0$, i.e., in region
${\cal F}$), only
$\Delta\tau_{nm}^+=\Delta\tau_n^++i2m\pi$ solve Eq.~(\ref{E111}). (Here it is
assumed that $ b>\ln2$ and the self-consistent case with $b=2\pi$
obviously satisfies this condition.) All other $\Delta\tau$'s in
Eq.~(\ref{E113}) are not roots of Eq.~(\ref{E111}), though they solve
Eq.~(\ref{E112}). Therefore the poles are (where $i\epsilon$ is returned)
\begin{eqnarray}
   \Delta\tau=
   \left\{\begin{array}{lr}
   \Delta\tau_n^\pm+i\epsilon,&{\rm in}~~{\cal R};\\
   \Delta\tau_n^++i2m\pi+i\epsilon,&{\rm in}~~{\cal F}. 
   \end{array}  
   \right.
   \label{E115}
\end{eqnarray}
Obviously in region ${\cal R}$ all poles are in the upper-half plane of complex
$\Delta\tau$. Therefore
\begin{eqnarray}
   {d{\cal F}\over dT}=0, 
   \label{E115a}
\end{eqnarray}
when the particle detector is in region ${\cal R}$.
So the particle detector sees nothing while it is in region ${\cal R}$.

In region ${\cal F}$, only the poles with $m<0$ are in the lower-half plane of
complex $\Delta\tau$. The residues of the integrand at poles
$\Delta^+_{nm}+i\epsilon$ are
\begin{eqnarray}
   {\rm Res}(\Delta\tau=i2m\pi+i\epsilon,n=0)={ir_0\Delta E\over
   4\pi^2}e^{2m\pi\Delta Er_0},
   \label{E117}
\end{eqnarray}
\begin{eqnarray}
   &&{\rm Res}(\Delta\tau=\Delta\tau_n^++i2m\pi+i\epsilon,n\not=0)
   ={1\over16\pi^2L\sinh{\Delta\tau_n^+\over2}}\times\nonumber\\
   &&{\alpha_1(1+L^4-2L^2\cosh\Delta\tau_n^+)e^{2m\pi\Delta E
   r_0-i\Delta E r_0\Delta\tau_n^+}\over
   -\alpha_1L(L^2-1)\cosh{\Delta\tau_n^+\over2}
   +(\alpha_2+n b)(L^2\cosh\Delta\tau_n^+-1)},
   \label{E118}
\end{eqnarray}
where $\alpha_1={\rm
Arccosh}{L^2-1\over\sqrt{1+L^4-2L^2\cosh\Delta\tau_n^+}}$ and
$\alpha_2= {\rm Arccosh}{L^2-\cosh\Delta\tau_n^+\over
\sqrt{1+L^4-2L^2\cosh\Delta\tau_n^+}}$.
By the residue theorem, we have that $d{\cal F}_0/dT$ has the
same value as that 
in Eq.~(\ref{E101}), which represents Hawking radiation with
the Gibbons-Hawking temperature; the contribution of all
$n\not=0$ terms (note that $\Delta_n^+=-\Delta_{-n}^+$) is
\begin{eqnarray}
   &&{d\over dT}\sum_{n\not=0}{\cal F}_n={1\over4\pi^2(e^{2\pi
   r_0\Delta E}-1)}\sum_{n=1}^\infty{\sin(\Delta 
   Er_0\Delta\tau_n^+)\over L\sinh
   {\Delta\tau_n^+\over2}}\times\nonumber\\
   &&{\alpha_1(1+L^4-2L^2\cosh\Delta\tau_n^+)\over\alpha_1
   L(L^2-1)\cosh{\Delta\tau_n^+\over2}
   -(\alpha_2+nb)(L^2\cosh\Delta\tau_n^+-1)},
   \label{E119}
\end{eqnarray}
which represents a ``grey-body'' Hawking radiation. As $T\rightarrow\infty$
(or $L\rightarrow\infty$), ${d\over dT}\sum_{n\not=0}{\cal F}_n$
exponentially drops to zero; therefore, 
at events far from the Cauchy horizon in ${\cal F}$, the
particle detector only perceives pure Hawking radiation (the same as that
in case 2). As $L\rightarrow1$ (approaching the Cauchy horizon), we
also have ${d\over dT}\sum_{n\not=0}{\cal F}_n\rightarrow0$.
Thus as the Cauchy horizon is approached from the side of region
${\cal F}$, the particle detector comoving in the steady-state
coordinate system perceives pure Hawking radiation with
Gibbons-Hawking temperature.

From the above discussion, we find that in our multiply connected de
Sitter space with the adapted Rindler vacuum, region ${\cal R}$ is cold
(where the temperature is zero) but
region ${\cal F}$ is hot (where the temperature is $T_{\rm G-H}$). Similarly,
region ${\cal L}$ is cold but ${\cal P}$ 
is hot, the above results can be easily 
extended to these regions. This gives rise to an arrow of increasing
entropy, from a cold region to a hot region (Fig.~\ref{f5}).

\subsection{Classical Stability of the Cauchy Horizon and the Arrow of Time}
In classical electromagnetic theory, it is well known that both the
retarded potential $\phi_{\rm ret}$ and the advanced potential
$\phi_{\rm adv}$ (and any part-retarded-and-part-advanced potential
$a\phi_{\rm ret}+b\phi_{\rm adv}$ with $a+b=1$) are solutions of
Maxwell's equations. But from our experience, we know that all the electromagnetic
perturbations we see are propagated only by the {\em retarded} potential. (For
example, if at some time and some place, a light signal is emitted, it
can only be received by a receiver at another place sometime {\em
later}). This indicates that there is an {\em arrow of time} in the
solutions of Maxwell's equations, though Maxwell's equations themselves are
time-symmetric. This arrow of time is sometimes called the electromagnetic
arrow of time, or the causal arrow of time. How this arrow
of time arises is a mystery. Many people have tried to 
solve this problem by attributing it
to a boundary condition of the Universe
\cite{whe45,gol62,hog62,hoy64} (for review of the arrows of time, see
\cite{zeh92,sch97}). In this subsection we argue that the principle of
self-consistency \cite{nov83,fri90a}
naturally gives rise to an arrow of time in our multiply connected de~Sitter space. 

First let us consider the arrow of time
in Misner space. Suppose at an event E in region F in
Misner space [by boost
and translation, assume we have moved E to $(t=t_0,x=0,y=0,z=0)$], a
spherical pulse of electromagnetic wave is created. If the potential
is retarded [here ``retarded'' and ``advanced'' are defined relative to 
the direction of $(\partial/\partial t)^a$ ($t$ is the time
coordinate in the global Cartesian coordinates of the
covering space --- Minkowski space)],
the pulse will propagate in the future direction
as a light cone originating from E. At any point on the light cone, the
energy-momentum tensor of the wave is
\begin{eqnarray}
   T^{ab}=\mu k^ak^b,
   \label{E122}
\end{eqnarray}
where $\mu\equiv\mu(t)$ is a scalar function and $k^a=k^0(\partial/\partial
t)^a+k^1(\partial/\partial x)^a+k^2(\partial/\partial y)^a+
k^3(\partial/\partial z)^a$ is a null vector tangent to the light
cone, and the energy density measured by an observer with four-velocity vector
$(\partial/\partial t)^a$ (whose ordinary three-velocity is zero) is
\begin{eqnarray}
   \rho=T_{ab}\left ({\partial\over\partial t}\right)^a\left({\partial\over\partial
   t}\right)^b=\mu(k^0)^2.
   \label{E124}
\end{eqnarray}
(Thus $\mu$ measures the energy density of the electromagnetic wave.)
By Einstein's equations, the back-reaction of $T_{ab}$ on $R$ and
$R_{ab}R^{ab}$ (where $R_{ab}$ is the Ricci tensor and $R=R_a^{~a}$
is the Ricci scalar curvature) is $\delta R\sim T_a^{~a}$, $\delta
(R_{ab}R^{ab})\sim T_{ab}T^{ab}$. The Riemann tensor can be decomposed
as $R_{abcd}=C_{abcd}+Q_{abcd}$, where $C_{abcd}$ is the Weyl tensor
and $Q_{abcd}$ is constructed entirely from the Ricci tensor 
\begin{eqnarray}
   Q_{abcd}=g_{a[c}R_{d]b}-g_{b[c}R_{d]a}-{1\over3}Rg_{a[c}g_{d]b},
   \label{wel}
\end{eqnarray}
where square brackets denote antisymmetrization \cite{wal84}. The Weyl
tensor describes the part of the curvature that is due to pure
gravitational field, whereas the Ricci tensor describes the part 
that, according to Einstein's equations, is directly due to the
energy-momentum tensor of matter \cite{pen89}. Therefore, in some
sense, the values of $T_a^{~a}$ and $T_{ab}T^{ab}$ determine
the influence of matter fields on the stability of the background spacetime.  
An infinite $T_a^{~a}$
or $T_{ab}T^{ab}$ implies that the spacetime is unstable
against this perturbation and a singularity may form; on the other hand, 
if $T_a^{~a}$
and $T_{ab}T^{ab}$ are finite, the spacetime may be 
stable against this perturbation.
Self-consistent solutions should require that $T_a^{~a}$ and
$T_{ab}T^{ab}$ do not blow up. If they did, the starting
geometry --- on the basis of which $T_a^{~a}$ and
$T_{ab}T^{ab}$ were calculated --- would be greatly perturbed and the $T_a^{~a}$ and
$T_{ab}T^{ab}$ calculation itself would be invalid, and thus it would not
be a self-consistent solution.
For electromagnetic fields we always have $T_a^{~a}=0$, so we need only
consider $T_{ab}T^{ab}$. For
$T_{ab}$ in Eq.~(\ref{E122}), we also have
\begin{eqnarray}
   T^{ab}T_{ab}=0.
   \label{E123}
\end{eqnarray}
Thus significant perturbations (indicated by a non-vanishing
$T_{ab}T^{ab}$) can only occur when the
light cone ``collides'' with its images under the boost transformation. 
At any point $p$
on the intersection of the light cone $L$ and its $n$-th 
image $L_n$ (suppose $n>0$), the
energy-momentum tensor is
\begin{eqnarray}
   T^{ab}=\mu k^ak^b+{\tilde \mu}{\tilde k}^a{\tilde k}^b,
   \label{E125}
\end{eqnarray}
where $k^a$ is the null vector tangent to the light cone $L$ at $p$,
$\tilde{k}^a$ is the null vector tangent to the light cone $L_n$ at
$p$; $\mu$ measures the energy density in light cone $L$,
$\tilde{\mu}$ measures the energy density in light cone
$L_n$. From Eq.~(\ref{E125}) we have
\begin{eqnarray}
   T^{ab}T_{ab}=[2\mu\tilde{\mu}(k^a\tilde{k}_a)^2]_p,
   \label{E126}
\end{eqnarray}
the index $p$ denotes that the quantity is evaluated at the point $p$.

Since the point $p$ on $L_n$ is obtained from some point $p^\prime$ on $L$
by boost transformation, $p$ and $p^\prime$ must have the same
timelike separation from the origin $(t=0,x=0,y=0,z=0)$ (remember that
$p$ is on the intersection of $L$ and $L_n$, see Fig.~\ref{f6}a). If we take the
$\tilde{k}^a$ at $p$ being transported from the ${k^\prime}^a$ at
$p^\prime$, we have $\tilde{\mu}_{p\in L_n}=\mu_{p^\prime\in
L}$. Because the light cone $L$ is spherically symmetric, 
we have $t_p$=$t_{p^\prime}$. Therefore we have $\mu_{p^\prime\in
L} =\mu_{p\in L}$ and at $p$ we have $\tilde{\mu}=\mu$. Under the boost
transformation $B$, we have
\begin{eqnarray}
   (\tilde{k}^a)_p&=&B[({k^\prime}^a)_{p^\prime}]={k^\prime}^0\left[\cosh
   nb~\left({\partial\over\partial t}\right)^a+\sinh nb~\left(
   {\partial\over\partial x}\right)^a\right]+
   \nonumber\\
   &&{k^\prime}^1\left[\cosh
   nb~\left({\partial\over\partial x}\right)^a+\sinh
   nb~\left({\partial\over\partial t}\right)^a\right]+
   {k^\prime}^2\left({\partial\over\partial y}\right)^a+{k^\prime}^3\left({\partial
   \over\partial z}\right)^a,
   \label{E127}
\end{eqnarray}
where $({k^\prime}^a)_{p^\prime}={k^\prime}^0(\partial/\partial
t)^a+{k^\prime}^1(\partial/\partial x)^a+{k^\prime}^2(\partial/\partial y)^a+
{k^\prime}^3(\partial/\partial z)^a$. Due to the spherical symmetry,
we have ${k^\prime}^0=k^0$. Define
$(r,\theta,\phi)$ by $x=r\cos\theta$,
$y=r\sin\theta\cos\phi$, and 
$z=r\sin\theta\sin\phi$. Then we have
$r^\prime=r$, 
$\theta^\prime=\pi-\theta$,
$\phi^\prime=\phi$ (``$\prime$'' means ``at $p^\prime$''), and
\begin{eqnarray}
   k^1=k^0\cos\theta,~~~k^2=k^0\sin\theta\cos\phi, 
   ~~~k^3=k^0\sin\theta\sin\phi,
   \label{E128}
\end{eqnarray}
and 
\begin{eqnarray}
   &&{k^\prime}^1={k^\prime}^0\cos\theta^\prime=-k^0\cos\theta=-k^1,
   ~~~{k^\prime}^2={k^\prime}^0\sin\theta^\prime\cos\phi^\prime=
   k^0\sin\theta\cos\phi=k^2, \nonumber\\
   &&{k^\prime}^3={k^\prime}^0\sin\theta^\prime\sin\phi^\prime=
   k^0\sin\theta\sin\phi=k^3.
   \label{E129}
\end{eqnarray}
Then
\begin{eqnarray}
   (k^a\tilde{k}_a)_p=(k^0)^2[-(1+\cos^2\theta)\cosh nb+
   2\cos\theta\sinh nb+\sin^2\theta],
   \label{E130}
\end{eqnarray}
and
\begin{eqnarray}
   T^{ab}T_{ab}=2\rho^2(t_p)[-(1+\cos^2\theta)\cosh nb+
   2\cos\theta\sinh nb+\sin^2\theta]^2. 
   \label{E131}
\end{eqnarray}
It is easy to find that $T^{ab}T_{ab}$ reaches a maximum at
$\theta=0$ and
\begin{eqnarray}
   (T^{ab}T_{ab})_{\max}=8\rho^2(t_p)e^{-2nb},
   \label{E132}
\end{eqnarray}
where $\rho(t_p)$ is the energy density from $L$ as measured in a frame
at event $p$ with ordinary velocity $v_x=v_y=v_z=0$. $(T^{ab}T_{ab})_{\max}$ 
is always finite [less than $8\rho^2(t_p)$] since $n$ is
positive. If $n<0$ we have $(T^{ab}T_{ab})_{\max}=
8\rho^2(t_p)e^{2nb}<8\rho^2(t_p)$. So if we have a retarded potential in
region F, even considering the infinite number of images, 
$T_{ab}T^{ab}$ is always finite.

If the potential is advanced however, the pulse wave will propagate
backward in the past direction as a light cone originating from E. And, within
a finite time, it will hit the Cauchy horizon. By an analysis similar to
the above arguments, we find that in this case
\begin{eqnarray}
   T^{ab}T_{ab}=2\rho^2(t_p)[(1+\cos^2\theta)\cosh nb
   +2\cos\theta\sinh nb-\sin^2\theta]^2, 
   \label{E133}
\end{eqnarray}
which reaches a maximum at $\theta=0$ and
\begin{eqnarray}
   (T^{ab}T_{ab})_{\max}=8\rho^2(t_p)e^{2\vert n\vert b}.
   \label{E134}
\end{eqnarray}
Since $\rho(t_p)$ is finite (the past light cone from E at $\theta=0$
hits the Cauchy horizon in a finite affine distance), 
thus $(T^{ab}T_{ab})_{\max}\rightarrow\infty$ as
$n\rightarrow\pm\infty$. As $n\rightarrow\pm\infty$, $L$ and $L_n$ collide at
the Cauchy horizon [as $n\rightarrow\pm\infty$ the point $p(\theta=0)$
approaches the Cauchy horizon] (see Fig.~\ref{f6}c). 
Thus $(T^{ab}T_{ab})_{\max}$ diverges as the
Cauchy horizon is approached and
the Cauchy horizon may be destroyed. Therefore the advanced potential
is {\em not}
self-consistent in region F of Misner space. It is easy to see that any
part-retarded-and-part-advanced potential is also {\em not}
self-consistent in F.
The {\em only} self-consistent potential
in region F is the {\em retarded} potential.

Similarly, in region P the only self-consistent potential is the
{\em advanced} potential (see Fig.~\ref{f6}a). [Note that here ``advanced'' and
``retarded'' are defined relative to the global time direction
in Minkowski spacetime (the covering space). An observer in P will regard it as
``retarded'' relative to his own time direction.]

In region R, by boost and translation, we can always move the event E
(where a spherical pulse of electromagnetic waves is emitted) to
$(t=0,x=x_0,y=0,z=0)$. Either pure retarded or pure advanced potentials are
self-consistent in this region because the light cone never ``collides'' with the
images of itself and thus we always have $T^{ab}T_{ab}=0$ 
(see Fig.~\ref{f6}b). But, for a
part-retarded-and-part-advanced potential, the retarded light cone ($L^+$)
propagates
forward while the advanced light cone ($L^-$)
propagates backward,
both originating from E. The forward part of the light cone will collide
with images of the backward part of the light cone and {\sl vice
versa} (see Fig.~\ref{f6}d). We
find that at a point $p$ on the intersection of $L^+$ and $L^-_n$ (or $L^-$
and $L^+_n$)
\begin{eqnarray}
   T^{ab}T_{ab}=2\rho(t)\rho(-t)[(1+\cos^2\theta)\cosh nb-
   2\cos\theta\sinh nb+\sin^2\theta]^2, 
   \label{E135}
\end{eqnarray}
where $\rho(t)$ is the energy density from $L^+$ observed in a frame
on $L^+$ with time coordinate $t$ and with ordinary velocity
$v_x=v_y=v_z=0$ and $\rho(-t)$ is the energy density from $L^-$ seen
in a frame on $L^-$ with time coordinate $-t$ and with ordinary velocity
$v_x=v_y=v_z=0$. $T^{ab}T_{ab}$  
reaches a maximum at $\theta=\pi$, and
\begin{eqnarray}
   (T^{ab}T_{ab})_{\max}=8\rho(t)\rho(-t)e^{2\vert n\vert b},
   \label{E136}
\end{eqnarray}
where $t$ is the global time coordinate in the covering Minkowski
space. As $p$ approaches the Cauchy horizon, where
$n\rightarrow\pm\infty$, $\rho(t)$ and $\rho(-t)$ are both finite, since in the
$\theta=\pi$ direction the future and past light cones of E both
hit the Cauchy horizon in a finite affine distance.
Thus $(T^{ab}T_{ab})_{\max}\rightarrow\infty$ as $p$ approaches the
Cauchy horizon (where $n\rightarrow\pm\infty$). Therefore in region R
both the retarded and the advanced potential are self-consistent, but the
part-retarded-and-part-advanced potential is {\em not} self-consistent. This
conclusion also holds for region L. Furthermore, there must be
a correlation between time arrows in region L
and region R: if we choose the retarded
potential in R, we must choose the advanced
potential in L (see Fig.~\ref{f6}b); if we choose the advanced potential in R,
we must choose the retarded
potential in L. Otherwise the collision of light
cones from R and light cones
from L will destroy the Cauchy horizon. 

As another treatment for perturbations in Misner space, consider that
at an event E in region F two photons are created \cite{mis69} 
[we choose E to be at
$(t=t_0,x=0,y=0,z=0)$ as before]. One photon runs to the right along
the $+x$ direction, the other photon runs to the left along the
$-x$ direction. They have the same frequency (thus the same energy). The
tangent vector of the null geodesic of the right-moving photon
is chosen to be $_rk^a={q\over v}
({\partial\over\partial u})^a\equiv({\partial\over\partial\lambda_r})^a$, where
$\lambda_r$ is an affine parameter of the geodesic, $q$ is a constant
and $u=t+x$, $v=t-x$. The tangent vector of
the null geodesic of the left-moving photon
is chosen to be $_lk^a={q\over u}
({\partial\over\partial v})^a\equiv({\partial\over\partial\lambda_l})^a$, where
$\lambda_l$ is an affine parameter of that geodesic. 
The null vectors $_rk^a$ and $_lk^a$
are invariant under boost transformations. At any point where a photon with
null wave-vector $k^a$ is passing by, the frequency of the photon
measured in a frame of reference passing by the same point with the four-velocity
$v^a$ is $\omega=-k^av_a$. If $v^a=(\partial/\partial t)^a$ (i.e., the
frame of reference has ordinary three-velocity $v_x=v_y=v_z=0$) and 
$k^a=~_rk^a$ or $_lk^a$, we have $\omega_r=\omega_l=q/2t_0
\equiv\omega_0$ (thus
$q$ measures the frequency of the photon). At any point where the $n$-th
image of the right-moving (left-moving) photon is passing by, using the boost
transformation we can always
find a frame of reference in which the frequency of the photon is
$\omega_0$. But at a point $p$ where the right-moving (left-moving) photon
passes the $n$-th image of the left-moving (right-moving) photon,
we cannot find a frame of reference such that the two ``colliding'' photons
both have frequency $\omega_0$. In such a case we should analyze it
in the center-of-momentum frame.
The four-velocity of the center-of-momentum frame is 
$v^a=\gamma(_rk^a+~_lk^a)$ where $\gamma^2=-
[(_rk^a+~_lk^a)
(_rk_a+~_lk_a)]^{-1}=uv/q^2=\tilde{\eta}^2/q^2$
where $\tilde{\eta}=(t^2-x^2)^{1/2}$ is
the proper time separation of $p$ from the origin
$(t=0,x=0,y=0,z=0)$. 
Therefore the total energy
of the two oppositely directed photons in the center-of-momentum frame is
\begin{eqnarray}
   {\cal E}=\omega_1+\omega_2=-_rk^av_a-~_lk^av_a=
   {1\over\gamma}={2t_0\over\tilde{\eta}}\omega_0.
   \label{E137}
\end{eqnarray}
(For all other frames the total energy would be greater.)
If the potential is retarded, so
photons move in the future direction, all points
where photons and their images ``collide'' are in the future of the hypersurface
$t^2-x^2=t_0^2$. Therefore we have $\tilde{\eta}\geq\tilde{\eta}_0=t_0$ and
${\cal E}\leq2\omega_0$, so the total energy in the 
center-of-momentum frame is always bounded.
But, if the potential
is advanced, photons move in the past direction; thus all
points where photons and oppositely directed
image photons ``collide'' are in the past of the
hypersurface $t^2-x^2=t_0^2$. In particular, the right-moving (left-moving)
photon collides with the $\infty$-th ($-\infty$-th) image of the left-moving
(right-moving) photon at the Cauchy horizon, where $\tilde{\eta}=0$ and
thus ${\cal E}\rightarrow\infty$. Thus, the Cauchy horizon may be destroyed by
these photon pairs. Therefore in agreement with our earlier argument,
the advanced potential 
is {\em not} self-consistent in region F. The {\em retarded} potential is
self-consistent
in region F. Similarly, the {\em advanced} potential is self-consistent in
region P.
In region R and region L, both the retarded potential and the advanced
potential are self-consistent, because the photons and 
their images will not collide with each other and at any point a photon
is passing by we can always find a frame for whom the frequency of this
photon is $\omega_0$. And, the potentials in region R and region L
must be
correlated in the following way: If the potential in R is retarded,
the potential in L must be advanced; if the potential in R is advanced,
the potential in L must be retarded (we would call
them ``anti-correlated''). Otherwise the
photons from L and photons from R passing in opposite directions
would be measured to have infinite energy in center-of-momentum
frames as the Cauchy horizon is approached and this
may similarly destroy the Cauchy
horizon. These conclusions are consistent with those obtained from the
analysis of the perturbation of a pulse wave discussed above.

Our multiply connected de~Sitter space is conformally related to
Misner space via Eqs.~(\ref{E58}-\ref{E61}). Because light cones and
chronological relations are conformally invariant \cite{sac77} 
(thus regions ${\cal F}$,
${\cal P}$, ${\cal R}$, and ${\cal L}$ in multiply connected de~Sitter 
space correspond respectively to regions F,
P, R, and L in Misner space under the conformal map, 
as discussed in section \ref{IX.B}), Maxwell's equations are
also conformally invariant \cite{wal84,sac77}, so it is easy to generalize the results
from Misner space to our multiply connected de~Sitter
space. Under the conformal transformation
$g_{ab}\rightarrow\Omega^2g_{ab}$, the energy-momentum tensor of
the electromagnetic field is transformed as $T_a^{~b}\rightarrow\Omega^{-4}
T_a^{~b}$ \cite{wal84}. Thus $T^{ab}T_{ab}$ is transformed as
$T^{ab}T_{ab}\rightarrow\Omega^{-8} T^{ab}T_{ab}$. 
From the above discussion of $T^{ab}T_{ab}$ in Misner space, we know
that $T^{ab}T_{ab}$ is zero everywhere except at the intersection of
two light cones. Thus, in multiply connected de Sitters pace,
$T^{ab}T_{ab}$ is also zero everywhere except at the intersection of
two light cones. At the intersection of two light cones in multiply
connected de Sitter space, it is easy to show that the maximum value
of $T^{ab}T_{ab}$ is at the points with $\theta=0$ or $\theta=\pi$ on
the intersection. From Eq.~(\ref{E60}) and Eq.~(\ref{E60a}) we find
that for $\theta=0$ or $\theta=\pi$, $\Omega^2$ is non-zero except at
the points with $\theta=0$ on the Cauchy horizon (where $r=r_0$ or $\tilde{t}=r_0$).
Also because $\Omega^2$ is finite everywhere on the
Cauchy horizon (i.e. it is never infinite),
we have that: (1) if $T^{ab}T_{ab}$ diverges on the Cauchy horizon in
Misner space, the corresponding $T^{ab}T_{ab}$ also diverges on the
Cauchy horizon in our multiply connected de~Sitter space; (2) if
$T^{ab}T_{ab}$ is finite in some region (except at the Cauchy horizon) in
Misner space, the corresponding $T^{ab}T_{ab}$ is also finite in the
corresponding region (not at the Cauchy horizon) in the multiply
connected de~Sitter space; (3) if $T^{ab}T_{ab}$ is zero   
in some region (not a single point) in
Misner space, the corresponding $T^{ab}T_{ab}$ is also zero in the
corresponding region in the multiply
connected de~Sitter space. 
Under the conformal transformation $g_{ab}\rightarrow\Omega^2g_{ab}$,
the affine parameter of a null geodesic is transformed as
$\lambda\rightarrow\tilde{\lambda}:
d\tilde{\lambda}/d\lambda=C\Omega^2$ where $C$ is a constant
\cite{wal84} and thus the null vector $k^a=(\partial/\partial\lambda)^a$ is
transformed as $k^a\rightarrow C^{-1}\Omega^{-2}k^a$. Then 
$\gamma=[-
(_rk^a+~_lk^a)
(_rk_a+~_lk_a)]^{-1/2}$ is transformed as $\gamma\rightarrow
C\Omega\gamma$ and the total energy of the photon pairs in the 
center-of-momentum frame is transformed as ${\cal E}\rightarrow
C^{-1}\Omega^{-1}{\cal E}$ and the constant $C^{-1}$ can be absorbed into $\omega_0$.
Therefore, we can transplant the above results
for Misner space directly to our multiply connected de
Sitter space: {\em In region ${\cal F}$ the only self-consistent potential is
the retarded potential; in region ${\cal P}$ the only self-consistent potential
is the advanced potential; in regions ${\cal R}$ and ${\cal L}$ 
both the retarded
potential and the advanced potential are self-consistent, but they
must be anti-correlated} (Fig.~\ref{f6}).

The Cauchy horizon \cite{haw73} separating a region with CTCs from
that without closed causal curves is also called a chronology 
horizon \cite{tho93}. A chronology horizon 
is called a {\em future}
chronology horizon if the region with CTCs lies to the
future of the region without closed causal curves; a chronology horizon 
is called a {\em past}
chronology horizon if the region with CTCs is in the
past of the region without closed causal curves. It is generally believed
that a future chronology horizon is classically unstable unless there
is some diverging effect near the horizon \cite{tho93,li97}. 
The argument says that a wave packet propagating
in the future direction in this spacetime will pile up on the future
chronology horizon and destroy the horizon due to the effect of
the infinite blue-shift of the frequency (and thus the energy) seen by
a timelike observer near
a closed null geodesic on the
horizon \cite{mis69,tho93}. 
But if there is some diverging mechanism (like the diverging
effect of a wormhole in a spacetime with CTCs
constructed from a wormhole \cite{mor88})  
near the horizon, the amplitude of the
wave packet will decrease with time due to this mechanism, and this
may cancel the effect of the blue-shift of the frequency, making the
energy finite and thus rendering the
future chronology horizon classically stable. Unfortunately, in our
multiply connected de~Sitter spacetime (as also in Misner space)
there is no such diverging
mechanism. A light ray propagating in de~Sitter space will focus rather
than diverge. This can be seen from the focusing equation \cite{mis73}
\begin{eqnarray}
   {d^2{\cal A}^{1/2}\over d\lambda^2}=-\left(\sigma^2+{1\over2}R_{ab}
   k^ak^b\right){\cal A}^{1/2},
   \label{E121}
\end{eqnarray}
where ${\cal A}$ is the cross-sectional area of the bundle of rays,
$\lambda$ is the affine parameter along the central ray, the null
vector $k^a$ is $k^a=(\partial/\partial\lambda)^a$, and $\sigma$ is
the magnitude of the shear of the rays. For de~Sitter space we have
$R_{ab}k^ak^b=\Lambda g_{ab}k^ak^b=0$ 
and thus we have ${d^2{\cal A}^{1/2}\over
d\lambda^2}\leq0$, so the ray will never diverge. (In fact this always
holds if the spacetime satisfies either the
weak energy condition or the strong
energy condition and it is called the focusing theorem \cite{mis73}.) 
Hawking \cite{haw92b} has given a general proof along the above lines 
that any {\em future}
chronology horizon is classically unstable unless light rays are
diverging when they
propagate near the chronology horizon. 
You could cause this instability by
shaking an electron in the vicinity of the future chronology
horizon. The retarded wave would then propagate to the future
causing the instability.

However, in Hawking's proof \cite{haw92b}, 
if we replace a future chronology horizon
with a {\em past} chronology horizon, then the proof breaks down
because, in such a case, a wave packet propagating
toward the future near the past
chronology horizon will suffer a red-shift instead of
a blue-shift. Therefore a {\em past} chronology horizon, according to
Hawking's argument, is
classically stable in a world with retarded potentials.
If the universe started with a region of CTCs, but there are no CTCs
now, that early region of CTCs would be bounded to
the future by a past chronology horizon, and that horizon would be
classically stable in a world with retarded potentials --- which is
what we want. In our multiply connected de~Sitter space, this
is realized, since the arrow of time in region ${\cal F}$ is in the future
direction and the arrow of time in region ${\cal P}$ is in the past
direction [here ``future'' and ``past'' are defined globally by the
direction of $(\partial/\partial\tau)^a$, where $\tau$ is the time
coordinate in the global coordinate system $(\tau,\chi,\theta,\phi)$ of the
de~Sitter covering space]. ${\cal F}$ and ${\cal R}$ can have retarded
potentials, while ${\cal P}$ and ${\cal L}$ have advanced potentials,
as we have noted. In this case the Cauchy horizons separating ${\cal
F}$ from ${\cal R}$ and ${\cal P}$ from ${\cal L}$ are classically
stable, as indicated by our detailed study of $T^{ab}T_{ab}$ as these Cauchy 
horizons are approached. What about the Cauchy horizons separating 
${\cal P}$ from ${\cal R}$ and ${\cal F}$ from ${\cal L}$? 
In region ${\cal P}$, 
the potentials are advanced, so Hawking's instability does not 
arise as one approaches the Cauchy horizon separating it from 
${\cal R}$. In region ${\cal R}$, the potentials are retarded, so  by 
Hawking's argument, one might think that there would be an instability
as the Cauchy horizon separating ${\cal R}$ from ${\cal P}$ 
is approached from the ${\cal R}$ side. But, as we have shown, with
retarded potentials in ${\cal R}$, $T^{ab}T_{ab}$ does not diverge as
the Cauchy horizon separating ${\cal R}$ from ${\cal P}$ is approached
from the ${\cal R}$ side, indicating no instability. Why?
Because one can always find frames where the passing photon energies are
bounded as the Cauchy horizon is approached. Hawking's argument works only
if one can pick a particular frame like the frame of a timelike observer
crossing the Cauchy horizons and observe the blow up of the energy in that frame.
(Thus Hawking's approach
is observer-dependent, while our approach with $T^{ab}T_{ab}$
is observer-independent.)
Hawking's timelike observer would be killed by these photons. But, as we 
have shown, ${\cal R}$ is in a pure vacuum state in our model, so there are
no timelike observers in this region, and no preferred frame. If there were 
timelike particles of positive mass crossing from ${\cal P}$ to ${\cal R}$
through the Cauchy horizon, we have shown (Li and Gott \cite{li97}) that these would
cause a classical instability; but there are none. There are, as we shall
show in the next subsection, no real particles in regions ${\cal L}$ and ${\cal R}$
(because these are vacuum states) and no real particles in region ${\cal F}$
and ${\cal P}$ until the vacuum state there decays by forming bubbles at 
a timelike separation $|\tau|>\tau_0$ from the origin ($\tau_0$ will be
given in the next subsection).
Thus, there are no particles crossing the Cauchy horizons separating
${\cal P}$ from ${\cal R}$ and ${\cal F}$ form ${\cal L}$. Thus, there
is no instability caused by particles crossing the Cauchy horizons;
and since there are no timelike observers in region ${\cal R}$ to be
hit by photons as the Cauchy horizon separating ${\cal R}$ from ${\cal
P}$ is approached, there is no instability, as indicated by the fact
$T^{ab}T_{ab}$ does not blow up as that Cauchy horizon is
approached. As indicated in Fig.~\ref{f4}b, region ${\cal F}+{\cal R}$
is one causally connected region which can be pictured as partially
bounded to the future by the future light cone of an event E$^\prime$
and bounded to the past by the future light cone of an event E; but E
and E$^\prime$ are identified by the action of the boost, so these two
light cones are identified, creating a periodic boundary condition for
region ${\cal F}+{\cal R}$. As our treatment using $T_{ab}T^{ab}$ with
images indicates, retarded photons created in ${\cal F}+{\cal R}$
cause no instability. Particles with timelike worldlines crossing the
Cauchy horizons separating ${\cal F}+{\cal R}$ from ${\cal P}+{\cal
L}$ would cause instability by crossing an infinite number of times
between the future light cones of E and E$^\prime$, thus making an
infinite number of passages through the region ${\cal F}+{\cal R}$
(also ${\cal P}+{\cal L}$) shown in Fig.~\ref{f4}b. However, as we
have shown, there should be no such particles with timelike worldlines
crossing the Cauchy horizons separating ${\cal F}+{\cal R}$ from ${\cal P}+{\cal
L}$, and no photons crossing these horizons either, since the
potentials in ${\cal F}+{\cal R}$ are retarded, while the potentials
in ${\cal P}+{\cal L}$ are advanced. Thus, we expect ${\cal F}+{\cal
R}$ and ${\cal P}+{\cal L}$ to both be stable, and causally
disconnected from each other.  
(See further discussion in the next subsection). 

Thus, the principle of self-consistency \cite{nov83,fri90a} 
produces classical stability of the Cauchy horizons and 
naturally gives rise to an arrow of time in our model of the
Universe. 

\subsection{Bubble Formation in the Multiply Connected de~Sitter
Space}
From the above discussion we find that in the multiply connected de
Sitter space region ${\cal F}$ and region ${\cal P}$ are causally
independent in physics: the self-consistent potential in ${\cal F}$ is
the retarded potential, while the self-consistent potential in ${\cal
P}$ is the advanced potential, thus an event in ${\cal F}$ can never
influence an event in ${\cal P}$, and {\em vice versa}. ${\cal F}$ and
${\cal P}$ are physically disconnected though they are mathematically
connected. If we choose the potential in ${\cal R}$ to be retarded,
then the potential in ${\cal L}$ must be advanced. (Note that here
``advanced'' and ``retarded'' are defined relative to the global time
direction in de~Sitter space --- the covering space of our multiply
connected de~Sitter space.) Then region ${\cal
F}+{\cal R}$ (including the Cauchy horizon separating ${\cal F}$ from
${\cal R}$) forms a causal unit, and region ${\cal
P}+{\cal L}$ (including the Cauchy horizon separating ${\cal P}$ from
${\cal L}$) forms another causal unit. (See Fig.~4b, where the two null
surfaces partially bounding the grey ${\cal F}+{\cal R}$ region to the
past and future are identified. Similarly for the null surfaces
partially bounding the ${\cal P}+{\cal L}$ region.) An event in ${\cal
F}+{\cal R}$ and an event in  ${\cal
P}+{\cal L}$ are always causally independent in physics: they can
never physically influence each other though they may be
mathematically connected by some causal
curves (null curves or timelike curves). Though ${\cal
F}+{\cal R}$ and ${\cal
P}+{\cal L}$ are connected in mathematics, they are disconnected in
physics. They are separated by a Cauchy horizon. When we consider
physics in ${\cal F}+{\cal R}$, we can completely forget region ${\cal
P}+{\cal L}$ (and {\em vice versa}). Though in such a case the Cauchy
horizon separating ${\cal F}+{\cal R}$ from ${\cal P}+{\cal L}$ is a null
spacetime boundary, we do not need any boundary condition on it
because the topological multi-connectivity in ${\cal F}+{\cal R}$ has
already given rise to a periodic boundary condition (which is a kind of
self-consistent boundary condition). (In Fig.~4b this is shown by the
fact that the null curves partially bounding ${\cal F}+{\cal R}$ to the past and
future are identified.) This periodic boundary condition
(the self-consistent condition) is sufficient to fix the solutions of
the universe. For example, in our multiply connected de~Sitter
space model, the stability of the Cauchy horizon requires
that the regions with CTCs (${\cal R}$ and ${\cal L}$) must be confined
in the past and in these regions all quantum fields must be in vacuum
states (as we have already remarked, the appearance of any real particles there seems to destroy the
Cauchy horizon \cite{li97}). This gives rise to an arrow of
time and an arrow of entropy in this model. 

${\cal F}+{\cal R}$ is a
Hausdorff manifold with a null boundary, and thus ${\cal
F}+{\cal R}$ is geodesically incomplete to the 
past. But, the geodesic incompleteness of
${\cal F}+{\cal R}$ may {\em not} be important in physics because in
the inflationary scenario all real particles are created during the
reheating process after inflation within bubbles created 
in region ${\cal F}$ and these particles emit
only retarded photons which never run off the spacetime because
here the geodesic
incompleteness takes place only in the past direction. On the other hand,
we can smoothly extend ${\cal F}+{\cal R}$ to ${\cal P}+{\cal L}$ so that
the total multiply connected de~Sitter space $dS/B$ is geodesically
complete but at the price that it is not a manifold at a
two-sphere (section \ref{IX.A}). This model describes two physically
disconnected but mathematically connected universes. [The analogy
between the causal structures in region ${\cal F}+{\cal R}$ and region
${\cal P}+{\cal L}$ might motivate us to identify antipodal points in our
multiply connected de~Sitter space, as we did for the simply connected
de~Sitter space (section \ref{VIII}). The spacetime so obtained is a
Hausdorff manifold everywhere. It is geodesically complete but not
time orientable. For computing the energy-momentum tensor of vacuum
polarization, we must take into account the images of antipodal points
in addition to the images produced by the boost transformation. 
Further research is needed to
find a self-consistent vacuum for this spacetime.]

Now we consider formation of bubbles in ${\cal F}+{\cal R}$ in
multiply connected de~Sitter space. [The results (and the arguments for
${\cal F}+{\cal R}$ in the previous paragraph) also apply to region
${\cal P}+{\cal L}$, except that while in ${\cal F}+{\cal R}$ bubbles expand
in the future direction, in ${\cal P}+{\cal L}$ they expand in the
past direction; here ``future'' and ``past'' are defined with respect
to $(\partial/\partial\tau)^a$ where $\tau$ is the time coordinate in
the global coordinates of de~Sitter space.] Region 
${\cal R}$ (for its fundamental cell
see Fig.~\ref{f4}) which is multiply connected has a finite four-volume
$V_{\rm I}={4\over3}\pi br_0^4$ (here $b=\beta/r_0$, $\beta$ is the de
Sitter boost parameter). If
the probability of forming a bubble per volume $r_0^4$ in de~Sitter
space is $\epsilon$, then the total probability of forming a bubble in
$V_{\rm I}$ is $P_{\rm I}={4\over3}\pi b\epsilon$. 

Region ${\cal F}$ (its fundamental cell is shown 
in Fig.~\ref{f4}) has an infinite
four-volume and thus there should be an infinite number of bubble
universes formed \cite{got82,got86}. The metric in region ${\cal F}$
is given by Eq.~(\ref{E50}) with $0<\tau<\infty$, $0\leq l<\beta$,
$0<\theta<\pi$, and $0\leq\phi<2\pi$ (see Fig.~4a); it is multiply
connected (periodic in $l$ with period $\beta$).  
In order that the inflation proceeds and the bubbles
(which expand to the future --- as expected with the
retarded potential in region ${\cal F}$) do not
percolate, it is required that $\epsilon<\epsilon_{\rm cr}$ where
$5.8\times 10^{-9}<\epsilon_{\rm cr}<0.24$ \cite{gut83}. Gott and
Statler \cite{got84} showed that in order that we on 
earth today should not have witnessed another
bubble colliding with ours within our past
light cone (with $95\%$ confidence) $\epsilon$ must be less 
than $7.60\times10^{-4}$
for $\Omega=0.1$ (for $\Omega=0.4$ Gott \cite{got97} found
$\epsilon<0.01$). In our multiply connected de~Sitter space,
for inflation to proceed, there should be the additional requirement
that bubbles do not
collide with images of themselves (producing percolation).
A necessary condition for a bubble
formed in ${\cal F}$ not to collide with itself is that from time
$\tau$ when the bubble forms to future infinity
($\tau\rightarrow\infty$) a light signal moving along the $l$
direction [where
$\tau$ and $l$ are defined in Eq.~(\ref{E47}) and Eq.~(\ref{E49})]
propagates a co-moving distance less than $\beta/2$, which leads to the
condition that $\tau>\tau_0\equiv r_0\ln{e^{b/2}+1\over e^{b/2}-1}$. In fact
this is also a sufficient condition, which can be shown by the
conformal mapping between region ${\cal F}$ in the multiply connected de
Sitter space and region F-O in Misner space defined by
Eqs.~(\ref{E59a}-\ref{E61}). If the collision of two light cones in F
occurs beyond the hyperbola $t^2-x^2-y^2-z^2=1~(t>0)$ in Misner space (i.e., in the
region O), the corresponding
two light cones (and thus the bubbles formed inside these light cones) 
in ${\cal F}$ will never collide because
$t^2-x^2-y^2-z^2=1$ in F corresponds to $\tau\rightarrow\infty$ in
${\cal F}$. It is easy to show that the condition for a light cone not
to collide with its images within F-O is that
$e^b(t^2-x^2)-y^2-z^2>1$, where $(t,x,y,z)$ is the event where the
light cone originates. By Eq.~(\ref{E59a}) this condition corresponds to
$e^b[({{\tilde t}\over r_0})^2-1]>1+({{\tilde t}\over
r_0})^2-2{{\tilde t}\over r_0}\cos\theta$. Since ${\tilde t}>r_0$ and
$-1\leq\cos\theta\leq1$, a {\em sufficient} condition is
$e^b[({{\tilde t}\over r_0})^2-1]>1+({{\tilde t}\over
r_0})^2+2{{\tilde t}\over r_0}$, i.e. $e^b({{\tilde t}\over r_0}-1)>
{{\tilde t}\over r_0}+1$ which is equivalent to $\tau>\tau_0= 
r_0\ln{e^{b/2}+1\over e^{b/2}-1}$. Therefore all bubbles formed after
the epoch
$\tau_0$ in ${\cal F}$ in the multiply connected de~Sitter space will
never collide with themselves. The $0<\tau<\tau_0$ part of the
fundamental cell in ${\cal F}$ has a finite four-volume
$V_{\rm II}=V_{\rm I}(\cosh^3{\tau_0\over r_0}-1)$. The total probability of
forming a bubble in $V_{\rm II}$ is $P_{\rm II}={4\over3}\pi b\epsilon(
\cosh^3{\tau_0\over r_0}-1)$. For $b=2\pi$ we have
$\tau_0\simeq0.086r_0$, $V_{\rm II}\simeq0.011V_{\rm II}$, and thus
$P_{\rm II}\simeq0.011P_{\rm I}$. 

For the case of $b=2\pi$, in order that there be less than a
$5\%$ chance that a bubble forms in $V_{\rm I}$ (and thus less than
$0.05\%$ chance in
$V_{\rm II}$), $\epsilon$ should be less than $2\times10^{-3}$. This should
be no problem because we expect that this tunneling probability
$\epsilon$ should be
exponentially small. Thus it would
not be surprising to find region ${\cal R}$ and region ${\cal F}$ for epochs
$0<\tau<\tau_0=0.086r_0$ clear of bubble formation events (and clear
of real particles), which is all we
require. 

Also note that there may be two epochs of inflation, one at the Planck
scale caused by $\langle T_{ab}\rangle_{\rm
ren}=-g_{ab}/960\pi^2r_0^4$ [Eq.~(\ref{E75})] which later decays in region
${\cal F}$ at
$\tau\gg\tau_0$ into an inflationary metastable state at the GUT scale
produced by a potential $V(\phi)$, which, still later, forms bubble universes.

\section{Baby Universe Models}
Inflationary universes can lead to the formation of baby universes
in several different scenarios. If one of these baby universes simply
turns out to be the original universe that one started out with, we
have a multiply connected solution in many ways similar to our
multiply connected de~Sitter space. There would be a multiply
connected region of CTCs bounded by a past
Cauchy horizon which would be stable because of the self-consistency
requirement as in the previous section, and this would also engender
pure retarded potentials. Thus, in a wide class of scenarios,
the epoch of CTCs would be long over by now,
as we would be one of the many later-formed bubble universes.
Also, the model might either be geodesically complete to the past
or not. This might not be a problem in physics since we would
in any case have a periodic boundary condition; and because
with its pure retarded potentials, no causal signals could be
propagated to the past in any case. There are several
different baby universe scenarios --- any one of which could
accommodate our type of model.

First, there is the Farhi, Guth, and Guven \cite{far90} method of creation of
baby universes in the lab. At late times in an open
universe, for example, an advanced civilization might implode a mass
(interestingly, it does not have to be a large mass --- a few kilograms
will do) with enough energy to drive it up to the GUT energy scale,
whereupon it might settle into a metastable vacuum, creating a small
spherical bubble of false vacuum with a $V=\Lambda/8\pi$
metastable vacuum inside. This could be done either by just driving
the region up over the potential barrier, or by going close to the
barrier and tunneling through. The inside of this vacuum bubble would
contain a positive cosmological constant with a positive energy
density and a negative pressure. This bubble could be created
with an initial kinetic energy of expansion with the bubble wall
moving outward. But the negative pressure would pull it inward, and it
would eventually reach a point of maximum expansion (a classical
turning point), after which it would start to collapse and would form
a black hole. But occasionally, (probability $P=10^{-10^{18}}$ for
typical GUT scales \cite{far90}) when it reaches its
point of maximum expansion it tunnels to a state of equal energy but a
different geometry, like a doorknob, crossing the
Einstein-Rosen bridge \cite{ein35}. The ``knob'' itself would be the the interior
of the bubble, containing the positive cosmological constant, and
sitting in the metastable vacuum state with
$V=\Lambda/8\pi$. The ``knob'' consists of more than a
hemisphere of an initially static $S^3$ closed de~Sitter universe, where the
bubble wall is a surface of constant ``latitude'' on this sphere. At
the wall, the circumferential radius is thus decreasing as one moves
outward toward the external spacetime. Just outside the wall is the
Einstein-Rosen neck which reaches a minimum circumferential radius at
$r=2M$, and then the circumference increases to join the
open external solution. This ``doorknob'' solution
then evolves classically. The knob inflates to form a de~Sitter space
of eventually infinite size. It is connected to the original spacetime
by the narrow Einstein-Rosen bridge. But an observer sitting at $r=2M$
in the Einstein-Rosen bridge will shortly hit a singularity in the
future, just as in the Schwarzschild solution. So the connection only
lasts for a short time. The interior of the ``knob'' is hidden from an
observer in the external spacetime by an event horizon at
$r=2M$. Eventually the black hole evaporates via Hawking radiation \cite{haw75},
leaving a flat external spacetime (actually part of an open Big Bang
universe) with simply a coordinate singularity at $r=0$ as seen from
outside. (See Fig.~\ref{f7}.)

From the point of view of an observer sitting at the center of
$V=\Lambda/8\pi$ bubble, he would see himself, just after the
tunneling event, as sitting in a de~Sitter space that was initially
static but which starts to inflate. Centered on this observer's
antipodal point in de~Sitter space, he would see a bubble of
ordinary $V=0$ vacuum surrounding a black hole of mass
$M$. The observer sees his side of the Einstein-Rosen bridge and an
event horizon at $r=2M$ which hides the external
spacetime at late times from him. From the point of view of the de~Sitter observer,
the black hole also evaporates by Hawking radiation, eventually
leaving an empty $V=0$ bubble in an ever-expanding de
Sitter space. This infinitely expanding de~Sitter space, which
begins expanding at the tunneling event, is a perfect starting point (just
like Vilenkin's tunneling universe) for making an infinite number
of bubble universes, as this de~Sitter space has a finite beginning and
then expands forever. Now
suppose {\em one} of these open bubble universes simply turns out
to be the {\em original} open universe where that advanced
civilization made the baby de~Sitter universe 
in the first place (Fig.~\ref{f7}). Now
the model is multiply connected, with no earliest event. There is a
Cauchy horizon (${\cal CH}$, see Fig.~\ref{f7}) separating the region of CTCs from
the later region that does not contain them. This Cauchy horizon is
generated by ingoing closed null geodesics that represent signals that
could be sent toward the black hole, which then tunnel 
across the Euclidean tunneling
section jumping across the Einstein-Rosen bridge 
and then continuing as ingoing signals to enter the de~Sitter
space and reach the open single bubble in the de~Sitter space (that
turns out to be the original bubble in which the tunneling event
occurs). A retarded photon traveling around one of those closed null
geodesics will be red-shifted more and more on each cycle, thus not
causing an instability. Another novel effect is that although these
null generators are converging just before the tunneling event, they
are diverging just after the tunneling event, having jumped to the
other side of the Einstein-Rosen bridge. Thus, converging rays are
turned into diverging rays (as in the wormhole solution) during the
tunneling event without violating the weak energy condition. 
These closed null geodesics need not be
infinitely extendible in affine distance toward the past. It would
seem that it can be
arranged that the renormalized energy-momentum tensor does not blow up
on this Cauchy horizon so that a self-consistent solution is
possible. Using the method of images, note that the
$N$-th image is from $N$ cycles around the multiply connected
spacetime. The path connecting an observer to the $N$-th image will
have to travel $N$ times through the hot Big Bang phase which occurs in the
open bubble after the false vacuum with $V(\phi<\phi_0)=\Lambda/8\pi$
dumps its false vacuum energy into thermal radiation as it falls off
the plateau and reaches the true vacuum $V(\phi=\phi_0)=0$. Thus, to
reach the $N$-th image one has to pass through the hot optically thick
thermal radiation of the hot Big Bang $N$ times. And this will cause
the contribution of the $N$-th image to the renormalized energy-momentum
tensor to be exponentially damped by a factor $e^{-N\tau}$ where
$\tau\equiv nz\sigma_t\gg 1$ (where $n$ is the number density of target
particles, $z$ is the thickness of hot material, $\sigma_t$ is the
total cross-section). Li \cite{li96} has calculated the renormalized
energy-momentum tensor of vacuum polarization with the effect of
absorption. Li
\cite{li96} has estimated the fluctuation of the metric of the
background spacetime caused by vacuum polarization with absorption,
which is a small number in most cases. If the absorption is caused by
electron-positron pair production by a photon in a
photon-electron collision, the maximum value of the metric fluctuation
is $(\delta g_{\mu\nu})_{\rm max}\sim l_{\rm P}^2/(r_{\rm e}L)$, where
$l_{\rm P}$ is the Planck length, $r_{\rm e}$ is the classical radius
of electron, $L$ is the spatial distance between the identified points
in the frame of rest relative to the absorber \cite{li96}. If we take
$L$ to be the Hubble radius at the recombination epoch
($\sim10^{23}$cm), we have $(\delta g_{\mu\nu})_{\rm max}\sim10^{-76}$.
Thus, we expect that the renormalized energy-momentum
tensor will not blow up at the Cauchy horizon \cite{li96}, so that a
self-consistent solution is possible. 

The tunneling event is shown as the epoch indicated by the dashed line
in Fig.~\ref{f7}. During the tunneling event, the trajectory may be approximated as a
classical space with four spacelike dimensions solving Einstein's
equations, with the potential inverted, so that this Euclidean section
bridges the gap between the two classical turning points. (In such a
case, the concepts of CTCs and closed null curves should be
generalized to contain a spacelike interval. Thus, there are neither
closed null geodesics nor closed timelike geodesics with the
traditional definitions. According to Li \cite{li94}, this kind of
spacetime can be stable against vacuum polarization.)  

As Farhi, Guth, and Guven \cite{far90} note, the probability for forming such
a universe is exponentially small, so an exponentially large number
of trials would be required before an intelligent civilization would
achieve this feat. If the metastable vacuum is at the Planck
density, the number of trials required is expected to be not too large;
but if it is at the GUT density which turns out to
be many orders of magnitude lower than the Planck density, then
the number of trials becomes truly formidable ($P\sim10^{-10^{18}}$) 
\cite{far90}. Thus, Farhi,
Guth, and Guven \cite{far90} guess that it is unlikely that the human race
will ever succeed in making such a universe in the lab at the GUT scale. Gott
\cite{got93}, applying the Copernican principle to estimate our future
prospects, would come to similar conclusions. However,
if our universe is open, it has an infinite number of galaxies,
and it would likely have some super-civilizations powerful enough
to succeed at such a creation event, or at least have so many
super-civilizations (an infinite number) that even if they each
tried only a few times, then some of them (again an infinite
number) would succeed. In fact, if the probability for a
civilization to form on a habitable planet like the Earth and
eventually succeed at creating a universe in the lab is some
finite number greater than zero (even if it is very low), then
our universe (if it is an open bubble universe) should spawn an
infinite number of such baby universes.

This notion has caused Harrison \cite{har95} to speculate that our universe was
created in this way in the lab by some super-civilization in a previous
universe. He noted
correctly that if super-civilizations in a universe can create many
baby universes, then baby universes created in this way should greatly
outnumber the parent universes, and that you (being not special) are
simply likely to live in one of the many baby universes, because there
are so many more of them. Here he is using implicitly the formulation
of Gott \cite{got93} that according to the Copernican principle, out of all
the places for intelligent observers to be, there are, by definition,
only a few special places and many non-special places, and you are
simply more likely to be in one of the many non-special places. Thus,
if there are many baby universes created by intelligent
supercivilizations in an infinite open bubble universe, then you are
likely to live in a baby universe created in this way. Harrison uses
this idea to explain the strong anthropic principle. The strong
anthropic principle as advanced by Carter \cite{car74} says that the laws of
physics, in our universe at least, must be such as to allow the
development of the intelligent life. Why? Because we are here. It is
just a self-consistency argument. This might lead some to believe,
particularly with inflationary cosmologies that are capable of
producing an infinite number of bubble universes, that these different
universes might develop with many different laws of physics, given a
complicated, many-dimensional inflationary potential with many
different minima, and many different low energy laws of physics. If
some of these did not allow the development of intelligent life and
some of these did, well, which type of universe would you expect to
find yourself in? --- one that allowed intelligent observers, of
course. (By the same argument, you are not surprised to find yourself
on a habitable planet --- Earth --- although such habitable planets may
well be outnumbered by uninhabitable ones --- Mercury, Venus, Pluto, etc.)
Thus, there may be many more universes that have laws of physics that
do not allow intelligent life --- you just would not find yourself
living there. It has been noticed that there are various coincidences
in the physical constants --- like the numerical value of the fine
structure constant, or the ratio of the electron to proton mass, or the
energy levels in the carbon nucleus --- which, if they were very
different, would make intelligent life either impossible, or much less
likely. If we observe such a coincidence, according to Carter \cite{car74}, it
simply means that if it were otherwise, we would not be here. Harrison \cite{har95}
has noted that if intelligent civilizations made baby universes they
might well, by intelligent choice, make universes that purposely had
such coincidences in them in order to foster the development of
intelligent life in the baby universes they created. If that were the
case, then the majority of universes would have laws of physics
conducive to the formation of intelligent life. In this case, the
reason that we observe such coincidences is that a previous
intelligent civilization made them that way. One might even speculate
in this scenario that if they were smart enough, they could have left
us a message of sorts in these dimensionless numbers (a theme that
resonates, by the way, with part of Carl Sagan's thesis in {\em
Contact}). However, it is unclear whether any super-civilization
would be able to control the laws of physics in the
universes they created. All, they might reasonably be able to do would be
to drive the baby universe up into a particular metastable vacuum \cite{far90}. But
then, such a metastable vacuum inflates in the knob, and an infinite
number of bubble universes form later, with perhaps many different
laws of physics depending on how they tunnel away from the metastable
vacuum and which of the many potential minima they roll down
into. Controlling these
phase transitions would seem difficult. Thus, it would
seem difficult for the super-civilization that made the metastable
state that later gave rise to our universe to have been able to
manipulate the physical constants in our universe. Harrison's model
could occur in many generations, making it likely that we were
produced as great, great, ..., great grandchildren universes from a
sequence of intelligent civilizations. Harrison \cite{har95} was able to explain
all the universes by this mechanism except for the first one! For
that, he had to rely on natural mechanisms. This seems to be an
unfortunate gap. In our scenario, suppose that ``first'' universe
simply turned out to be one of the infinite ones formed later by
intelligent civilizations. Then the Universe --- note capital U ---
would be multiply connected, and would have a region of
CTCs; all of the individual universes would owe their birth
to some intelligent civilization in particular in this picture.
 
All this may overestimate the importance of intelligent
civilizations. It may be that bubbles
of inflating metastable vacuum are simply produced at late times in
any Big Bang cosmology by natural processes, and that baby universes
produced by natural processes may vastly outnumber those produced by
intelligent civilizations. Such a mechanism has been considered by
Frolov, Markov, and Mukhanov \cite{fro90}. They considered the hypothesis
that spacetime curvature is limited by quantum mechanics and that as
this limit is approached, the curvature approaches that of de~Sitter
space. Then, as any black hole collapses, the curvature increases as
the singularity is approached; but before getting there it will convert
into a collapsing de~Sitter solution. This can be done in detail in
the following way. Inside the horizon, but outside the collapsing star
the geometry becomes Schwarzschild which is a radially collapsing but
stretching cylinder. This can be matched onto a radially collapsing
and radially shrinking cylinder in de~Sitter space as described by the
metric in Eq.~(\ref{E50}) with the
time $\tau$ being negative
and the coordinate $l$ being unbounded rather than
periodic. Both surfaces are cylinders with identical intrinsic
curvature, but with different extrinsic curvature. This mismatch is
cured by introducing a shell of matter which converts the stretching
of the Schwarzschild cylinder to collapsing as well which then
matches onto the collapsing de~Sitter solution. This phase
transition may occur in segments which then merge as noted by Barabes
and Frolov \cite{bar96a,bar96b}. 
The de~Sitter solution then bounces and becomes an expanding de
Sitter solution which can in turn spawn an infinite number of open
bubble universes. This all happens behind the event horizon of the
black hole. Within the de~Sitter phase, one finds a Cauchy horizon
like the interior Cauchy horizon of the Reisner-Nordstrom solution,
but this inner Cauchy horizon is not unstable because the curvature is
bounded by the de~Sitter value so the curvature is not allowed to blow
up on the inner horizon. (This is an argument that one could also rely on to
produce self-consistent multiply connected de~Sitter phases with
CTCs --- if needed.) This model thus produces,
inside the black hole, to the future, and behind the event horizon, an
expanding de~Sitter phase that has a beginning, just like
Vilenkin's tunneling universe. If one of those bubble universes
simply turns out to be the original one in which the black hole
formed, then the solution is multiply connected with a region of
CTCs. This would make every black hole produce an
infinite number of universes. This would be the dominant mechanism for
making new bubble universes, since the number of black holes in our
universe would appear to greatly outnumber the number of baby
universes ever produced by intelligent civilizations, since the
tunneling probability for that process to succeed is exceedingly
small. 

Smolin \cite{smo92a,smo97} has proposed that this type of mechanism works and
furthermore that the laws of physics (in the bubble universes) are
like those in our own but with small variations. Then, there would be
a Darwinian evolution of universes. Universes that produced many black
holes would have more children that would inherit their
characteristics --- with some small variations. Soon, most universes
would have laws of physics that were fine-tuned to produce the maximum
number of black holes. Smolin \cite{smo92a,smo97} 
points out that this theory is testable,
since we can calculate whether small changes in the physical constants
would decrease the number of black holes formed. In this picture we
should be near a global maximum in the black hole production rate. One
problem is that the laws of physics that maximize the number of black
holes and those that simply maximize the number of main sequence stars
may be rather similar, and the laws that maximize the number of main
sequence stars might well simply maximize the number of intelligent
observers, and the anthropic principle alone would suggest a
preference for us observing such laws, even if no baby universes were created in
black holes. Another possible problem with this model, pointed out by
Rothman and Ellis \cite{rot93}, 
is that if the density fluctuations in the early universe had been
higher in amplitude, this would form many tiny primordial black holes
(presumably more black holes per comoving volume than in our
universe), so, we well might wonder why the density fluctuations in
our universe were so small. One way out might be that tiny black holes
do not form any baby universes, but this seems a bit forced since the de
Sitter neck formed can be as small as the Planck scale or GUT scale and it
would seem that even primordial black holes could be large enough to
produce an infinite number of open bubble universes. 

Another possibility is the recycling universe of Garriga and 
Vilenkin \cite{gar97}. In
this model there is a metastable vacuum with cosmological constant
$\Lambda_1$, and a true lowest energy vacuum with a cosmological
constant $\Lambda_2$. $\Lambda_1$ is at the GUT or Planck energy scale,
while $\Lambda_2$ is taken to be the present value of $\Lambda$ (as
might be the case in a flat-$\Lambda$ model). As long as $\Lambda_2>0$,
then Garriga and Vilenkin assert that there is a finite (but small) probability
per unit four volume that the $\Lambda_2$ state could tunnel to form a
bubble of $\Lambda_1$ state, which could therefore inflate, decaying
into bubbles of $\Lambda_2$ vacuum, which could recycle forming
$\Lambda_1$ bubbles, and so forth. They point out that depending on the
coordinate system, a bubble of $\Lambda_2$ forming inside a
$\Lambda_1$ universe could also be seen as a $\Lambda_1$ bubble
forming inside of a $\Lambda_2$ universe. Take two de~Sitter spaces,
one with $\Lambda_1$ and one with $\Lambda_2$, and cut each along a
vertical slice ($W=W_0$) in the embedding space. They can
then be joined along an appropriate hyperbola of one sheet
representing a bubble wall, with the $\Lambda_2$ universe lying to the
$W<W_0$ side and the $\Lambda_1$ universe lying to the
$X>W_0$ side. Slicing along hyperplanes with $V+W={\rm
constant}$ gives a steady-state coordinate system for a $\Lambda_1$
universe in which a bubble of $\Lambda_2$ vacuum appears. Slicing along
hyperplanes with $V-W={\rm constant}$, however, gives a steady-state
coordinate system for a $\Lambda_2$ universe in which a bubble of
$\Lambda_1$ appears. So, one can find a steady-state coordinate system
in which there is a $\Lambda_1$ universe, with bubbles of $\Lambda_2$
inside it, and bubbles of $\Lambda_1$ inside these $\Lambda_2$
bubbles, and so forth. If the roll down is slow, within the $\Lambda_2$
bubble as it forms, as in Gott's open bubble universe \cite{got82}, then it
will have at least 67 $e$-folds of inflation with
$\Lambda\simeq\Lambda_1$ before it falls off the plateau into the
absolute minimum at $\Lambda_2$, and this will be an acceptable
Big Bang model which will have the usual Big Bang properties except
that it will eventually be dominated by a lambda term
$\Lambda_2$. Being bubble universes, they will all be open with
negative curvature as in Gott's model \cite{got82} but they will be
asymptotically open de~Sitter models at late times with
$a(t)=r_0\sinh(t/r_0)$ and $\Lambda=\Lambda_2$. Garriga and Vilenkin 
\cite{gar97} wondered whether such a recycling
model could be geodesically complete toward the
past. Such a outcome, they pointed out, would violate no known theorems
and should be investigated. They hoped to find such a
geodesically-complete-to-the-past model so that it
could be eternal without a need for a beginning. However, in the special
case, where $\Lambda_1=\Lambda_2$, one can show that the recycling
steady state solution becomes a simple single de~Sitter space geometry
with $\Lambda_1$ and the usual steady-state coordinate system in a
single de~Sitter space is not geodesically complete to the past.

Now take this recycling model where it turns out that one of the
$\Lambda_1$ bubbles formed inside an $\Lambda_2$ bubble inside a
$\Lambda_1$ region is, in fact, the $\Lambda_1$ region that one
started out with. In this case, we would have a multiply connected
model such as we are proposing which would include a region of CTCs
(Fig.~\ref{f8}). (If $\Lambda_1=\Lambda_2$, this model is just
the multiply connected de~Sitter space we have considered.) 
If our multiply connected model was geodesically complete to the past,
so would the covering space (a simply connected Garriga-Vilenkin
model) be. If our multiply connected model was geodesically 
incomplete to the past, so would the covering space (a simply 
connected Garriga-Vilenkin model) be also. In
our model, there would be a strong self-consistency reason for pure
retarded potential, whereas in the Garriga-Vilenkin recycling model, there
would be no such strong reason for it. With pure retarded potentials
throughout, the issue of whether the spacetime was geodesically
complete to the past is less compelling, as we have argued above, and
our model, having a periodic boundary condition, would not need
further boundary conditions, unlike a simply connected recycling model
that was geodesically incomplete to the past. 

Thus, there are a number of models in which baby universes are created
which can be converted into models in which the Universe creates
itself, if one of those created baby universes turns out to be the
original universe that one started with. Since these models are all
ones in which there are an infinite number of baby universes created,
this multiply connected outcome must occur unless the probability for
a particular multiple connectivity to exist is exactly zero. In other words,
it should occur, unless it is forbidden by the laws of physics. Given
quantum mechanics, it would seem that such multiple connectivities would
not be absolutely forbidden, particularly in the Planck foam era.   

We should note here that, in principle, there might even be solutions
that are simply connected in which there was an early region of CTCs
bounded to the future by a Cauchy horizon followed by
an inflationary region giving rise to an infinite number of bubble
universes. The models considered so far have all obeyed the weak
energy condition, and these models have all been multiply connected; in
other words, they have a genus of $1$, like a donut, since one of the
later baby universes is connected with the original one. Consider an
asymptotically flat spacetime with two connected wormhole mouths that
are widely separated. The existence of the wormhole connection
increases the genus by one. Instead of a flat plane, it becomes
a flat plane with a handle. To do this, the wormhole solution
must violate the weak energy condition \cite{mor88}. 
It must have some negative
energy density material, for it is a diverging lens
(converging light rays entering one wormhole mouth, diverge upon
exiting the other mouth). For a compact two dimensional surface,
the integrated Gaussian curvature over the surface divided by
$4\pi$ is equal to $1$ minus the genus. Thus, the integrated
Gaussian curvature over a sphere (genus=0) is $4\pi$, while the
integrated Gaussian curvature over a donut (genus=1) is zero,  and the integrated
Gaussian curvature over a figure 8 pretzel (genus=2) is $-4\pi$. Negative
curvature is added each time the genus is increased. Conversely,
positive curvature can be added to reduce the genus by $1$. When a donut is
cut, so that it resembles a letter ``C'', the ends of the
letter ``C'' are sealed with positive curvature (two spherical
hemispherical caps would do the job, for example). Our solutions
are already multiply connected, so they might in principle
be made simply connected by the addition of some extra
positive mass density, without violating the weak energy
condition. An example of this
is seen by comparing Grant space \cite{gra93} with Gott's two-string
spacetime \cite{got91a}. Grant space is multiply connected, has 
$T_{ab}=0$ everywhere, and includes CTCs. It can be pictured as a
cylinder. Gott's two-string spacetime is simply connected, but is
identical to Grant space at large distances from the strings. It
also contains CTCs. It can be pictured as a
cardboard cylinder that has been stepped on and then stapled shut
at one end, like an envelope. There are two corners at the closed
end, representing the two strings, but the cylinder continues
outward forever toward its open end (so it is like a test tube,
a cylinder closed on one end). The two strings provide positive
energy density (i.e. they do not violate the weak energy condition).
CTCs that wrap around the two strings far out in
the cylinder (which is identical to a part of Grant space; see
Laurence \cite{lau94}) can be shrunken to points by slipping them through the
strings --- but they become spacelike curves during this process.
Thus, Gott space represents how a multiply connected spacetime with
CTCs (Grant space) can be converted into a simply
connected spacetime with CTCs by adding to the
solution material that obeys the weak energy condition. A similar
thing might in principle be possible with these cosmological models.
Since our multiply connected versions already obey the weak energy
condition, so would the associated simply connected versions.

\section{Conclusions}
The question of first-cause has been a troubling one for cosmology.
Often, this has been solved by postulating a universe that has
existed forever in the past. Big Bang models supposed that the
first-cause was a singularity, but questions about its almost, but
not quite, uniformity remained. Besides, the Big Bang singularity
just indicated a breakdown of classical general relativity, and 
with a proper theory-of-everything, one could perhaps push through to earlier times.
Inflation has solved some of these problems, but Borde and Vilenkin have shown that
if the initial inflationary state is metastable, then it must have
had a finite beginning also. Ultimately, the problem seems to be how to
create something out of nothing. 

So far, the best attempt at this
has been Vilenkin's tunneling from nothing model and the similar
Hartle-Hawking no-boundary proposal. 
Unfortunately, tunneling is, as the name suggests, usually a process that involves
tunneling from {\em one} classical state to {\em another}, thus, with
the Wheeler-DeWitt potential and ``energy'' $E=0$ that 
Hartle and Hawking adopted,
the Universe, we argue, should really start not as nothing 
but as an $S^3$ universe of radius
zero --- a point. A point is as close to nothing as one can get, but
it is not nothing. Also, how
could a point include the laws of physics? In quantum cosmology,
the wave function of the Universe is treated
as the solution of a Schr\"{o}dinger-like equation 
(the Wheeler-DeWitt equation), where the three-sphere $S^3$
radius $a$ is the abscissa and there is a potential $U(a)$ with 
a metastable minimum at $U(a=0)=0$, and a barrier with $U(a)>0$ for $0<a<a_0$, 
and $U(a)<0$ for $a>a_0$. Thus, the evolution can be seen as a particle, 
representing the universe, starting as a point, $a=0$, at the bottom 
of the metastable potential well, with $E=0$. Then it tunnels through
the barrier and emerges at $a=a_0$ with $E=0$, whereupon it becomes a 
classically inflating de~Sitter solution. It can then decay via the 
formation of open single bubble universes \cite{got82,got86}. The 
problem  with this model is that it ignores the ``zero-point-energy''. 
If there is a conformal scalar field $\phi$, then the ``energy'' levels should
be $E_n=n+{1\over2}$. Even for $n=0$ there is a ``zero-point-energy''. The 
potential makes the system behave like a harmonic oscillator in the 
potential well near $a=0$. A harmonic  oscillator cannot 
sit at the bottom of the potential well --- the uncertainty principle 
would not allow it. There must be some zero-point-energy and the 
particle must have some momentum, as it oscillates within the potential 
well when the field $\phi$ is included. Thus, when the
``zero-point-energy'' is considered, we see that the 
initial state is not a point but a tiny oscillating ($0\leq 
a\leq a_1$) Big Bang
universe, that oscillates between Big Bangs and Big Crunches (though
the singularities at the Big Bangs and Big Crunches might be smeared by
quantum effects). This is
the initial classical state from which the tunneling occurs. It is 
metastable, so this oscillating universe could not have existed forever:
after a finite half-life,  it is likely to decay. It reaches maximum 
radius $a_1$, and then tunnels to a classical de~Sitter state at minimum
radius $a_2$ where $a_2<a_0$. 
The original oscillating universe could have formed by a similar 
tunneling process from a contracting de~Sitter phase, but such a
phase would have been much more likely to have simply classically 
bounced to an expanding de~Sitter phase instead of tunneling into the
oscillating metastable state at the origin. In this
case, if one found oneself in an 
expanding de~Sitter phase, it would be much more likely that it was 
the result of classical bounce from a contracting de~Sitter phase, 
rather than the result of  a contracting de~Sitter phase that had 
tunneled to an oscillating phase and then back out to an expanding
de~Sitter phase. Besides, a contracting de~Sitter phase would be 
destroyed by the formation of bubbles which would percolate  
before the minimum radius was ever reached.

In this paper, we consider instead the notion that the Universe did not arise
out of nothing, but rather created itself. One of the
remarkable properties
of the theory of general relativity is that in principle it allows
solutions with CTCs. Why not apply
this to the problem of the first-cause? Usually the beginning of the
Universe is viewed like the south pole. Asking what is before that is
like asking what is south of the south pole, it is said. But as we have
seen, there remain unresolved problems with this model. If instead
there were a region of CTCs in the early universe,
then asking what was the earliest point in the Universe would be like
asking what is the easternmost point on the Earth. There is no
easternmost point --- you can continue going east around and around
the Earth. Every point has points that are to the east of it. If
the Universe contained an early region of CTCs,
there would be no first-cause. Every event would have events to its
past. And yet the Universe would not have existed eternally in the past
(see Fig.~\ref{f1}). Thus, one of the most remarkable properties of general
relativity --- the ability in principle to allow CTCs
--- would be called upon to solve one of the most perplexing
problems in cosmology. Such an early region of CTCs
could well be over by now, being bounded to the future by a Cauchy
horizon. We construct some examples to show that vacuum states can be
found such that the renormalized energy-momentum tensor does not blow
up as one approaches the Cauchy horizon. For such a model to work the Universe has
to reproduce at some later time 
the same conditions that obtained at an earlier time. Inflation is
particularly useful in this regard, for starting with a tiny piece
of inflating state, at later times a huge volume of inflating state
is produced, little pieces of which look just like the one we started
with. Many inflationary models allow creation of baby inflationary universes
inside black holes, either by tunneling across the Einstein-Rosen
bridge, or by formation as one approaches the singularity. If one of
these baby universes simply turns out to be the universe we started
with, then a multiply connected model with early CTCs
bounded by a Cauchy horizon is produced. Since any closed null
geodesics generating the Cauchy horizon must circulate through the
optically thick region of the hot Big Bang phase of the universe after
the inflation has stopped, the renormalized energy-momentum tensor
should not blow up as the Cauchy horizon is approached.

As a particularly simple example we consider a multiply connected de~Sitter
solution where events E$_i$ are topologically identified with events
E$^\prime_i$ that lie inside these future light cones
via a boost transformation. If the boost $b=2\pi$, we show that we
can find a Rindler-type vacuum where the renormalized energy-momentum
tensor does not blow up as the Cauchy horizon is approached but rather produces a
cosmological constant throughout the spacetime which self-consistently
solves Einstein's equations for this geometry. Thus, it is possible to
find self-consistent solutions. When analyzing classical fields in this model,
the only self-consistent solution without a blow up as the Cauchy
horizon is approached
occurs when there is a pure retarded potential in the causally
connected region of the model. Thus, the multiply connected nature of
this model and the possibility of waves running into themselves, ensure
the creation of an arrow of time in this model. This is a remarkable
property of this model. Interestingly, this model, although having no
earliest event and having some timelike geodesics that are infinitely
extendible to the past, is nevertheless geodesically incomplete to the
past. This is not a property we should have thought desirable, but
since pure retarded potentials are established automatically in this
model, there are no waves propagating to the past and so there may be
no problem in physics with this, since there are never any waves that
run off the edge of the spacetime. 
The region of CTCs has a
finite four-volume equal to $4\pi br_0^4/3$ and should be in a
pure vacuum state containing no real particles or Hawking
radiation and no bubbles. After
the Cauchy horizon for a certain amount of proper time (depending on
the bubble formation probability per four volume $r_0^4$) no bubbles
(or real particles) form, but eventually this model expands 
to infinite volume, creating
an infinite number of open bubble universes, which do not percolate.
At late times in the de~Sitter phase a particle detector would find
the usual Hawking radiation just as in the usual vacuum for de~Sitter
space.

There are a number of problems to be solved in this model. The chronology 
projection  conjecture proposes that the laws of physics conspire so as to 
prevent the formation of CTCs. 
This conjecture was motivated by Hiscock and Konkowski's 
result that the energy-momentum tensor of the 
adapted Minkowski vacuum in Misner space diverges as 
the Cauchy horizon is approached. But as we have shown \cite{li97}, the adapted 
Rindler vacuum for Misner space has $\langle T_{ab}\rangle_{\rm ren}=0$ 
throughout the space if $b=2\pi$; thus, 
this is a self-consistent vacuum for this spacetime since it solves 
Einstein's equations for this geometry. It's true that $\langle T_{ab}
\rangle_{\rm ren}$ remains 
formally ill-defined on the Cauchy horizon itself [$\xi=0$ in
Eq.~(\ref{E69}) with $b=2\pi$],  a set of measure zero. 
But it is not clear that this  creates a problem for physics, since 
continuity might require that this formally ill-defined quantity be 
defined to be zero on this set of measure zero as well, since it 
is zero everywhere else. In fact, a treatment in the Euclidean section
shows this is the case, for in the Euclidean section, if $b=2\pi$,
$\langle T_{ab}\rangle_{\rm ren}=0$ everywhere, including at $\xi=0$.
Other counter-examples to the chronology
protection conjecture have also been found, as discussed in section
V. Hawking himself has also admitted that the back-reaction of vacuum
polarization does not enforce the chronology protection conjecture.

One of the remarkable 
properties of general relativity is that it allows, in principle, the 
formation of event horizons. This appears to be realized in the case 
of black holes. Just as black hole theory introduced singularities at 
the end, standard Big Bang cosmology introduced singularities at the 
beginning of the universe. Now, with inflation, we see that event 
horizons should exist in the early  universe as well \cite{got82}. 
Inflationary ideas prompt the 
suggestion that baby universes may be born. If one of the baby universes 
simply turns out to be the one we started with, then we get a model 
with an epoch of CTCs that is over by now, bounded toward 
the future by a Cauchy horizon. We have argued that the divergence of the 
energy-momentum tensor as one approaches the Cauchy horizon 
does not necessarily occur, particularly when the Cauchy horizon crosses 
through a hot Big Bang phase where absorption occurs. 

If the energy-momentum
tensor does not diverge as the Cauchy horizon is approached, 
other problems must still be tackled. The classical instability of a Cauchy 
horizon to the future (a future chronology horizon) 
in a spacetime with CTCs 
is one. But this problem is solved in a world with retarded potentials for 
a Cauchy horizon that occurs to our past 
(a past chronology horizon) and which ends an epoch of 
CTCs. It thus seems easier to have a Cauchy horizon 
in the early universe. At the microscopic level, quantum 
mechanics appears to allow acausal behavior. Indeed the creation and annihilation 
of a virtual positron-electron pair can be viewed as creation of a small 
closed loop, where the electron traveling backward in time to complete 
the loop appears as a positron. So, why should the laws of physics forbid 
time travel globally? Indeed one of the most remarkable properties of the 
laws of physics is that although they are time (CPT) symmetric, the 
solutions we observe have an arrow of time and retarded potentials.  
Without this feature of the solutions, acausal behavior 
would be seen all the time. Interestingly, in our model, the multiply 
connected nature of the 
spacetime geometry forces an arrow of time and retarded potentials. 
Thus, it is the very presence of the initial region of CTCs 
that produces the strong causality that we observe later on. 
This is a very interesting and unexpected property. An entropy arrow of 
time is automatically produced as well, with the region of CTCs
in the simplest models sitting automatically in a cold 
vacuum state, with the universe becoming heated after the Cauchy 
horizon. Recently, Cassidy and Hawking \cite{cas97b} have proposed yet another 
supposed difficulty for CTCs, in that the formally defined entropy appears 
to diverge to negative infinity as the Cauchy horizon is approached. 
Yet, in the early universe this may turn out to be an advantage, since 
to produce the ordinary entropy arrow of time we observe in the universe 
today, we must necessarily have some kind of natural low-entropy 
boundary condition in the early universe \cite{pen79,pen89}. 
This could occur on the Cauchy 
horizon that ends the period of CTCs.

New objections to spacetimes with CTCs can continue 
to surface, as old problems are put to rest, so it might seem that disproving 
the chronology protection conjecture would be a tall order. But, proving 
that there are no exceptions to the chronology protection conjecture, ever, 
would seem an equally daunting task. This is particularly true since we 
currently do not have either a theory of quantum gravity or a 
theory-of-everything.

Perhaps the most obvious problem with the model we have proposed is that 
the simplest solutions we have obtained so far are not geodesically 
complete to the past. But we may need no boundary condition since we 
have a periodic boundary condition instead. This thus may not be a 
problem in physics if retarded potentials are the only ones allowed. 
Alternatively, as Garriga and Vilenkin have indicated, it would violate no known 
theorems for some type of recycling universe (making bubble universes 
within bubble universes {\em ad infinitum}) to exist that was geodesically 
complete to the past. If such solutions exist, it might be possible to
find a solution in which there was an early epoch of CTCs
that would be geodesically 
complete to the past as well by simply identifying an 
earlier bubble with a later one. 

Thus, a number of important questions remain, and we would not minimize 
them. The models presented here, however, do have some interesting and 
attractive properties, suggesting that this {\em type} of model should be 
investigated further, and that we {\em should} ask the question: 

{\em Do the laws of physics prevent the Universe from being its own mother?}
  
\acknowledgments
This research was supported by NSF grant AST95-29120 and NASA grant NAG5-2759.

\begin{figure}
\caption{A self-creating Universe scenario. Four inflating baby universes 
are shown --- A, B, C, and D --- from left to right. Universe A and D 
have not created any baby universes so far. Universe C has created 
universe D. Universe B has created three universes: universe A, universe C 
and itself. The toroidal --- shaped region at the bottom is a region
of CTCs (closed timelike curves).
The region is bounded to the future by a 
Cauchy horizon, after which, there are no CTCs. Universes A, C, and D, for 
example, are formed after the Cauchy horizon when the epoch of CTCs
is already over.}\label{f1}
\end{figure}

\begin{figure}
\caption{The potential function in the Wheeler-DeWitt equation in the
minisuperspace model. The horizontal axis is the scale factor of the
universe. If the conformally coupled scalar field is in the ground
state, it has a ``zero-point-energy'' $1/2$. If this
``zero-point-energy'' is considered, the quantum behavior of the
universe is like a particle of unit mass with total energy $1/2$
moving in the potential $U(a)$. Regions $0<a<a_1$ and $a>a_2$ are
classically allowed, region $a_1<a<a_2$ is classically forbidden. The
left dark disk is a tiny radiation-dominated closed 
oscillating universe, which oscillates
between Big Bangs and Big Crunches. The smoothness of the potential at
$a=0$ may indicate that any Big Bang and Big Crunch singularities are removed by
quantum theory. This tiny oscillating universe has a small but
non-zero probability to tunnel through the barrier out to become a
de~Sitter-type inflating universe, which is represented by the dark
disk on the right. The circle inside the barrier is a Euclidean
bouncing space. If the ``zero-point-energy'' $1/2$ were neglected (as
Hartle and Hawking did), the left classically allowed region would
shrink to a point. The grey
disk represents a contracting and re-expanding de
Sitter universe. If the ``zero-point-energy'' is neglected, the
Universe could start out at the metastable minimum as a point
with $a=0$, tunneling through the barrier out to become a
de~Sitter universe. In this paper we argue that
we have no reason to neglect the ``zero-point-energy'' so that it is
the tiny oscillating universe initial state that applies.}\label{f2}
\end{figure}

\begin{figure}
\caption{(a) The solution of the Euclidean Einstein's equations
representing the tunneling regime (open circle) in Fig.~\ref{f2}. This
is a solution to the Euclidean Einstein's equations with a positive
cosmological constant and a conformally coupled scalar field in its
ground state. This is a Euclidean space bouncing between the state
with maximum radius $a_2$ and the state with (non-zero) 
minimum radius $a_1$. One ``copy'' of
this Euclidean bouncing solution 
is shown in this diagram, which has two boundaries with
minimum radius $a_1$. (b) This is the case when the ``zero-point-energy'' of
the conformally coupled scalar field is neglected, as Hartle and
Hawking did. In this case the minimum radius is zero, and thus one
copy of the bouncing Euclidean solution is a four-sphere. This 
four-sphere has no-boundary, which is the basis of Hawking's quantum
cosmology. But we argue that since the ``zero-point-energy'' of the
conformally coupled scalar field cannot be neglected, the true
solution should be that given by diagram 3a, which does not enforce
Hartle and Hawking's no-boundary proposal.}\label{f3}
\end{figure}

\begin{figure}
\caption{Penrose diagrams of our multiply connected de~Sitter
space mapped onto its universal covering space (de~Sitter 
space). Under a boost transformation, points with the same symbols
(squares, disks, triangles, or double-triangles) are identified. 
Our multiply connected de~Sitter space is
divided into four regions ${\cal R}$, ${\cal L}$, ${\cal F}$, and
${\cal P}$, which are separated by Cauchy horizons ${\cal CH}$. 
The shaded regions represent fundamental cells of the multiply
connected de~Sitter space. (Fig.~\ref{f4}a and Fig.~\ref{f4}b represent two
different choices of the fundamental cells, but they are equivalent.)
The fundamental cells ${\cal R}$ and ${\cal L}$ have a finite
four-volume, whereas the fundamental cells ${\cal F}$ and ${\cal P}$
(which extend infinitely to the future and the past, respectively)
have an infinite four-volume. In Fig.~\ref{f4}a the left and 
right boundaries of ${\cal F}$ are 
identified, likewise for ${\cal P}$; the upper and lower boundaries 
of ${\cal R}$ are identified, likewise for ${\cal L}$. In 
Fig.~\ref{f4}b region ${\cal F}+{\cal R}$ is partially
bounded by two null  surfaces, the lower one is the future light 
cone of an event E, and the upper one is the future future light 
cone of an event E$^\prime$ which is identified with E under the action of 
a boost. These two future light cones are identified creating a 
periodic boundary condition for the causally connected 
region ${\cal F}+{\cal R}$. 
${\cal R}$ and ${\cal F}$ are separated by a Cauchy horizon 
${\cal CH}$. Self-consistency 
(non-divergence of $T^{ab}T_{ab}$ as ${\cal CH}$ is approached) 
requires retarded potentials 
in ${\cal R}$ and ${\cal F}$. Region ${\cal P}+{\cal L}$ is 
partially bounded by the past light cone of an 
event F and the past light cone of of an event F$^\prime$ which is identified 
with F under the action of a boost. These two surfaces are identified 
creating a periodic boundary condition for ${\cal P}+{\cal L}$, 
where self-consistency as ${\cal CH}$ separating ${\cal P}$ from
${\cal L}$ is approached requires advanced potentials.
}\label{f4}
\end{figure}

\begin{figure}
\caption{With our adapted conformal Rindler vacuum, our multiply
connected de~Sitter space is cold (with zero temperature) in ${\cal R}$
and ${\cal L}$, but hot (with the Gibbons-Hawking temperature) in ${\cal
F}$ and ${\cal P}$. The arrows indicate the direction of
increasing entropy.}\label{f5}
\end{figure}

\begin{figure}
\caption{Self-consistency near the Cauchy horizons in a spacetime
with CTCs naturally gives rise to an arrow of time. 
Grey thick lines represent light
cones of electromagnetic waves or photons emitted from event E. 
(a) This diagram shows that in
${\cal F}$  the retarded potential is
self-consistent. The ``collision'' of an electromagnetic wave
with its images cannot destroy the Cauchy horizon, since the
proper time from the ``collision'' (event $p$) to the
origin is always bigger than the proper time from E to the
origin. Likewise the advanced potential in region ${\cal P}$ would not
destroy the Cauchy horizon. (b) This diagram shows that a retarded potential in ${\cal
R}$ and an advanced potential in ${\cal L}$ (or {\em vice versa}) are
self-consistent. But the potentials in ${\cal R}$ and ${\cal L}$
cannot be both retarded or both advanced, otherwise the ``collision''
of two waves from ${\cal R}$ and ${\cal L}$
respectively will destroy the Cauchy horizon. (c) This diagram shows
that the advanced potential in ${\cal F}$ is not self-consistent,
since the collision of an electromagnetic wave with its
images will destroy the Cauchy horizon. (as $n\rightarrow\pm\infty$,
the collision event $p$ approaches the Cauchy horizon.) (d) This
diagram shows that a part-advanced-and-part-retarded potential in ${\cal
R}$ (or ${\cal L}$) is also not self-consistent, as $T^{ab}T_{ab}$ would
also diverge as the Cauchy horizon is approached. }\label{f6}
\end{figure}

\begin{figure}
\caption{A schematic Penrose diagram of a self-creating Universe based
on the baby universe model of Farhi, Guth, and Guven. We identify $M_1N_1$
with $M_2N_2$, to obtain a model of the Universe creating itself.
($M_1N_1$ is the future light cone of event $N_1$.) In
this model the closed null curves generating the Cauchy horizon
(${\cal CH}$)
pass through a hot Big Bang region, where the dense absorber can
make the Cauchy horizon stable against vacuum polarization effects.
The metastable de~Sitter phase is shown in grey. It decays along a 
hyperboloid $H^3$ near the bottom of the figure to form a single open 
bubble universe with a hot Big Bang phase and an epoch of recombination 
which is also shown. After recombination a super-civilization creates, 
at the right, an expanding bubble of de~Sitter metastable vacuum. 
This reaches a point of maximum expansion at which point it tunnels 
to a doorknob-shaped configuration. The tunneling epoch is shown 
by the dashed line: just below the dashed line is how the spacetime 
appears just before tunneling, and just above the dashed line is how 
the spacetime appears just after the tunneling. Just after the 
tunneling, the geometry (just above the dashed line) from left 
to right goes from infinite radius (where future null infinity 
${\cal I}^+$ meets the dashed line) to a minimum radius $r=2M$ at the neck 
in the Einstein-Rosen bridge (where the inside and outside black 
hole event horizons meet just below the word ``Black Hole") then 
to a radius $r>2M$ at the surface of the de~Sitter bubble, 
reaching a maximum radius at the equator of the de~Sitter bubble 
``knob" and finally decreasing to $r=0$ at the center of the bubble 
at the extreme right. The de~Sitter bubble expands forever. 
$M_2$ is at $t=\infty$. To the left of $M_2$ is another open bubble 
universe forming out of the metastable de~Sitter vacuum. It 
is diamond-shaped --- the bottom two lines representing the 
expanding bubble wall and the top two lines representing 
future null infinity for that bubble. Within this bubble the de~Sitter
vacuum decays to a hot Big Bang phase along a hyperboloid $H^3$ shown
as a curved line crossing the diamond. Another open bubble 
universe forms to the right of $M_1$. Recall, $M_1 = M_2$. These 
two bubble universes both form after the Cauchy horizon 
${\cal CH}$ as do an infinite number of others. The black hole 
singularity is shown, as well as the fact that the black hole 
evaporates. $N_1N_2$ is a CTC; CTCs occur 
on the $N_1N_2$ side of the Cauchy horizon 
${\cal CH}$. After ${\cal CH}$, there are no CTCs.}\label{f7}
\end{figure}

\begin{figure}
\caption{A self-creating Universe model based on 
Garriga and Vilenkin's recycling Universe. 
In a region of cosmological constant $\Lambda_1$, a bubble $B$ of 
cosmological constant $\Lambda_2$ is formed by tunneling at the 
epoch $BB_1$. The expanding bubble wall is represented by 
$BB_2$. At a later time, within bubble $B$ a bubble $A$ forms 
at epoch $AA_1$ by tunneling. The expanding bubble wall is 
shown by $AA_2$. Inside bubble $A$ the cosmological constant is $\Lambda_1$.
In the limit where $\Lambda_1=\Lambda_2$ we can plot this in 
a single de~Sitter space. Now we identify the two
hypersurfaces denoted by
$A_1AA_2$ to obtain a model of the Universe creating
itself. The Cauchy horizon bounding the region of CTCs is indicated by
${\cal CH}$. After ${\cal CH}$ there are no CTCs. 
If $\Lambda_1=\Lambda_2$, this reduces to our multiply
connected de~Sitter model ${\cal F}+{\cal R}$ shown in Fig.~\ref{f4}b.}\label{f8}
\end{figure}

\end{document}